\documentclass[12pt,a4paper]{article}
\usepackage{elliptic}
\usepackage[dvipsnames]{xcolor}
\usepackage{amscd,amsmath,amssymb,amsfonts,xspace,mathrsfs,mathtools,scalefnt}
\usepackage[utf8]{inputenc}
\usepackage{bbm}
\usepackage{graphicx} 
\usepackage{tabu}
\usepackage[color,matrix,arrow]{xy}
\usepackage{wasysym}
\usepackage{stmaryrd}
\usepackage[size=footnotesize]{caption,subcaption}
\usepackage[textsize=tiny]{todonotes}

\newcommand{\cpo}{\mathbb P^1(\mathbb C)}
\newcommand{\jsw}{\mathcal J(u,\bfm,\Lambda)}

\newcommand{\lad}{\CL^{\text{AD}}} 
\newcommand{\ubp}{u_{\text{bp}}}
  
\newcommand{\bfmad}{\bfm_{\text{AD}}} 
\newcommand{\mad}{m_{\text{AD}}} 
\newcommand{\uad}{u_{\text{AD}}} 

\newcommand{\pma}{P^{\text{M}}}
\newcommand{\fd}{\backslash\mathbb H}

\newcommand{\tad}{\tau_{\text{AD}}}
\newcommand{\ord}{\text{ord}} 
\newcommand{\vol}{\text{vol}}

\newcommand{\ind}{\text{ind}\,} 
\newcommand{\wt}{\text{wt}\,}

\newcommand{\psl}{\text{PSL}(2,\mathbb Z)} 
\newcommand{\slz}{\text{SL}(2,\mathbb Z)}

\newcommand{\nstar}{\CN=2^*}

\newcommand{\jt}{\vartheta}

\newcommand{\be}{\begin{equation}} 
\newcommand{\ee}{\end{equation}} 
\newcommand{\bes}{\begin{equation*}}
\newcommand{\ees}{\end{equation*}}

\DeclareMathOperator{\im}{{\mathring \imath}}

\newcommand{\CB}{\mathcal{B}}  
\newcommand{\CC}{\mathcal{C}}

\newcommand{\CF}{\mathcal{F}}

\newcommand{\CI}{\mathcal{I}}
\newcommand{\CJ}{\mathcal{J}}

\newcommand{\CL}{\mathcal{L}} 
  
\newcommand{\CN}{\mathcal{N}}
\newcommand{\CO}{\mathcal{O}}

\newcommand{\CR}{\mathcal{R}}
\newcommand{\CS}{\mathcal{S}}
\newcommand{\CT}{\mathcal{T}}

\newcommand{\CW}{\mathcal{W}}

\newcommand{\BC}{\mathbb{C}}
\newcommand{\BH}{\mathbb{H}}

\newcommand{\BZ}{\mathbb{Z}}
\newcommand{\BQ}{\mathbb{Q}}

\newcommand{\bfm}{{\boldsymbol m}}

\newcommand{\Tr}{\text{Tr}}

\newcommand{\SL}{\text{SL}(2,\BZ)}

\newtheorem{theorem}{Theorem}

\title{Cutting and gluing with running couplings in $\CN=2$ QCD}

\abstract{ 
	
\vspace{30pt}\noindent\today
 
}

\author{Johannes Aspman$^a$, Elias Furrer$^b$, Jan Manschot$^c$ \\
{\it School of Mathematics, Trinity College, Dublin 2, Ireland\\
\it Hamilton Mathematical Institute, Trinity College, Dublin 2 \vspace{20pt}

 $^a$\href{mailto:aspmanj@maths.tcd.ie}{aspmanj@maths.tcd.ie}\\
 $^b$\href{mailto:furrere@maths.tcd.ie}{furrere@maths.tcd.ie}\\
 $^c$\href{mailto:manschot@maths.tcd.ie}{manschot@maths.tcd.ie}

}}

\abstract{We consider the order parameter $u=\left<{\rm Tr}\phi^2\right>$ as function of the running coupling constant $\tau \in \mathbb{H}$ of asymptotically free $\mathcal{N}=2$ QCD with gauge group $SU(2)$ and $N_f\leq 3$ massive hypermultiplets. If the domain for $\tau$ is restricted to an appropriate fundamental domain $\mathcal{F}_{N_f}$, the function $u$ is one-to-one. We demonstrate that these domains consist of six or less images of an ${\rm SL}(2,\mathbb{Z})$ keyhole fundamental domain, with appropriate identifications of the boundaries. For special choices of the masses, $u$ does not give rise to branch points and cuts, such that $u$ is a modular function for a congruence subgroup $\Gamma$ of ${\rm SL}(2,\mathbb{Z})$ and the fundamental domain is $\Gamma\backslash\mathbb{H}$. For generic masses, however, branch points and cuts are present, and subsets of $\mathcal{F}_{N_f}$ are being cut and glued upon varying the mass. We study this mechanism for various phenomena, such as decoupling of hypermultiplets, merging of local singularities, as well as merging of non-local singularities which give rise to superconformal Argyres-Douglas theories.
\vspace{.4cm}\\

\vspace{30pt}\noindent

}

\preprint{}

\begin{document}  
\maketitle

\section{Introduction}
A manifestation of $S$-duality or strong-weak coupling duality is the equivalent dynamics of a quantum field theory at distinct values of its coupling constant \cite{Montonen:1977sn, Sen:1994fa, Seiberg:1994rs, Vafa:1994tf, Verlinde:1995mz}. A natural question for such a quantum field theory is the determination of a domain for the coupling constant parametrizing inequivalent quantum field theories. We address this question for asymptotically free $\CN=2$ Yang-Mills theories with gauge group $SU(2)$ and $N_f\leq 3$ fundamental hypermultiplets. To this end, we consider
the order parameter for the Coulomb branch, which is a function of the
running coupling $\tau$ invariant under $S$-duality \cite{Seiberg:1994rs,
  Seiberg:1994aj,Klemm:1994qs,Klemm:1994qj,
  Klemm:1995wp,Danielsson:1995is}. 
We put forward a fundamental domain $\CF_{N_f}$ for $\tau$
such that the function is one-to-one. Part of our motivation is
the $u$-plane integral \cite{Moore:1997pc, LoNeSha}, which is a physical approach to Donaldson
invariants and other topological gauge-theoretic invariants of smooth
compact four-manifolds. This approach involves an integral over the
Coulomb branch of the theory. Recently, 
the change of variables from $u$ to $\tau$ has been instrumental for
the evaluation of the integral for generic four-manifolds
\cite{Malmendier:2008db,Griffin:2012kw,Malmendier:2012zz,Malmendier:2010ss, Korpas:2017qdo, Moore:2017cmm, Korpas:2018dag, Korpas:2019ava,
  Korpas:2019cwg, Manschot:2021qqe,mm1998,Marino:1998rg,Hyun:1995hz}, which suggests a potential fundamental role for this parametrization of the Coulomb branch.

The Coulomb branches of the rank 1 theories mentioned above are complex one-dimensional, and parametrised by the Higgs vacuum
expectation value $u=\frac{1}{16\pi^2} \langle \Tr\phi^2\rangle$, $\phi$
being the complex scalar of the $\CN=2$ vector multiplet
\cite{Seiberg:1994aj} (see \cite{Klemm:1997gg,DHoker:1999yni} for a review). In general, these order parameters are
functions of the running coupling $\tau$, the masses $m_i$ of the
hypermultiplets and the dynamical scales  $\Lambda_{N_f}$ generated by the
renormalisation group flow.  

Before discussing the results of this paper, let us briefly recall relevant aspects of the pure theory, i.e. $N_f=0$.
The duality group  of this theory is isomorphic to $\Gamma^0(4)$
and acts on $\tau$ through linear fractional transformations \cite{Seiberg:1994aj}. Its
Coulomb branch can be parametrized as the corresponding fundamental domain, $\Gamma^0(4)\backslash \BH$
\cite{Seiberg:1994aj, Moore:1997pc}. See Figure \ref{fig:fundgamma0(4)}. The cusps of
$\Gamma^0(4)\backslash \BH$ correspond to the strong coupling
singularities of the theory. Moreover, the order parameter $u$ is a
weakly holomorphic modular function for this group. 

For the massive theories with $N_f\leq 3$, we find a number of new
phenomena. To study these theories, 
we consider their order parameters as roots of certain degree six
polynomials constructed from the Seiberg-Witten (SW) curves. These polynomials in turn
encode many of the interesting structures of the Coulomb
branches. For example, their ramification loci include the Argyres-Douglas (AD) theories, where the
curves degenerate, as well as branch points. 
We show that the fundamental
domain of $u$ can be described as six or less copies of the corresponding
fundamental domain of the full modular group $\SL$ as displayed in
Figure \ref{fig:sl2z}. The cusps of these
domains  correspond to the singularities of the physical theory
and the width of each cusp to the number of hypermultiplets becoming massless
there. The images of the $\SL$ fundamental domains under the map
$u(\tau)$ provide intriguing partitions of the $u$-plane. See for
example Figures \ref{fig:purePartitions}, \ref{fig:nf1domains},
\ref{fig:nf1partition} and \ref{fig:nf2partition}.
 
Since the polynomials are order six in $u$ it is in general not
possible to find the roots, and solve for $u$ in terms of the
coefficients. Only for special configurations of the masses,
e.g., equal masses in $N_f=2$ and one non-zero mass in
$N_f=3$, the polynomial splits over the field of modular functions for
a congruence subgroup of $\SL$, and we can thus find explicit closed
expressions for $u$ in terms of known modular forms, reproducing and
extending previous results \cite{matone1996, Nahm:1996di, Ito:1995ga, Kanno:1998qj,
  Huang:2009md}.

For generic choices of the masses, the function $u(\tau)$ gives rise to
branch points $\tau_{\rm bp}$, where $u-u(\tau_{\rm bp})\sim
\sqrt{\tau-\tau_{\rm bp}}+\dots$ does not
return to itself as $\tau$ encircles $\tau_{\rm bp}$. While the branch points, and the
inevitable branch cuts, obstruct the identification of $\CF_{N_f}$ as
a quotient $\Gamma\backslash \mathbb{H}$ with $\Gamma$ a congruence
subgroup, they provide a mechanism for $\CF_{N_f}$ to evolve as function of
the mass. More precisely, the branch points move in the domain $\CF_{N_f}$
upon varying the masses, and the domain $\CF_{N_f}$ is literally cut and glued along the branch
cuts. This provides a way to analyze how the domain evolves as
function of the masses. We have studied this phenomena in detail in the following limits:

\begin{itemize}
\item {\it Decoupling of a hypermultiplet}:\\ A hypermultiplet
  decouples in the limit that its mass goes to infinity, $m\to \infty$. We demonstrate that in this
  situation, a branch cut disconnects (or cuts) the strong
  coupling cusp associated to this hypermultiplet from the rest of the
  domain. At the same time, the sides of the branch cut are identified
  to the sides of another branch cut. In this way, the strong coupling
  cusp is glued back to the weakly coupled cusp, near $i\infty$, where these
  branch points and cuts disappear in the limit $m\to \infty$.
As a result, the periodicity at $i\infty$ increases by 1 in the limit, while the cusp
  has disappeared from the strongly coupled region. This is displayed for $N_f=1$ 
  in Figures \ref{fig:nf1partition_m>mAD} and \ref{fig:nf1CutsAlt}.
\item {\it Merging of local singularities}:\\
For a generic choice of masses, the theory with $N_f$ hypermultiplets
has $N_f+2$ distinct strong coupling singularities in the $u$-plane, where dyons become
massless and the effective field theory breaks down. By tuning the
masses to special values, the singularities for $l$ mutually local dyons
can merge in the $u$-plane. We demonstrate that such cases give rise to a cusp with
width $l>1$ in $\CF_{N_f}$. Moreover, when perturbing away from such a special value of the
masses, we find that two branch cuts develop from the cusp, which disconnect the
singularity in $\CF_{N_f}$. This is displayed for $N_f=2$ in Figure
\ref{fig:nf2m1m2partition}. 
\item {\it Merging of non-local singularities (AD theories):}\\
The dynamics is quite different if we tune the masses to special values where
singularities corresponding to non-local dyons collide in the $u$-plane. Such
singularities give rise to superconformal Argyres-Douglas (AD) field theories
\cite{Argyres:1995jj, Argyres:1995xn, 
  Eguchi:1996vu,Eguchi:1996ds,Seo:2012ns,Xie:2012hs}. In such a
situation, we find that two branch points in $\CF_{N_f}$ typically come together and annihilate at
the pre-image $\tau_{\rm AD}$ of the AD singularity $u_{\rm AD}$. The two
branch cuts join at $\tau_{\rm AD}$ in the interior of $\CF_{N_f}$, and disconnect a region from
$\CF_{N_f}$ with the ``non-local'' cusps.\footnote{If we view $\CF_{N_f}$ as a sphere with $3+N_f$
punctures, the region is pinched off rather than cut off.}  Thus $\CF_{N_f}$ consists
then of $<6$ copies of the $\SL$ fundamental domain, and the AD point is in a sense a remnant of the disconnected region. On the other hand if we take the appropriate scaling
limit near the CFT point \cite{Argyres:1995xn}, we find that the disconnected region is a fundamental
domain for the order parameter of the AD theory. If no other branch points remain in $\CF_{N_f}$, the order parameters become 
modular functions for a congruence subgroup.  This is displayed in
Figures \ref{fig:nf1mBP}, \ref{fig:fundu2m} and  \ref{fig:fundu3m} for $N_f=1,2,3$.
\end{itemize}

Let us briefly return to the $u$-plane integral. The change of
variables from $u$ to $\tau$, gives rise to the factor $du/d\tau$ in
the integrand. Interestingly, $du/d\tau$ can be expressed in case of
the $N_f=0$ theory in terms of the discriminant $\Delta$ and $du/da$ \cite{Moore:1997pc},
which is a consequence of a relation between the prepotential and $u$
\cite{Matone:1995rx, matone1996}. Up to numerical constants, $\Delta$
and $du/da$ are precisely the two gravitational couplings of the topological theory 
\cite{Witten:1995gf,Manschot:2019pog}, such that $du/d\tau$ is naturally included.
Extending previous work on the massless $N_f\leq3$
\cite{Kanno:1998qj}, we derive a further generalisation for all cases
$N_f\leq 4$ with generic masses. We also discuss how this relation encodes interesting
information on the special points of the Coulomb branch, specifically
the branch points.

We have organised the paper in the following way. In Section
\ref{sec:mod_appr}, we review the SW solutions for the $N_f\leq 3$
theories and then derive a sextic polynomial for $u$ which is crucial 
for the construction of fundamental domains for the coupling
$\tau$. Section \ref{sec:uplanetiling} discusses partitions of the
$u$-plane induced by $\CF_{N_f}$. Section \ref{sec:matone} derives a generalisation
of Matone's relation for the generic mass cases, as well as giving a
discussion on what information it encodes for the Coulomb branch. We
then go on to analyse explicit examples in Sections
\ref{sec:nf1}--\ref{sec:nf3}. We conclude with a brief discussion on
our findings as well as some possible further directions in Section
\ref{sec:discussion}. As a by-product of our analyses we also propose
an expression for the beta functions of the massive $N_f\leq3$
theories, generalising the results of \cite{Dolan:2005mb,
  Dolan:2005dw}.

\section{Fundamental domains for $SU(2)$ SQCD}\label{sec:mod_appr}
In this Section we develop techniques to determine a fundamental domain for the effective coupling 
of the asymptotically free $\CN=2$, $SU(2)$ SQCD theories. 

\subsection{The SW solutions} 
We recall a few essential aspects of the SW solutions for these
theories, which we use to analyze $u$ as function of $\tau$. The gauge group $SU(2)$ is spontaneously broken to $U(1)$ on
the Coulomb branch. The order parameter for this branch is the vev
$u$, defined as 
\be
u=\frac{1}{16\pi^2}\left< \mathrm{Tr}\phi^2 \right>_{\mathbb{R}^4}\in \CB_{N_f},
\ee
where the trace is in the 2-dimensional representation of $SU(2)$. Topologically, $\CB_{N_f}$ is the complex plane $\mathbb{C}$ minus $2+N_f$ singular
points (for generic masses).  

The scalar field related to the photon in the low energy effective
field theory is $a$, while $a_D$ is related to the dual photon. The SW
solution identifies these fields as periods of a specific
differential $\lambda$ over two dual cycles, $A$ and $B$, of an elliptic curve with complex structure $\tau$,
\be 
\label{aaDperiods}
a=\int_{\gamma} \lambda,\qquad a_D=\int_{\gamma_D} \lambda.
\ee

To list the SW curves of the theories with $N_f\leq 3$
hypermultiplets, let $\Lambda_{N_f}$ be the scale of the theory with $N_f$
hypermultiplets, and $m_j$, $j=1,\dots,N_f$ be the masses of the
hypermultiplets. The SW curves of the theories are given
by \cite{Seiberg:1994aj}\footnote{There are other formulations of the
  SW curve. For example the class $S$ form is
  $x^2=p_{N_f}(z,u,\Lambda_{N_f},\bfm)$ \cite{Witten:1997sc,
    Klemm:1996bj, Gaiotto:2009hg}. This has the advantage that the SW
  differential is canonically determined as $\lambda=x\,dz$. The analysis in the present paper still holds for these formulations.} 
\begin{equation}\label{eq:curves}  
\begin{aligned}  
N_f=0:\quad y^2&=x^3-ux^2+\frac{1}{4}\Lambda_0^4x, \\
N_f=1:\quad y^2&=x^2(x-u)+\frac{1}{4}m\Lambda_1^3x-\frac{1}{64}\Lambda_1^6, \\
N_f=2:\quad y^2&=(x^2-\frac{1}{64}\Lambda_2^4)(x-u)+\frac{1}{4}m_1m_2\Lambda_2^2x-\frac{1}{64}(m_1^2+m_2^2)\Lambda_2^4, \\
N_f=3:\quad y^2&=
x^2(x-u)-\frac{1}{64}\Lambda_3^2(x-u)^2-\frac{1}{64}(m_1^2+m_2^2+m_3^2)\Lambda_3^2(x-u)\\&\quad
+\frac{1}{4}m_1m_2m_3\Lambda_3x-\frac{1}{64}(m_1^2m_2^2+m_2^2m_3^2+m_1^2m_3^2)\Lambda_3^2.
\end{aligned}
\end{equation} 
The family of SW curves are Jacobian rational elliptic surfaces with singular fibres 
\cite{Malmendier:2008yj,shioda1972,schuett2009,maier2006}. Rational in
this context means that $g_2$ and $g_3$ are polynomials in $u$ of
degree at most $4$ and $6$, respectively \cite{miranda1995}. 

\begin{figure}[h]\centering
	\includegraphics[scale=1]{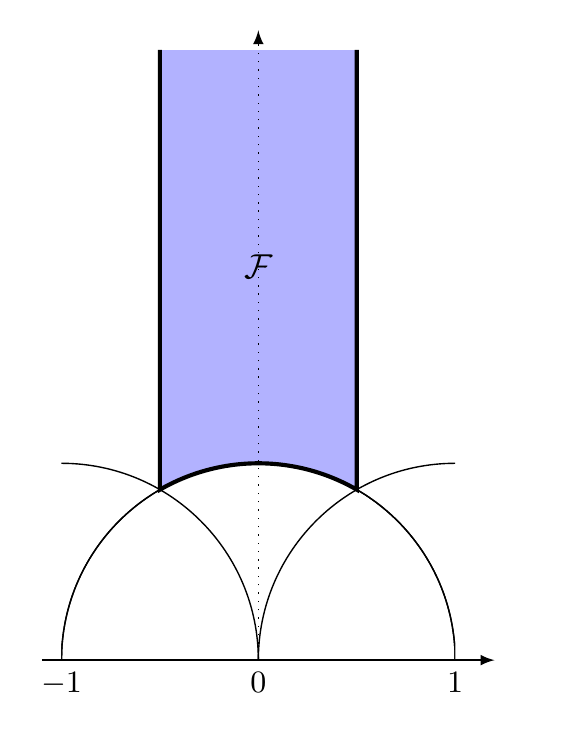}
	\caption{The key-hole fundamental domain $\CF$ of $\SL$. The
          vertical sides are identified, as well as the two halfs of
          the boundary arc on the unit circle.}\label{fig:sl2z}
\end{figure}

Decoupling a hypermultiplet corresponds to the following double
scaling limit
\cite{Eguchi1999}
\begin{equation}
m_j\to \infty, \quad \Lambda_{N_f}\to 0, \quad m_j \Lambda_{N_f}^{4-N_f}=\Lambda_{N_f-1}^{4-(N_f-1)}
\end{equation} 
One can directly decouple more than one hypermultiplet, where the scales of the low energy theories are defined as
\begin{equation}
\Lambda_0^2=m\Lambda_2,\qquad  
\Lambda_0^4=m^3\Lambda_3, \qquad 
\Lambda_1^3=m^2\Lambda_3,   
\end{equation} 
and $m$ is the equal mass of the hypermultiplets being decoupled.  
These curves are constructed in such a way that their mathematical
discriminants will, up to an overall normalisation, correspond to the
\emph{physical discriminant}. This we define as the monic polynomial,
\begin{equation}
\label{PhysDisc}
	\Delta_{N_f}\coloneqq \prod_{i=1}^{N_f+2}(u-u_i),
\end{equation}
with $u_i$ being singular points of the effective theory, where hypermultiplets become massless. It is a polynomial of degree $\deg\Delta_{N_f}=N_f+2$ in $u$.\footnote{One important note is that in \cite{Seiberg:1994rs} another convention is used for the curve of the pure theory. This gives the duality group $\Gamma(2)$ rather than $\Gamma^0(4)$ as in the above. The $\Gamma(2)$-convention, however, turns out to not be suitable for the discussion in this paper due to multiplicities of the singularities of the curve.}
To see this, we bring the SW curves (\ref{eq:curves}) into
Weierstra{\ss} form by shifting $x\to x+\frac{u}{3}+\frac{\Lambda_3^2}{192} \delta_{3,N_f}$, and rescaling $y\to y/2$,
\begin{equation}\label{weierstrassequation}
\CW: \quad y^2=4\, x^3-g_2\, x-g_3, 
\end{equation} 

where $g_2=g_2(u,\bfm,\Lambda_{N_f})$ and $g_3=g_3(u,\bfm,\Lambda_{N_f})$ are polynomials in
$u$, $\bfm=(m_1,\dots,m_{N_f})$ and the scale $\Lambda_{N_f}$. The discriminant $\Delta_{N_f}$
is unchanged for this change of variables, and equals
\be\label{deltag2g3}
\Delta_{N_f}=(-1)^{N_f} \Lambda_{N_f}^{2N_f-8}(g_2^3-27\,g_3^2),
\ee
where the last factor is the ``mathematical'' discriminant.
The functions $g_2$ and $g_3$ can be combined to an absolute
invariant $\CJ$,
\begin{equation}\label{jinvg2g3} 
\mathcal J=12^3\frac{g_2^3}{g_2^3-27g_3^2}. 
\end{equation}
As opposed to $g_2$ and $g_3$, $\CJ $ is invariant under admissible
changes of variables. Two curves are isomorphic if and only if they
have the same absolute invariant $\CJ$.  
Since $g_2(u,\bfm,\Lambda)$ and $g_3(u,\bfm,\Lambda)$ are polynomial functions of $u$, $\bfm$ and $\Lambda$ for
the SW curves, $\CJ$ is naturally a rational function $\CJ(u,\bfm,\Lambda)$ of these variables. On the other hand, the modular Weierstra{\ss} form
expresses $\CJ$ in terms of the complex structure $\tau$, namely as
the modular $j$-invariant $j(\tau)$ (see \eqref{je4e6} for a definition). 
\begin{equation}
\label{CJj}
\mathcal J(u,\bfm,\Lambda)=j(\tau).
\end{equation}
This allows to obtain $u$ as function of $\tau$, which is physically the
effective coupling constant. 
Cusps are points where $j(\tau)=\infty$, which correspond to $\tau\in \{\im\infty\}\cup \mathbb Q$.  
The $j$-function has fundamental domain $\CF=\SL\backslash\mathbb H$, which is typically taken to be
the key-hole fundamental domain displayed in Figure \ref{fig:sl2z}. In
other words, the function $j:\CF\to \mathbb{C}$ is a bijective map.

\subsection{Partitioning the upper half-plane}\label{sec:sexticeqn}
We are interested in determining the fundamental domains $\CF_{N_f}$ for the
effective coupling $\tau$ for a theory with $0\leq N_f<4$. Let us  consider $u$ as a function,
\be\label{ufnfbnf}
u:\mathbb H\longrightarrow \mathcal{B}_{N_f}, 
\ee
and study the analytic properties of this map.   
We will discuss later the dependence of
$\CF_{N_f}$ on the masses $\bfm$, which we will 
make manifest in the notation as $\CF_{N_f}(\bfm)$ or more compactly $\CF(\bfm)$. We find that for $N_f\geq1$ and generic masses the duality group does not act on $\tau$ by fractional linear transformations. This prevents us from defining a fundamental domain as is customary for a congruence subgroup $\Gamma$ of $\slz$: 
For any point $\tau\in\mathbb H$ there exists a $g\in\Gamma$ such that $g\cdot\tau\in \Gamma\backslash\mathbb H$, and no two distinct points $\tau$, $\tau'$ in $\Gamma\backslash\mathbb H$ are equivalent to each other under $\Gamma$. Rather, we can compare if points $\tau$, $\tau'$ are equivalent under \eqref{ufnfbnf}: If we define the equivalence relation
\begin{equation}
\tau\sim \tau'\,\, \Longleftrightarrow \,\, u(\tau)=u(\tau'),
\end{equation}
then the quotient set $\mathbb H/{\sim}$ is a fundamental domain
$\CF_{N_f}$  for the function $u$. Upon plotting $\CF_{N_f}$ as a
domain in $\mathbb{H}$, we will have to introduce identifications along co-dimension 1 segments as for
$\CF$ in Figure \ref{fig:sl2z}. 

To determine $\CF_{N_f}$, we bring
(\ref{jinvg2g3}) into a more convenient form. 
We multiply (\ref{jinvg2g3}) by $\Delta_{N_f}$ and bring all terms
to one side. This gives the polynomial,
\begin{equation}\label{sexticpolynomial} 
\begin{split}
	P_{N_f}(X)&\coloneqq
        \left(g_2(X,\bfm,\Lambda)^3-27g_3(X,\bfm,\Lambda)^2\right)j-12^3g_2(X)^3\\
&= a_6\, X^6+a_{5}\,X^{5}+\ldots+a_1\,X+a_0,
\end{split}
\end{equation}
where the coefficients $a_i=a_i(\bfm,\Lambda,j)$ are polynomial functions of
$\bfm$, $\Lambda$, and the $j$ function, $a_i(\bfm,\Lambda,j)\in
\mathbb{C}[\bfm,\Lambda,j]$. The polynomials \eqref{sexticpolynomial}
can thus be viewed as polynomials over the field $\mathbb C[\bfm,\Lambda, j]$.

We see that \eqref{jinvg2g3} is equivalent to $P_{N_f}(u)=0$ for
$\Delta_{N_f}\neq 0$, or in other words, away from the singular locus of the
theory. The roots of $P_{N_f}$ can therefore be identified with the order parameter of the corresponding SW curve.  
Recall that we can assign $U(1)_{\CR}$ charges $[u: m_i: x:
y]=[4:2:4:6]$ to the quantities of the Seiberg-Witten curves
\cite{Seiberg:1994aj}. Since $g_2$ and $g_3$ are polynomials in $u$ by
construction, by bringing the SW curves to the Weierstra{\ss} form and
using that $[u]=4$ we have that the degrees of $g_2$ and $g_3$ as
polynomials in $u$ must be $\deg(g_2)=2$ and $\deg(g_3)=3$.
Therefore, $P_{N_f}$ is a sextic polynomial in $X$.

For generic masses $\bfm$, the sextic equation
$P_{N_f}=0$ gives rise to $n=6$ different solutions as functions of $j$,
while for special choices of $\bfm$, such as those giving rise to
superconformal (AD) theories, we have $2\leq n \leq 4$ different
$j$-dependent solutions
and $6-n$ $j$-independent solutions. Since $j:\CF\to \mathbb{C}$ is an
isomorphism, the $n\leq 6$ solutions provide a
multi-valued ($n$-valued) function over $\CF$. 

To
obtain $u$ as a single-valued function of the effective coupling, we choose a different copy of $\CF$  for each of the $n\leq 6$ branches, and
appropriately identify the boundaries of these domains. These are related to $\CF$ by an element of
$\SL$, and their union is
\begin{equation}\label{fbfm}
	\CF_{N_f}=\bigcup_{j=1}^{n}\alpha_j\CF,
	\end{equation}
with $\alpha_j\in\slz$. A priori, there is no canonical choice for the
$\alpha_j$, they are determined up to the action of the duality group
of the theory. However, some choices are more natural than others. If we demand
that $\CF_{N_f}$ is connected and take $\alpha_1={\mathbbm 1}\in \SL$,
there is only a finite number of choices for $\CF_{N_f}$. In some cases, $\CF_{N_f}$ is a modular curve
$\Gamma\backslash\mathbb H$ for a congruence subgroup $\Gamma\subseteq
\SL$. In such cases, $n$ equals the index of $\Gamma$ in  $\SL$
\cite{shimura1971} (see also Appendix \ref{sec:modularcurves} for the corresponding definitions for modular curves). For later use, we define the set of $\alpha_j$ as $\CC_{N_f}=\{\alpha_j, j=1,\dots,n\}$.

 For {\it generic} masses, $n=6$ and $\CF_{N_f}$ has $3+N_f$ cusps, corresponding to weak
coupling $\tau\to i\infty$ and the $2+N_f$ singularities of the theory. 
We find the widths of the cusps by expanding $j(\tau)=\mathcal
J(u,\bfm,\Lambda_{N_f})$ for $\tau$ near the cusp. For general $N_f\in\{0,1,2,3\}$, the cusp at infinity has width
$h_\infty=4-N_f$. This is because $q^{-1}\sim j(\tau)=\CJ\sim u^{4-N_f}$, which implies $u(\tau)\sim q^{-\frac{1}{4-N_f}}$ (where $q=e^{2\pi\im\tau}$). Thus for
large $\tau$, $u(\tau)$ is invariant under $T^{4-N_f}$, where $T:\tau\mapsto \tau+1$. Near any singularity $u_s$, it
is clear that $q^{-1}\sim \frac{1}{(u-u_s)^{h_s}}$, where $n_s$ is the
multiplicity of the singularity. Similarly, near $u_s$ one finds
$u(\tau)-u_s\sim q^{\frac{1}{h_s}}$. Locally, the function $u(\tau)$
has period $h_s$, giving the width $h_s$ of the cusp. The widths of all cusps then add up to 6, 
\be
h_\infty+\sum_{s} h_s=6.
\ee 

As mentioned above, the equation $P_{N_f}=0$ gives six different
solutions for $u$. A natural question that then arises is which of
these six to use as our $u$. In some sense this is of course
arbitrary, all of 
them correspond to the order parameter $u$ simply expressed in
different duality frames. On the other hand, the most natural solution
is the one corresponding to the weak coupling duality frame where
$|u|$ is large for $\tau\to i\infty$. Since the width of the cusp at
infinity is $4-N_f$ we see that there is still some ambiguity in this
choice as long as $N_f<3$, but for $N_f=3$ there is exactly one
choice. We show in Section \ref{sec:nf3} that this has $u\to-\infty$
for $\tau\to i\infty$, and it turns out that this choice can be taken
for all $N_f\leq 3$ theories, and is preserved under the decoupling of
hypermultiplets, we therefore make this choice throughout. Note that
this sign differs from the conventional choice in the
literature \cite{Seiberg:1994aj, Ito:1995ga, Kanno:1998qj}.

Different mass configurations can give different decompositions of 6. When singularities merge, their
cusps are identified under the duality group and their widths add
up. Moreover a cusp moves from the real axis to infinity upon
decoupling of a matter multiplet.

For {\it special} choices of the masses, not all singularities
correspond to cusps  $i\infty$ or the real line;
also singularities in the interior of the upper half-plane can
occur. The theories at these points are of superconformal or Argyres-Douglas type, and the
widths of all cusps add up to $n$. 

Yet another aspect of the parametrization by $\tau$ is that for special values of $\tau$ in the
interior of $\CF$, otherwise distinct solutions can coincide. These
are branch points of the solutions, where the function $u(\tau)$
ceases to be meromorphic in $\tau$. The branch points in $\CF_{N_f}$ emanate a branch cut. We will discuss these aspects in more detail in Section \ref{sec:ramifications}.

For generic masses the equation $P_{N_f}(X)=0$ furthermore defines a Riemann surface, which is a $6$-fold ramified covering over the classical modular curve $\slz\backslash\mathbb H$ \cite{Khovanskii2013}. On this Riemann surface, any root $u$ forms a \emph{meromorphic} map to the Coulomb branch. It would be interesting to study the topology of these surfaces in more detail. See also \cite{Park:2011is}. \\

\noindent{\it Example: pure $SU(2)$}\\
To give an example, we can study the well-known pure $SU(2)$ curve,
\cite{Seiberg:1994aj, matone1996, Nahm:1996di}. The absolute invariant
reads, 
\be
\CJ(u,\Lambda_0)=\frac{64}{\Lambda_0^8} \frac{(3\Lambda_0^4-4u^2)^3}{\Lambda_0^4-u^2},
\ee
and the sextic equation for this theory is, 
\be
P_0(X)=\Lambda_0^8(\Lambda_0^4-X^2)\,j(\tau)-64(3\Lambda_0^4-4X^2)^3.
\ee  

\begin{figure}[h]\centering
	\includegraphics[scale=1]{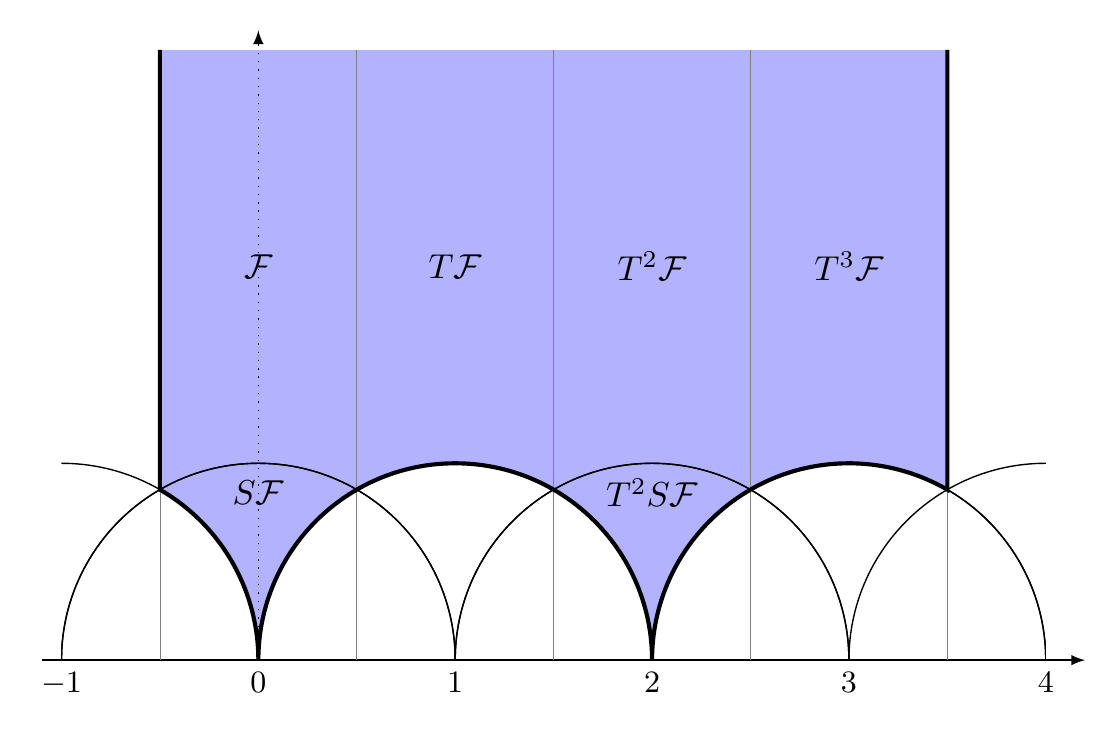}
	\caption{Fundamental domain of $\Gamma^0(4)$. This is the duality group of the pure $SU(2)$ theory. The two cusps on the real line correspond to the strong coupling singularities of the gauge theory, while the cusp at $\tau=\im\infty$ corresponds to weak coupling. }\label{fig:fundgamma0(4)}
\end{figure}

As discussed above, this equation has 6 independent solutions, and we pick the one that has $u\to -\infty$ for $\tau\to i\infty$. The cusp at
infinity has then width $h_\infty=4$, while the two strong coupling cusps
both have unit width corresponding to the multiplicity of the two singularities.
Moreover, since $P_0(X)$ is an even function of $X$, we naturally choose the fundamental
domain $\CF_0$ such that it is invariant under a shift by half the width, that
is $\tau\to \tau +2$. In this way, we arrive at the following union of six copies of the $\SL$
fundamental domain,  
\begin{equation}
	\CF_{0}=\CF \cup T\CF\cup T^2\CF\cup T^3\CF \cup S\CF\cup T^2 S\CF,
\end{equation}
which is displayed in Figure \ref{fig:fundgamma0(4)}. 
These copies form a fundamental domain for the congruence subgroup
$\Gamma^0(4)$, $\CF_0=\Gamma^{0}(4)\backslash \BH$. This demonstrates that the duality group of the theory is $\Gamma^0(4)$.

Since the duality group is $\Gamma^0(4)$, the order parameter $u$ can
be explicitly determined in terms of modular forms for this group. One finds
\begin{equation}\label{nf0parameter}
	\begin{aligned}
		\frac{u(\tau)}{\Lambda_0^2}=&-\frac{1}{2}\frac{\vartheta_2(\tau)^4+\vartheta_3(\tau)^4}{\vartheta_2(\tau)^2\vartheta_3(\tau)^2}=-1-\frac{1}{8}\left(\frac{\eta(\frac{\tau}{4})}{\eta(\tau)}\right)^8 \\
		=&-\frac{1}{8}(q^{-1/4}+20q^{1/4}-62q^{3/4}+216q^{5/4}+\CO(q^{7/4})),
	\end{aligned}
\end{equation}
with $q=e^{2\pi\im\tau}$. As mentioned above, we have made the (unconventional) choice for
the sign of $u$ with $u\to -\infty$ when $\tau\to i\infty$.

Even if the duality group is not a congruence subgroup of $\slz$, there is a procedure to find closed expressions for the order parameters in special cases. The sextic equation \eqref{sexticpolynomial} for fixed masses $\bfm$ and scale $\Lambda$ can be viewed as  a polynomial over the algebraic field $\mathbb C(\Gamma)$ of modular functions on $\Gamma=\slz$.
Such nontrivial polynomials define field extensions over $\mathbb
C(\Gamma)$. By the fundamental theorem of Galois theory, there is a
one-to-one correspondence between the Galois group of the field
extension and its intermediate fields. Intermediate fields can be
obtained by adjoining roots of the polynomial to the base field. Since
$P_{N_f}(X)$ is a sextic polynomial, for generic masses $\bfm$ it is
not possible to find exact expressions for the roots. However, if one
of the intermediate fields  is known, the polynomial factors over the
intermediate field into products of polynomials of lower degree. If
the resulting degree is less than or equal to 4, there are closed
formulas for the roots. 

We find below that in many cases, such as massive $N_f=2$ and $3$ with one mass parameter, $\mathbb C(\Gamma(2))$ for the principal congruence subgroup $\Gamma(2)$ (see Appendix \ref{sec:jacobitheta}) is an intermediate field. Since the function $\lambda=\frac{\jt_2^4}{\jt_3^4}$ is a Hauptmodul for the genus 0 congruence subgroup  $\Gamma(2)$, it is the root of a polynomial of degree $[\Gamma: \Gamma(2)]=6$ over $\mathbb C(\Gamma)$. More precisely, there exists a rational function $\CR$ with the property that $\CR(\lambda(\tau))=j(\tau)$. It is given by 
\begin{equation}
\CR(p)=2^8\frac{(1+(p-1)p)^3}{(p-1)^2p^2}.
\end{equation}
Instead of solving $\CJ(u,\bfm,\Lambda)=j(\tau)$ we can then rather solve $\CJ(u,\bfm,\Lambda)=\CR(\lambda(\tau))$. 
If $\mathbb C(\Gamma(2))$ is an intermediate field, the sextic equation corresponding to this equation  factors over $\mathbb C(\Gamma(2))$. In massive $N_f=2,3$ we find that it factors into three quadratic polynomials with coefficients depending on $\lambda$, which can be easily solved analytically. 
Such rational relations between the $j$-invariant and Hauptmoduln exist for any genus 0 congruence subgroup, which are classified. They allow to invert the equation $\CJ(u,\bfm,\Lambda)=j(\tau)$ for a large class of mass parameters, as we demonstrate in the following sections. See also \cite{Bourget:2017goy,Ferrari:2008rz,Ferrari:2009zh}.

\subsection{Ramification locus}\label{sec:ramifications}

The covering $\CF_{N_f}(\bfm)\to \CB_{N_f}$ is not 1-to-1 on a discrete
subset, namely at points of $\CF_{N_f}(\bfm)$ where the discriminant
$D(P_{N_f})$ vanishes.\footnote{The discriminant of a polynomial 
$p(X)=X^n+a_{n-1}X^{n-1}+\ldots+a_1X+a_0=\prod_{j=1}^n(X-r_j)$
is defined as $D(p)=\prod_{i<j}(r_i-r_j)^2$, in particular it
vanishes if and only if two roots coincide. Since we are interested in
finding the zeros of $D(p)$, we are not careful about  overall
normalisation factors.}  In all cases, $N_f=0,1,2,3$, we find that the discriminant of $P_{N_f}$ factorises as 
\begin{equation}\label{poldiscfactor0} 
D(P_{N_f})=j^4\,(j-1728)^3\,(D^{\text{AD}}_{N_f})^3\,D^{\text{bp}}_{N_f}.
\end{equation}
 
\begin{figure}[htbp]
\centering
	\includegraphics[scale=0.7]{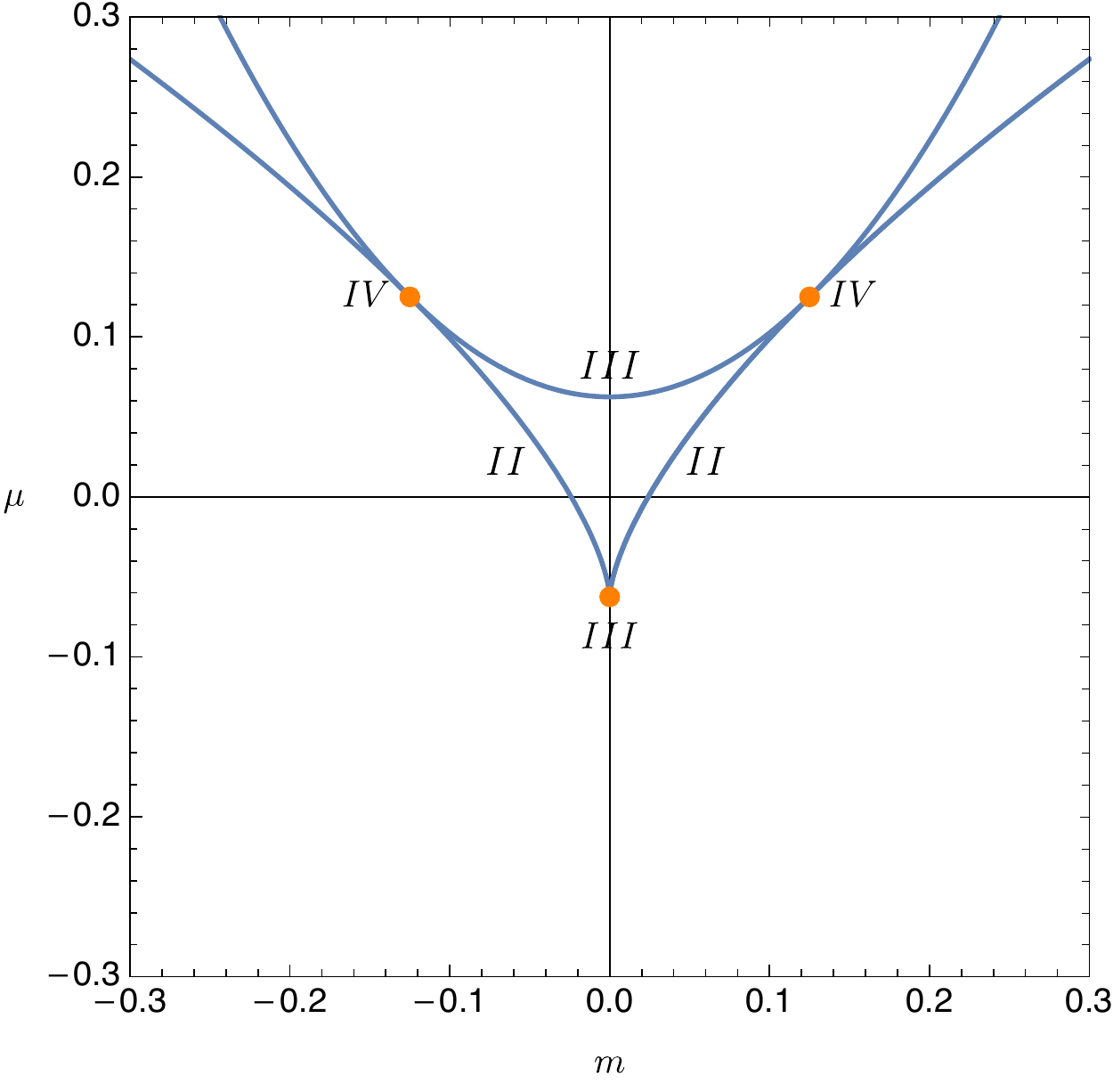}
	\caption{The AD locus $\lad_3$ for $N_f=3$ with masses $\bfm=(m,m,\mu)$ in the real $(m,\mu)$-plane, with units $\Lambda_3=1$. It is a union of three smooth lines,  two of them  generically describing type $II$ AD points and the third one type $III$. The two $II$ lines touch at a $III$ point, while both $II$ lines touch the $III$ line in a type $IV$ AD point.}\label{fig:ADlocusNf3mmu}
\end{figure}

We discuss each of the three factors:\\
\\
{\it The $\bfm$-independent factor}\\
The factor $j^4\,(j-1728)^3$ is independent of the masses $\bfm$ and can be understood from
 \eqref{sexticpolynomial}. It is immediate that when $j=12^3$, every
 root of $P_{N_f}$ has multiplicity at least 2, and if $j=0$ every root  has
 multiplicity at least 3. On $\mathbb H$ this occurs whenever $\tau\in
 \slz\cdot \im $ or $\tau\in \slz\cdot \omega_3$, with $\omega_j=e^{2\pi\im/j}$. On the modular
 curve $\slz\backslash \mathbb H$, these orbits collapse to a
 point and in fact the covering $\pi$ is ramified only over
 $\{\im\infty,\im,\alpha\}$, or $j\in \{0,1728,\infty\}$,
 respectively.  
This resembles the \emph{Belyi functions}, which are holomorphic maps
from a compact Riemann surface to $\cpo$ ramified over precisely these
three points \cite{schuett2009,He:2013eqa}. 
They can be described combinatorially by so-called \emph{dessins
  d'enfants}. Such dessins have also appeared in the context of SW
theory
\cite{Ashok:2006br,Ashok:2006du,He:2012kw}. For generic masses, the SW family of curves do not satisfy this definition, as there are additional ramification points. \\
\\
{\it The polynomial $D^{\text{\rm AD}}_{N_f}$}\\
The factor $D^{\text{AD}}_{N_f}$ corresponds to Argyres-Douglas (AD) loci,
where two or more singularities coincide \cite{Argyres:1995jj, Argyres:1995xn}. More precisely, the zero
locus of $D^{\text{AD}}_{N_f}$ corresponds to the masses for which the
Coulomb branch contains AD points. To see this, recall that the AD
points correspond to 
\be \label{g2g30}
g_2(u,\bfm,\Lambda)=g_3(u,\bfm,\Lambda)=0.
\ee
Since $g_2$ and $g_3$ are polynomials in $u$ of degrees 2 and 3,
respectively, we can eliminate $u$ from the above equations and
characterise $\CL^{\text{AD}}_{N_f}$ as the zero locus of a polynomial
$D^{\text{AD}}_{N_f}$ in $\bfm$, 
\begin{equation}
	\CL^{\text{AD}}_{N_f}=\{\bfm\in \BC^{N_f}|D^{\text{AD}}_{N_f}(\bfm)=0\}.
\end{equation}
These are precisely the polynomials appearing in \eqref{poldiscfactor0}. 
 From the SW curves we can easily find that they are given by 
\begin{equation}\label{ADpolys}
	\begin{aligned}
		D_0^{\text{AD}}=&\,\,1, \\
		D_1^{\text{AD}}=&\,\,27\Lambda_1^3-64m^3, \\
		D_2^{\text{AD}}=&\,\,\Lambda_2^6-12m_1m_2\Lambda_2^4+3(3m_1^4+3m_2^4-2m_1^2m_2^2)\Lambda_2^2-64m_1^3m_2^3, \\
		D_3^{\text{AD}}=& \,\,\Lambda_3^9-12\widetilde M_2\Lambda_3^7+168\widetilde M_3\Lambda_3^6-174\widetilde M_4'\Lambda_3^5+48\widetilde M_4\Lambda_3^5\\
		&+168\widetilde M_2\widetilde M_3\Lambda_3^4-372\widetilde M_3^2\Lambda_3^3+24\widetilde M_6'\Lambda_3^3-64\widetilde M_6\Lambda_3^3 \\
		&-24\widetilde M_3\widetilde M_4'\Lambda_3^2+96\widetilde M_3\widetilde M_4\Lambda_3^2+6\widetilde M_2\widetilde M_3^2\Lambda_3-27\widetilde M_8'\Lambda_3+8\widetilde M_3^3, 
	\end{aligned}
\end{equation} 
where for $N_f=3$ we have defined the symmetric combinations
\begin{equation}
	\begin{aligned}
		&\widetilde M_{2k}=2^{6k}\sum_{j=1}^3m_j^{2k}, \qquad \widetilde M_3=2^9\prod_{j=1}^3m_j, \\
		&\widetilde M_4'=2^{12}\sum_{i<j}m_i^2m_j^2, \quad \widetilde M_6'=2^{18}\sum_{i\neq j}m_i^2m_j^4,\quad \widetilde M_8'=2^{24}\sum_{i<j}m_i^4m_j^4.
	\end{aligned}
\end{equation}
The type of singularity that appears for specific masses on these loci are found by studying the order of vanishing of $g_2$, $g_3$ and $\Delta$ according to the Kodaira classification,
\begin{equation}\begin{aligned}\label{kodairaAD}
		II:\quad \ord(g_2,g_3,\Delta)&=(1,1,2)  \text{ or } (2,1,2),\\
		III:\quad \ord(g_2,g_3,\Delta)&=(1,2,3),\\
		IV: \quad\ord(g_2,g_3,\Delta)&=(2,2,4).
\end{aligned}\end{equation}
See Appendix \ref{sec:kodaira} for more details. The zero loci of the
AD polynomials can be understood as codimension $1$ loci in the space
$\mathbb C^{N_f}\ni \bfm$ \cite{Argyres:1995xn}. For $N_f=3$ such a
locus is shown in Fig. \ref{fig:ADlocusNf3mmu}. Argyres-Douglas loci
are studied for a more general class of SW theories in
\cite{Seo:2012ns}. 

In Section \ref{sec:sexticeqn}, we argued that the widths of the
different cusps of the $SU(2)$ theories always add up to $n\leq 6$.
We will now argue that $n<6$ if and only if $\bfm$ is a zero $\bfm_{\text{AD}}$ of $D^{\text{AD}}_{N_f}$. It is possible that some zero of $\Delta$ is also a
zero of $g_2$. Then the index is given by the degree of the numerator
of $j$, which can be smaller than 6. In Sections \ref{sec:nf1ad},
\ref{sec:adnf2} and \ref{31ADtheroy}-\ref{sec:nf3II} we study a few examples of AD
theories appearing in the $N_f=1,2,3$ theories, and demonstrate that
the curve degenerates to Kodaira types $II$, $III$ or $IV$. Each
singularity type is not exclusive to a specific number of flavours,
but appears on the discriminant divisor of the higher $N_f$ theories
as well  \cite{Argyres:1995xn}. See Sections \ref{sec:nf3III} and \ref{sec:nf3II} for two explicit examples of this. The three types of AD theories corresponds to
2, 3 or 4 mutually non-local states becoming massless at the AD
point. The cusps corresponding to the non-local states are
disconnected from the rest of the domain, and the branch points
collide at an elliptic point of the duality group. As a result, the
index is reduced by $\ord\,\Delta$, which equals the number of mutually
non-local states becoming massless, i.e., 2, 3, and 4 for the theories
$II$, $III$ and $IV$, respectively. Note that the order of vanishing
of the discriminant may be larger than zero for ordinary singularities
as well, so it is not enough to simply subtract $\ord\,\Delta$ from
six to get the index right but rather we should subtract the number of
mutually non-local states becoming massless at each cusp,
\be\label{n6MNLM}
n=6-\#\,(\text{mutually\,\,non-local\,\,massless\,\,dyons}).
\ee
This is because for the index to reduce it is necessary for $g_2$ and $\Delta$ to have a common root, such that due to \eqref{deltag2g3} it is also a root of $g_3$ and because of \eqref{g2g30} therefore an AD point. In the limit  $\bfm\to\bfm_{\text{AD}}$, the $6-n$ copies of $\CF_{N_f}(\bfm)$ corresponding to the regular singularities are removed from the fundamental domain. We have also found mass configurations whose corresponding Coulomb branch contains \emph{two} (type II) AD points. The correspondence \eqref{n6MNLM} nevertheless holds, for a similar argument as presented above. 
\\
\\
{\it The polynomial $D^{\text{\rm bp}}_{N_f}$}\\
The last factor $D^{\text{\rm bp}}_{N_f}$ corresponds to branch
points. These are values of $j$ for which two solutions of $P_{N_f}(X)=0$ coincide,
such that the map $u:\CF_{N_f}(\bfm)\to \CB_{N_f}$ is not
1-to-1 on these points. The identifications are different from the multiple images of $\CF$ in $\CB_{N_f}$, which identify the images of the boundary of $\CF$, $\alpha_j(\partial\CF)$, in $\CF_{N_f}(\bfm)$. 

The $D^{\text{bp}}_{N_f}$ are explicitly given by 
\begin{equation}\label{pbp}
	\begin{aligned}
		D_0^{\text{bp}}=&\,\,1,\\
		D_1^{\text{bp}}=&\,\,27j\Lambda_1^6-27\cdot 2^{14}m^3\Lambda_1^3+2^{20}m^6,\\
		D_2^{\text{bp}}=&\,\,(m_1^2-m_2^2)^2j^2\Lambda_2^8-128\Lambda_2^4\big(216(m_1^8+m_2^8)-288m_1^2m_2^2(m_1^4+m_2^4)\\
		&+16m_1^4m_2^4
		+240m_1^3m_2^3\Lambda_2^2-72m_1m_2(m_1^4+m_2^4)\Lambda_2^2+9(m_1^4+m_2^4)\Lambda_2^4\\
		&-42m_1^2m_2^2\Lambda_2^4-2m_1m_2\Lambda_2^6\big)j+2^{12}(16m_1m_2-\Lambda_2^2)^3P_2^{\text{AD}},
	\end{aligned}
\end{equation}
and we define $\CL_{N_f}^{\text{bp}}$ as the zero locus of $D_{N_f}^{\text{bp}}$.
The expression for $D_3^{\text{bp}}$ for generic masses is very long so we do not write it out here, but we can note that it is has degree three in $j$. For later reference we write it out for two special mass configurations
\begin{equation}
\begin{aligned}
		D_3^{\text{bp}}(m,m,m)&= 432m^4\Lambda_3^2j+(8m-\Lambda_3)^2(16m+\Lambda_3)^3(64m+\Lambda_3), \\
		D_3^{\text{bp}}(m,0,0)&= 16m^4\Lambda_3^2j+(8m-\Lambda_3)^3(8m+\Lambda_3)^3.
\end{aligned}
\end{equation}
To show that the zero locus of these polynomials really correspond to
branch points we will need some specific details of the corresponding
theory and we therefore hold off on this discussion until the
respective sections below. We can, however, note that by solving
$D_{N_f}^{\text{bp}}=0$ for $j$ and plugging it into
\eqref{sexticpolynomial} we get the corresponding solutions for
$u$. For example, in $N_f=1$ we find $u=\frac{4}{3}m^2$ and as we will
see, away from $m=\mad=\frac{3}{4}\Lambda_1$, this is not part of the
discriminant of the curve and therefore does not correspond to a
physical singularity of the theory. We denote a branch point of $u$ in $\CF_{N_f}$ by $\tau_{\rm bp}$,
and its image in $\CB_{N_f}$ as $u_{\rm bp}$. As explained in Section
\ref{sec:branchpoints}, for generic masses there are two branch points
$\tau_{\rm bp}$ and $\tau_{\rm bp}'$ with image $u_{\rm
  bp}=u(\tau_{\rm bp})=u(\tau'_{\rm bp})$. Since their image in
$\CB_{N_f}$ is the same, the points $\tau_{\rm bp}$
and $\tau_{\rm bp}'$ are identified in $\CF_{N_f}$, even though they appear as
distinct points in plots of $\CF_{N_f}$ in $\mathbb{H}$. A branch cut emanates from
each branch point; there can be a single cut connecting both branch
points, or two separate cuts which go to either $i\infty$ or to the
real axis.
\\ 
\\
{\it The genus of $\CF_{N_f}(\bfm)$}\\
For special choices of the masses $\bfm$,  $\CF(\bfm)$ coincides with the
modular curve $X(\Gamma)$ for a congruence subgroup
$\Gamma\in\slz$. Then the genus of $\CF(\bfm)$ is given by that of
$X(\Gamma)$, for which there is the formula \eqref{riemannhurwitz} in
terms of the index $n$, the number of elliptic points $\varepsilon_2$
and $\varepsilon_3$ and the number of cusps $\varepsilon_\infty$. 
 In all examples of such masses $\bfm$ discussed below, we find that
 $\CF(\bfm)$ is a genus zero Riemann surface. In the presence of
 branch points in $\CF(\bfm)$, Equation \eqref{riemannhurwitz} needs
 to be modified. First, we note that for an AD theory, $\tau_{\rm AD}$
 corresponds to an elliptic point. In fact, in all AD cases studied here,
 \eqref{n6MNLM} can be expressed as \footnote{The type IV AD point can be viewed as a collision of
  two elliptic fixed points of period $3$.}
\begin{equation}\label{n6e2e3}
n=6-2\varepsilon_3-3\varepsilon_2,
\end{equation}
For $\varepsilon_\infty$ there is no simple formula
since  for example it is not unique in some limit $\bfm\to
\bfm_{\text{AD}}$, but rather depends on the direction in mass space from which $\bfm_{\text{AD}}$ is approached. 
As the map $u:\CF(\bfm)\to \CB_{N_f}$ is between Riemann surfaces $\CF(\bfm)$ and $\CB_{N_f}$, we can consider the Riemann-Hurwitz formula \eqref{RHrs}, which relates their genera $g$. The inverse map $\tau: \CB_{N_f}\to \CF(\bfm)$ can be defined through $\tau=\frac{da_D}{da}=\frac{da_D}{du} / \frac{da}{du}$, with the periods $a$, $a_D$ given by \eqref{aaDperiods}. The dependence of $\tau$ on $u$ is holomorphic everywhere \cite{kodaira63II,kodaira63III}.
Then \eqref{RHrs} for the inverse map implies that $0=g_{\CB_{N_f}}\geq g_{\CF(\bfm)}$, such that 
\begin{equation}\label{gfm}
g_{\CF(\bfm)}=0.
\end{equation}
Applying this to the Riemann-Hurwitz formula for the ramified covering $\CF(\bfm)\to \CF$, we find the number of distinct branch points on $\CF(\bfm)$ for arbitrary $\bfm$ as
\begin{equation}\label{numberofbranchpoints}
\sum_{\tau_{\text{bp}}\in\CF(\bfm)}(e_{\tau_{\text{bp}}}-1) =\varepsilon_\infty-3+\varepsilon_2+\varepsilon_3.
\end{equation} 
This shows that $\CF(\bfm)$ is a  Riemann sphere with $\varepsilon_{\infty}$ cusps, $\varepsilon_2$, $\varepsilon_3$ elliptic points of periods 2 and 3 and $\varepsilon_\infty-3+\varepsilon_2+\varepsilon_3$ branch points. 
As an example, in massless $N_f=1$ (see Section \ref{sec:nf1masslesscurve}) we have $\epsilon_\infty=1+3=4$, while all singularities are on $\mathbb Q$ and thus $\varepsilon_2=\varepsilon_3=0$. There is one branch point in $\CF(0)$, which agrees with \eqref{numberofbranchpoints}.

\subsection{Partitioning the $u$-plane}\label{sec:uplanetiling}
\begin{figure}
	\begin{subfigure}{.49\textwidth}
		\centering
		\includegraphics[width=\linewidth]{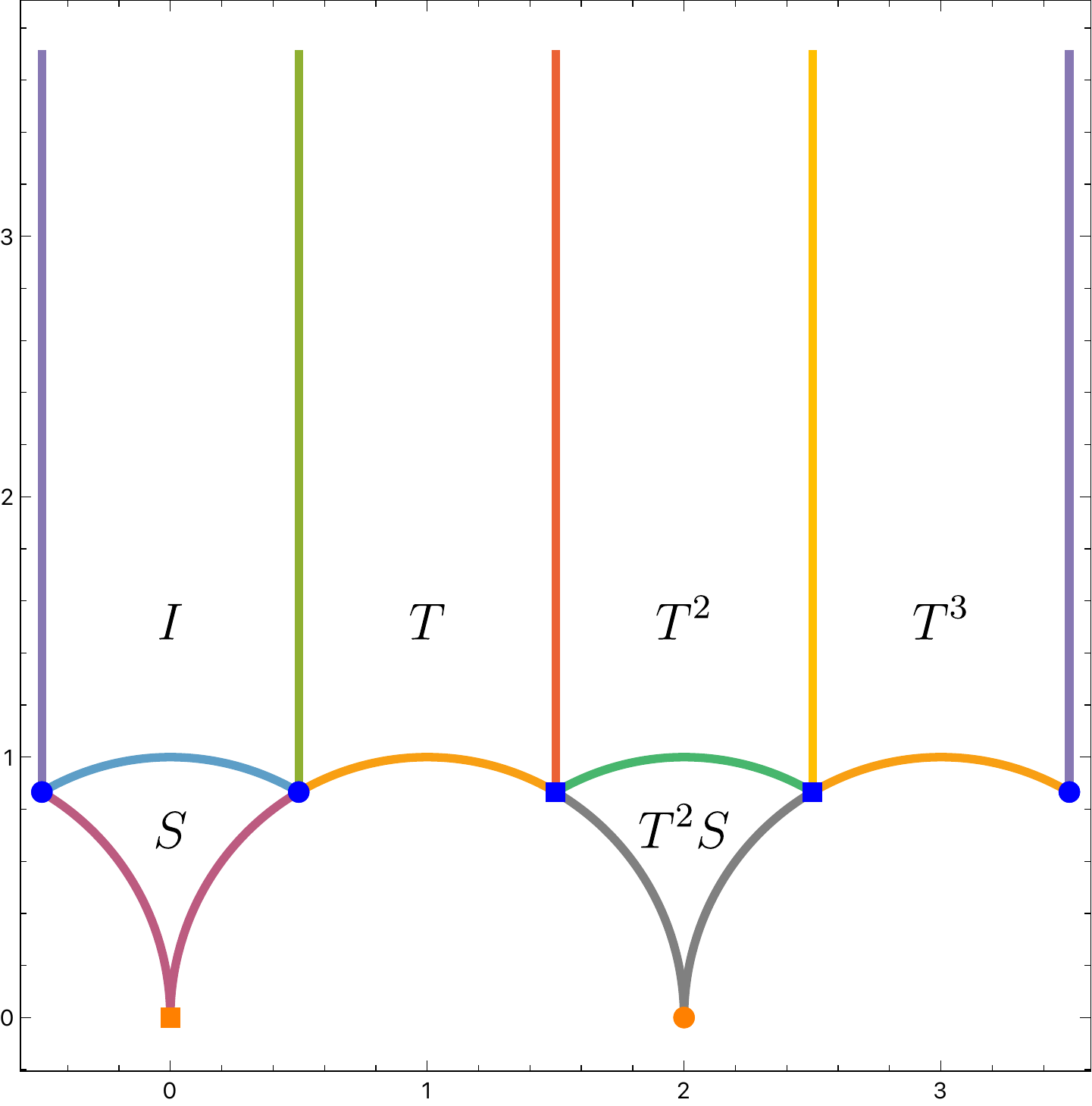}  
	\end{subfigure}
	\begin{subfigure}{.49\textwidth}
		\centering
		\includegraphics[width=\linewidth]{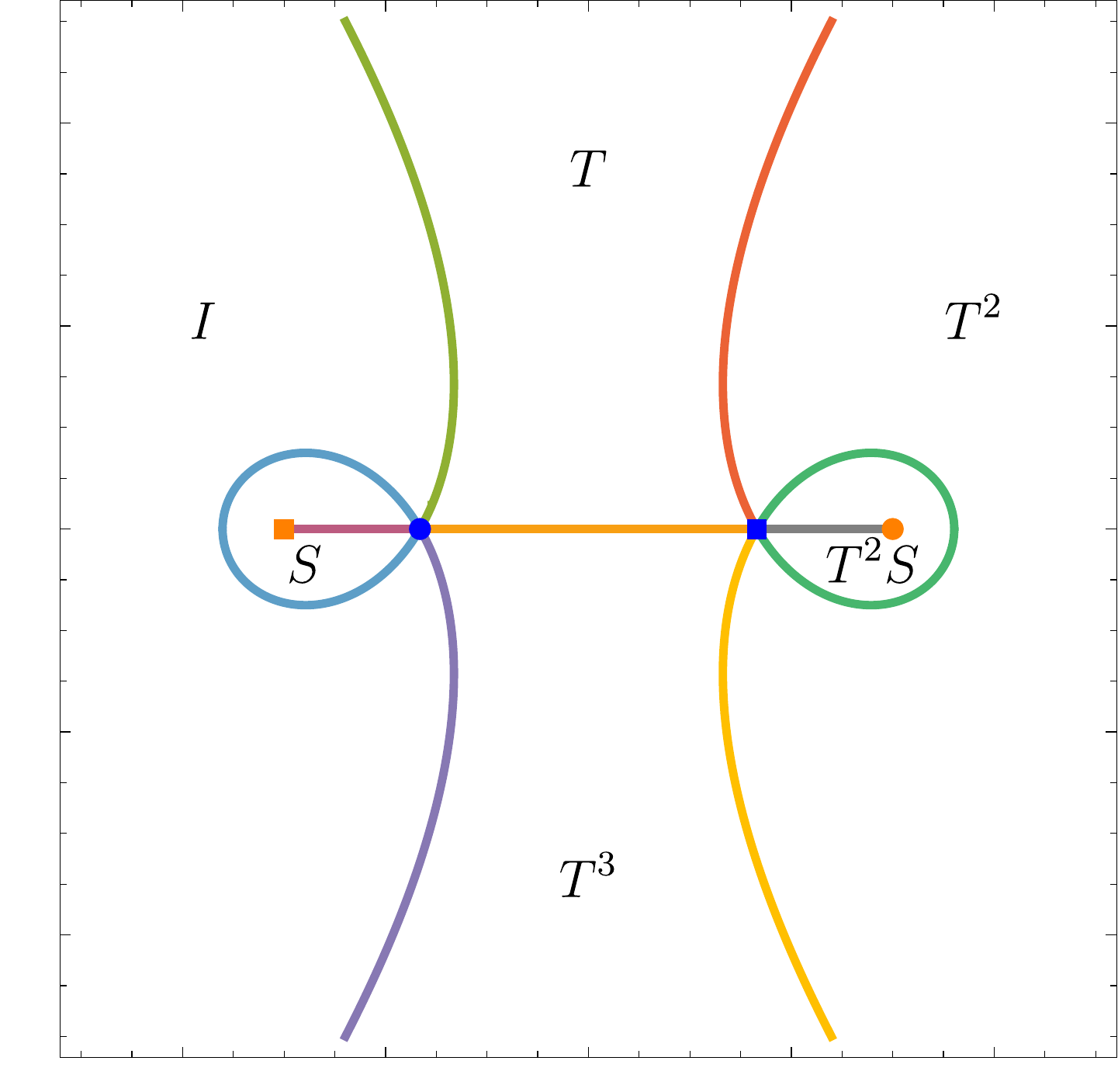}  
	\end{subfigure}
	\caption{Identification of the components of the partitioning $\CT$ in the pure theory. The $u$-plane $\CB_0$ is partitioned into 6 regions $u(\alpha\CF)$, with the $\alpha\in\slz$ given in both pictures.}
	\label{fig:purePartitions}
\end{figure}

An approach to better understand the $u$-plane geometry is to study the partitions that the map $u:\CF_{N_f}\to
\CB_{N_f}$ produces on the $u$-plane $\CB_{N_f}$. Let us study the
union \eqref{fbfm}. Now since $u(\CF_{N_f})=\CB_{N_f}$, it is natural
to ask what 
\begin{equation}\label{tiling}
\CT_\bfm=u\!\left(\bigcup_{j=1}^{n}\alpha_j\partial \CF\right)  \subseteq \CB_{N_f}
\end{equation}
describes.
The insight is that while $j:\CF\to \mathbb{C}$ is an isomorphism, it
surjects the boundary onto a half-line,\footnote{this is easy to
  prove. On the half-lines $\im[\tfrac{\sqrt3}{2},\infty)$ the
  $q$-series of $j$ is an alternating series with the same Fourier
  coefficients as $j$  and therefore real. On the arc
  $\{e^{\varphi\im}\,|\, \varphi\in
  (\tfrac{\pi}{3},\tfrac{2\pi}{3})\}$ the complex conjugate of
  $j(e^{\varphi\im})$ is equal to the value of $j$ at the
  $S$-transform of $e^{\varphi\im}$ and therefore equal to
  $j(e^{\varphi\im})$.} 
\begin{equation}\label{jboundary}
j(\partial\CF)=(-\infty,12^3]\subseteq \mathbb R\subseteq \mathbb{C}.
\end{equation}
The only other region in $\CF$ where $j$ is real are the $\slz$ images
of the  half-line $\im[1,\infty)$ on the imaginary axis. We can
directly apply this to the SW curves, whose $j$-invariant $\jsw$ is
identified with $j(\tau)$. The partitioning is then  
\begin{equation}\label{tmj}
\CT_\bfm=\{u\in \CB_{N_f}\,|\, \CJ(u,\bfm,\Lambda_{N_f})\in (-\infty,12^3]\}.
\end{equation}
It is included in the level set $\text{Im}\,  \CJ=0$.  Let us therefore study the curves
\begin{equation}\label{imj}
\text{Im} \,\CJ(u,\bfm,\Lambda_{N_f})=0,
\end{equation}
which contrary to \eqref{tmj} are algebraic curves. 
It turns out that some of the components of this equation do 
not belong to the partitioning \eqref{tmj}, and it is clear that they
correspond to components of curves with $j>12^3$. Due to the imaginary
part, it is instructive to choose coordinates $u/\Lambda_{N_f}^2=x+\im
y$. The equations \eqref{imj} are straightforward to
compute in terms of zero-loci of polynomials in $x$ and $y$. For
fixed $\bfm$, they define algebraic varieties 
\begin{equation}\label{tm0}
T_{\bfm}(x,y)=0.
\end{equation}
More specifically, they are an $N_f$-parameter family of affine algebraic plane curves.
For the pure $N_f=0$ theory,  one finds
\begin{equation}\nonumber
T_{0}=x y (81 - 288 x^2 + 336 x^4 - 128 x^6 + 288 y^2 - 352 x^2 y^2 - 
   128 x^4 y^2 + 336 y^4 + 128 x^2 y^4 + 128 y^6).
\end{equation}
The identification of this partitioning of the $u$-plane for the pure
theory is shown in Figure \ref{fig:purePartitions}. The defining
equations can be computed in full generality for any $N_f$, but they
are rather lengthy: The polynomials $T_\bfm$ for generic masses have
total degree $8+N_f$. For generic real masses, the polynomials
$T_\bfm$ have 30, 131, and 1081 terms in $N_f=1$, $2$ and $3$,
respectively. If we allow the masses to be complex, we can decompose
$m_i=\text{Re}\,m_i+\im \text{Im}\,m_i$ and the $T_m$ are then polynomials
in $x$, $y$, $\text{Re}\,m_i$ and $\text{Im}\, m_i$. For generic
(complex) masses in $N_f=1$, $2$ and $3$, $T_\bfm$ has 93, 1310 and
48754 terms, respectively.

The polynomials $T_\bfm$ are in general reducible. For instance, for $\bfm=(m,m)$ and $\bfm=(m,0,0)$, $T_\bfm$
factors into multiple nontrivial polynomials. It is straightforward to check that $T_\bfm$ for given $N_f$ flows into $T_\bfm$ for $N_f-1$ by decoupling one hypermultiplet. This allows to study the decoupling procedure of the fundamental domains in detail. 

The partitioning $\CT_\bfm$ is a finite union of smooth curves that
intersect. The tessellation of $\mathbb{H}$ in $SL(2,\mathbb{Z})$ images of $\CF$,
\begin{equation}\label{slztessellation}
\CT_{\mathbb H}=\bigcup_{\alpha\in\slz}\alpha(\partial\CF)=\left\{\tau\in\mathbb H\, | \, j(\tau)\leq 12^3\right\},
\end{equation}
has intersection points $\tau\in\slz\cdot e^{\frac{\pi\im}{3}}$, where
$j(\tau)=0$. From \eqref{CJj} we see that these intersection points
correspond to $\CJ(u,\bfm,\Lambda)=0$, whose only solutions are given
by $g_2(u,\bfm,\Lambda)=0$ (see \eqref{jinvg2g3}). Since $g_2$ is a
polynomial in $u$ of degree $2$ for all curves \eqref{eq:curves},
there are at most two intersection points  in $\CT_\bfm$ corresponding
to $\CJ=0$. As $g_2$ is strictly quadratic, there is also always at
least one such point. We find below that when the branch points (as
introduced in Section \ref{sec:ramifications}) belongs to $\CT_\bfm$,
they give further intersection points of $\CT_\bfm$.  

One can study how the partitioning is deformed upon varying the masses. For the cases where the branch points belong to $\CT_\bfm$, the complex $u$-plane is generically partitioned into 6 regions. When going to the AD locus two or more of these regions shrink to a point together with at least one branch point. At precisely $m=\mad$, the $u$-plane is then partitioned into $\leq 4$ regions,
giving an explanation for the discontinuous decrease in the index in
the limit $m\to\mad$. This can also be understood directly from the
polynomials $T_\bfm(x,y)$. For instance, at the point
$m=\mad=\frac34\Lambda_1$ in $N_f=1$, the polynomial $T_{\mad}(x,y)$
contains a factor $9 - 24 x + 16 x^2 + 16 y^2$. Its zero locus in
$\mathbb R^2$ is just a point $x+\im y=\frac34=\uad/\Lambda_1^2$, while the
massive deformation away from $\mad$ describes a curve that encloses a
region. For $\ubp\notin \CT_\bfm$ one needs to cut and glue interior
points of different regions and the $u$-plane is therefore partitioned
into less than 6 regions. See for example Fig. \ref{fig:nf1complexm}.

\section{Matone's relation for massive theories}
\label{sec:matone}

In pure $\CN=2$ supersymmetric gauge theories, there is a striking
expression for the derivative $du/d\tau$ in terms of the discriminant
$\Delta_0$ and $da/du$. The
relation reads \cite{Matone:1995rx, matone1996},
\begin{equation}\label{matonenf0} 
\frac{d u}{d\tau}=-4\pi\im\Delta_0\left(\frac{da}{d u}\right)^2.
\end{equation}
Since $u$ is proportional to $\partial F/\partial \Lambda_0$, this
equation is equivalent to a recursion relation for the prepotential $F$ \cite{Sonnenschein_1996,Eguchi:1995jh,Flume_2004}.
Moreover, as $\Delta_0$ and $da/du$ are both topological couplings,
this is a useful relation for evaluation of the $u$-plane integral
\cite{Moore:1997pc, Korpas:2017qdo, Korpas:2019ava}. Similar relations have also been obtained in the massless $N_f=1,2,3$ theories \cite{Kanno:1998qj}. We will refer to a relation of the type \eqref{matonenf0} as Matone's relation. 

In this Section, we derive a generalisation of \eqref{matonenf0} for massive $N_f=1,2,3$. Section \ref{SecPeriods} derives expressions for $da/du$ and
$\Delta_{N_f}$ as functions of $\tau$. Section \ref{MrelNf} derives
Matone's relation (\ref{matonepnf}) for generic $N_f\leq 3$.

\subsection{Periods and Weierstra{\ss} form}
\label{SecPeriods}

We proceed by deriving an expression for $da/du$. To this end, recall
that $a$ is given as a period integral (\ref{aaDperiods}), and that the derivative
of the SW differential $\lambda$ to $u$ is holomorphic \cite{Seiberg:1994aj}. Therefore, we can express
$da/du$ in terms of the variables $x$ and $y$ of
(\ref{weierstrassequation}) 
\be
\label{daduperiod}
\frac{da}{du}= \frac{\sqrt{2}}{4\pi} \int_{\gamma} \frac{dx}{y},
\ee
where $\gamma$ is one of the cycles of the elliptic curve. To determine this quantity for the theories with $N_f\leq 3$, we map the curve $\CW$ to the modular
Weierstra{\ss} form $\widetilde \CW$, $A: \CW\to \widetilde \CW$. See for example
\cite[Section 7.1]{Diamond}. The curve $\widetilde \CW$ reads
\begin{equation}\label{weierstrassequation2}
\widetilde\CW: \quad \tilde y^2=4\tilde x^3-\tilde g_2 \tilde x-\tilde g_3,
\end{equation}
with the variables related by the map $A$ as
\be
A: \left\{ \begin{array}{l} \tilde x=\alpha^{2}\, x=\wp(z),\\ \tilde
    y=\alpha^{3}\, y=\wp'(z),  \\ \tilde g_2=\alpha^4\, g_2,\\  \tilde g_3=\alpha^6\, g_3,\end{array}  \right.
\ee
where $\wp$ is the Weierstra{\ss} function and $z\in \mathbb{C}$ a
coordinate on the curve. Since $\widetilde \CW$ is the modular Weierstra{\ss} curve, the variables $\tilde g_2$ and $\tilde g_3$ equal
\be
\tilde g_2=\tfrac{4\pi^4}{3}E_4,\qquad  \tilde
g_3=\tfrac{8\pi^6}{27}E_6,
\ee
with $E_k$ the Eisenstein series defined in \eqref{Ek}. 
We note that
the variables for $\CW$ \eqref{weierstrassequation} have weight 0 under
modular transformations, while in \eqref{weierstrassequation2} the
weights are $\wt(\alpha,\tilde y, \tilde x, \tilde g_2, \tilde
g_3)=(1,3,2,4,6)$.
Using the two equations for $\tilde g_2$ and $\tilde g_3$, 
we can solve for $u$ and $\alpha$. 
 The relation
\begin{equation}\label{alphaE6E4}
\alpha=\frac{\sqrt{2}\pi}{3} \sqrt{\frac{g_2}{g_3} \frac{E_6}{E_4}},
\end{equation}
will be particularly useful for us in the next subsection. This
relation can also be derived using Picard-Fuchs equations
\cite{Brandhuber:1996ng}. 

Now it is straightforward to determine $da/du$ (\ref{daduperiod}) using the
Weierstra{\ss} representation of $(\tilde x,\tilde y)$,
\be
\label{alphadadu}
\frac{da}{du}=\frac{\sqrt{2}\,\alpha}{4\pi} \int_{\tilde \gamma}
\frac{d\tilde x}{\tilde y}= \frac{\sqrt{2}\,\alpha}{4\pi},  
\ee 
where $\tilde \gamma$ is the image of the $\gamma$ under the map
$A$, with the variable $z$ of $\tilde x(z)$ changing from 0 to 1.

We continue by studying the discriminants of $\CW$ and $\widetilde \CW$. 
Using $E_4^3-E_6^2=12^3\,\eta^{24}$ with $\eta$ as in (\ref{etafunction}), we find for the discriminant of
$\widetilde\CW$, $\tilde\Delta=(2\pi)^{12}\,\eta^{24}$. The discriminant of $\CW$,
$\Delta_{N_f}$ (\ref{deltag2g3}), on the other hand is a polynomial in $u$,
$\bfm$ and $\Lambda$ and therefore has weight $0$. The two discriminants are related by  
\begin{equation}\label{mathphysdiscriminant}
\tilde \Delta=\alpha^{12} (-1)^{N_f} \Lambda_{N_f}^{8-2N_f} \Delta_{N_f},
\end{equation}  
or substituting $\alpha$ in terms of $da/du$ (\ref{alphadadu}),
\begin{equation}\begin{aligned}\label{dadu12Delta}
\eta^{24}&= 2^6(-1)^{N_f}\Lambda^{2(4-N_f)}_{N_f}\left(\frac{da}{du}\right)^{12}\Delta_{N_f},
\end{aligned}\end{equation}
which holds for $0\leq N_f\leq 3$. Similar expression exist for
$N_f=4$ and $\nstar$ \cite{AFM:future, Manschot:2021qqe}.

Let us consider the case that $\CW$ or $\widetilde{\CW}$ is
singular. The curve $\widetilde{\CW}$ is only singular at the cusps
$\tau\in\{\im\infty\}\cup\BQ$, since $\tilde\Delta\sim \eta^{24}$ vanishes at the
cusps and is non-vanishing for $\tau$ in the interior of $\BH$. From
\eqref{mathphysdiscriminant} we see that, at the cusps of
$\widetilde{\CW}$ either $da/du$ or $\Delta_{N_f}$ must vanish. On the
other hand, for $\tau$ in the interior of $\BH$, $\tilde\Delta$ is
non-vanishing. This means that, if $\CW$ is singular
($\Delta_{N_f}=0$) for such values of $\tau$, $da/du$ should
diverge. This is exactly what happens at the AD points, 
\begin{equation}\label{ADconditions}
    \frac{du}{da}(\tad)=0,\qquad \Delta_{N_f}(u(\tad))=0,\qquad \tad\in\BH.
\end{equation}
We can further note that $\tfrac{du}{da}(\tau)=0$ is true also for
singularities that are cusps and not elliptic points, i.e.,
$\Delta_{N_f}=0$ for $\tau\in\BQ$. This is because if $u$ is not an
elliptic point then $g_2\neq0$ and $g_3\neq 0$, since otherwise, from
$\Delta=g_2^3-27g_3^2$, both would be zero, giving an elliptic
point. Then, from \eqref{alphadadu} we have that
$\left(\tfrac{du}{da}\right)^2$ is proportional to
$\tfrac{E_4}{E_6}$. This is a meromorphic modular form of weight $-2$
for $\SL$, and one can show using modular transformations that it
vanishes on $\BQ$. Therefore, we have that $\Delta_{N_f}=0$ implies
$\tfrac{du}{da}=0$.

\subsection{Matone's relation}
\label{MrelNf}
We will now give a generalisation of \eqref{matonenf0} that holds also
for the massive $N_f=1,2,3$ theories. Let us denote by ${}'$ the
derivative with respect to $u$ keeping $\bfm$ and $\Lambda_{N_f}$
fixed. The derivative with respect to $\tau$ is always
given explicitly. From the
explicit expression for $j$ as function of $\tau$
\eqref{je4e6}, it is easy to check that $\frac{d}{d\tau}j=-2\pi\im
\frac{E_6}{E_4}j$. Using the chain rule and (\ref{CJj}), we can express this as
$\frac{d}{d\tau}\CJ=\CJ'\,\frac{du}{d\tau}$. 
This gives the first important 
identity, 
\begin{equation}
\frac{du}{d\tau}=-2\pi\!\im\frac{E_6}{E_4}\frac{\CJ}{\CJ'},
\end{equation}
which holds for any SW curve. From \eqref{jinvg2g3} we can compute
$\CJ'$ in terms of $g_2'$ and $g_3'$. Using relations \eqref{alphaE6E4}
and \eqref{alphadadu}, we can substitute  $E_6/E_4$ in terms of $g_2$,
$g_3$ and $da/du$. This gives the exact relation
\begin{equation}\label{matoneg2g3}
\frac{du}{d\tau}=-72\,\pi \im\,\frac{g_3}{g_2}\frac{\CJ}{\CJ'}\left(\frac{da}{du}\right)^2=-\frac{8\pi\im}{3}\frac{g_2^3-27g_3^2}{2g_2g_3'-3g_2'g_3}\left(\frac{da}{du}\right)^2.
\end{equation}
An analogous formula for five-dimensional gauge theories was derived
from the Picard-Fuchs perspective in \cite[Eq. (4.23)]{Eguchi:2000fv}. Both factors on the rhs are only relative invariants, but their product is an absolute invariant of the curve $\CW$.
The numerator on the rhs is proportional to the physical
discriminant. The equation has modular weight 2, since both
$\frac{du}{d\tau}$ and $\left(\tfrac{da}{du}\right)^2$ are of weight 2.

For $0\leq N_f\leq3$,\footnote{We can in fact perform the same
  computations in the case of $N_f=4$, leading to a similar formula.}
we can compute the corresponding $g_i$, and one can rewrite
\eqref{matoneg2g3} as  
\begin{equation}\label{matonepnf}
\frac{du}{d\tau}=-\frac{16\pi\im}{4-N_f}\frac{\Delta_{N_f}}{\pma_{N_f}}\left(\frac{da}{du}\right)^2,
\end{equation}
where we substituted (\ref{deltag2g3}) for $\Delta_{N_f}$, and
defined the polynomial $\pma_{N_f}$,
\be\label{PMdefinition}
\pma_{N_f}=\,\frac{6}{4-N_f}\,(-1)^{N_f}\,\Lambda^{2N_f-8}_{N_f}\,(2g_2g_3'-3g_2'g_3).
\ee
The normalization is chosen such that $\pma_{N_f}$ is a monic
polynomial. Explicit computation gives, 
\begin{equation}\begin{aligned}\label{pnf}
\pma_0=\,&1, \\
\pma_1=\,&u-\tfrac43m_1^2, \\
\pma_2=\,&u^2-\tfrac32(m_1^2+m_2^2)u+2m_1^2m_2^2+\tfrac18m_1m_2\Lambda_2^2-\tfrac{1}{64}\Lambda_2^4, \\
\pma_3=\,&u^3-2 M_2 u^2+\left(3 M_4'+\tfrac34 M_3\Lambda_3-\tfrac{1}{64} M_2\Lambda_3^2\right)u+\tfrac{1}{256} M_3\Lambda_3^3 \\&-\tfrac14  M_2 M_3\Lambda_3+\tfrac{1}{32}( M_4- M_4')\Lambda_3^2-4 M_3^2,
\end{aligned}\end{equation}
where we defined
\begin{equation}\begin{aligned}\label{nf3masspoly}
 M_2&=m_1^2+m_2^2+m_3^2, \quad  & M_3&=m_1m_2m_3,\\
 M_4&=m_1^4+m_2^4+m_3^4, \quad & M_4'&=\sum_{i<j}m_i^2m_j^2.
\end{aligned}\end{equation}
We note that these polynomials appear in the Picard-Fuchs equations
for the periods of these theories and their zeros give regular
singular points of the differential equations \cite{Ohta:1996hq,
  Ohta_1997}.  In Appendix \ref{sec:matone2} we give an additional
proof of the identities \eqref{matonepnf}. \footnote{The identity
  \eqref{matonepnf} does in fact not depend on the specific form of
  the SW curves. Given a Jacobian rational elliptic surface, let
  $\omega=\int_{\gamma}\frac{dx}{y}$ be the period of the N\'eron
  differential on the elliptic curve. Then
  $\frac{du}{d\tau}=\frac{1}{3\pi\im}\omega^2
  \Delta/(2g_2g_3'-3g_2'g_3)$, with $u$ a coordinate on $\mathbb
  P^1(\mathbb C)$.}

\subsection{Branch points}\label{sec:branchpoints}
An important difference between $N_f=0$ and $N_f>0$ are the poles
where $\pma_{N_f}$ vanishes. To understand these poles  as well as 
zeros of $du/d\tau$, note that at such points the change of variables
between $u$ and $\tau$ is ill-defined. We have seen earlier that the
change of variables is ill-defined at the points where the
discriminant $D(P_{N_f})$ (\ref{poldiscfactor0}) vanishes. Indeed if
we substitute $\CJ(u,\bfm, \Lambda_{N_f})$ for $j(\tau)$ in
$D_{N_f}^{\rm bp}$, $\pma_{N_f}$ factors out.

The reason for this is the following. The discriminant of a polynomial $p$ vanishes if and only $p$ has a double root. It can be computed as the \emph{resultant} of the polynomial and its formal derivative, $D(p)\sim\text{Res}_X(p,p')$ (see also \cite{Eguchi:1996vu}).\footnote{The resultant of two polynomials over a commutative ring is a polynomial of their coefficients which vanishes if and only if the polynomials have a common root. It can be computed as the determinant of their Sylvester matrix.} The zero locus $D(P_{N_f})=0$ of  $P_{N_f}(X)$ is then given by the solutions to the two equations $P_{N_f}(X)=0$ and ${P_{N_f}}'(X)=0$. Since $\Delta_{N_f}\neq0$, all solutions can be found by solving the former for $j$ and inserting into the latter. It is straightforward to show that this gives
\begin{equation}\label{bpPD}
\frac{g_2^2g_3}{\Delta_{N_f}}P_{N_f}^\text{M}=0,
\end{equation}
which provides the decomposition \eqref{poldiscfactor0}: If $g_2=0$ but $g_3\neq0$, then  $j=0$. If $g_3=0$ but $g_2\neq 0$, then $j=12^3$. If both $g_2=g_3=0$, we are in $\CL^{\text{AD}}_{N_f}\subseteq \CL^{\Delta}_{N_f}$. 
Now since the sextic equation is only well-defined away from the physical discriminant locus  $\CL^{\Delta}_{N_f}$ where $\Delta_{N_f}=0$, the true branch point locus $\CL^{\text{bp}}$ is the difference of the Matone locus $\CL^{\text M}_{N_f}=\{u\in \mathbb \CB_{N_f}| P_{N_f}^{\text M}=0\}$ and the discriminant locus,
\begin{equation}
\CL^{\text{bp}}_{N_f}=\CL^{\text M}_{N_f}\setminus \CL^{\Delta}_{N_f}.
\end{equation}

On the Coulomb branch with $N_f$ hypermultiplets there are generically $2+N_f$ distinct singular points. For special mass configurations $\bfm$, some singularities can collide. Then $\Delta_{N_f}$ has a double root. From above it is clear that this is equivalent to $D(\Delta_{N_f})=0$, which in turn is equivalent to  $\Delta_{N_f}=0$ and $\Delta_{N_f}'=0$. We can again solve the former for $g_2$ and $g_3$ and insert into the latter to obtain $P^{\text{M}}_{N_f}\sim\frac{g_2'}{g_3}\Delta_{N_f}=0$. This implies that whenever $\Delta_{N_f}$ has a double root, it is also a root of $P^{\text{M}}_{N_f}$. It is also observed in all examples below. To be more precise, if $\Delta_{N_f}$ contains a root of $d>1$-th order, then $\Delta'_{N_f}$ has the same root but with multiplicity $d-1$. The excess factors can  be extracted by the operation $\gcd(\Delta_{N_f},\Delta'_{N_f})$, where $\gcd$ is the polynomial greatest common divisor. The multiple roots are  removed from the discriminant by the square-free factorisation \footnote{The polynomial gcd is unique only up to multiplication with invertible constants, we choose it such that $\widehat\Delta_{N_f}$ is again monic.}
\begin{equation}\label{gcd}
	\widehat\Delta_{N_f}= \frac{\Delta_{N_f}}{\gcd(\Delta_{N_f},\Delta'_{N_f})}.
\end{equation}
This \emph{reduced} discriminant $\widehat\Delta_{N_f}$  has single
roots only, concretely we map $\prod_s (u-u_s)^{n_s}$ to $
\prod_s(u-u_s)$. This quantity is also of importance for determining gravitational
couplings to Seiberg-Witten theory \cite{Huang:2011qx}. 
One can show that $\gcd(\Delta_{N_f},\Delta'_{N_f})$ always divides $P^{\text{M}}_{N_f}$, such that 
\begin{equation}\label{pmhat}
\widehat{P}^{\text M}_{N_f}\coloneqq \frac{\widehat\Delta_{N_f}}{\Delta_{N_f}}\pma_{N_f}
\end{equation}
is in fact a polynomial. The branch point equation \eqref{bpPD}  is then equivalent to  $\widehat{P}^{\text M}_{N_f}/\widehat \Delta_{N_f}=0$, which reduces to 
\begin{equation}\label{pma}
\widehat{P}^{\text M}_{N_f}=0.
\end{equation}
The Matone relation thus always takes the form
\begin{equation}\label{matonehat}
\frac{du}{d\tau}=-\frac{16\pi\im}{4-N_f}\frac{\widehat\Delta_{N_f}}{\widehat{P}^{\text M}_{N_f}}\left(\frac{da}{du}\right)^2,
\end{equation}
where both $\widehat\Delta_{N_f}$ and $\widehat{P}^{\text M}_{N_f}$ are polynomials. 
In the subsequent sections we show explicitly that the roots of the denominator $\eqref{pma}$ are precisely the branch points. We note that for generic masses the form \eqref{matonehat} does not differ from \eqref{matonepnf}, because $\,\,\widehat{}\,\, $ is trivial when all roots are distinct. 
 
As argued above, AD points correspond to points $\tad$ in the
upper half-plane. Since they lie on the discriminant locus, we exclude
them to define the  sextic polynomial $P_{N_f}$. We will discuss in
more detail below that, if the masses approach the AD locus, a branch point in the $u$-plane collides with two mutually
non-local singularities forming the AD point. The branch point under
consideration lifts, while the $N_f-1$ other branch points remain for
a generic point on the AD mass locus $\CL^{\rm AD}_{N_f}$. Thus for a generic point on the AD mass locus, AD
points are not branch points of $u(\tau)$. A
non-generic example is the most symmetric AD theory, the
$IV$ fibre in $N_f=3$, discussed in more detail in Section
\ref{31ADtheroy}. For this theory, $\tad$ corresponds to a singular
point of the theory as well as a branch
point. As a result, the domain for $\tau$ does not correspond to that
of a congruence subgroup of $\slz$. 
 
Since any branch point $\tau_{\rm bp}$ induces a non-trivial
monodromy, $u$ does not have a regular Taylor series at such a
point. For instance, if the $u$-plane contains one branch point
$u_{\text{bp}}=u(\tau_{\text{bp}})$, then we have
$u(\tau)-u_{\text{bp}}=\CO(\sqrt{\tau-\tau_{\text{bp}}})$ as $\tau\to
\tau_{\text{bp}}$. If the leading coefficient is nonzero, then
$\frac{du}{d\tau}$ diverges at $\tau_{\text{bp}}$. Away from the
discriminant locus, this can be understood from \eqref{matonehat}:
From \eqref{alphadadu} we see that $\frac{da}{du}$ is regular and
nonzero at a branch point, since none of $g_2$, $g_3$, $E_4$ and $E_6$
diverge or vanish. Thus the zeros of the denominator $\widehat
P_{N_f}^{\text M}$ correspond to the singular points of
$\frac{du}{d\tau}$, as observed.  

This can also be seen directly from the $\CJ$-invariant of the SW curve.  It is easy to show that 
\begin{equation}
\CJ'=36^3\frac{g_2^2 g_3}{\Delta_{N_f}^2} P_{N_f}^{\text M},
\end{equation}
which due to \eqref{bpPD}  vanishes at any branch point $u_{\text{bp}}$. Since for fixed mass and scale $\CJ(u)$ is rational in $u$, it is a meromorphic function on $\CB_{N_f}$. Away from the discriminant locus it thus has a Taylor series around $u_{\text{bp}}$, where the linear coefficient is missing. We therefore find 
\begin{equation}\label{Ju-Jubp}
\CJ(u)-\CJ(u_{\text{bp}})=\CO\left((u-u_{\text{bp}})^{n_{\text{bp}}}\right),
\end{equation}
with $n_{\text{bp}}\geq 2$. Now we identify $\CJ(u)=j(\tau)$, which relates the power series of $u$ and $\tau$. For a generic $\tau\in\mathbb H$, $j$ has a regular Taylor series at $\tau$ with non-zero linear coefficient. However if $\tau$ is in the $\slz$-orbit of $\im$ or $e^{\frac{\pi\im}{3}}$, $j$ has a zero of order $2$ or $3$. Let $n_{\tau_{\text{bp}}}\in\{1,2,3\}$ be this number for a given branch point $\tau_{\text{bp}}\in\mathbb H$. Then $\CJ(u)-\CJ(u_{\text{bp}})=\CO\left((\tau-\tau_{\text{bp}})^{n_{\tau_{\text{bp}}}}\right)$, such that from \eqref{Ju-Jubp} we conclude 
\begin{equation}
u(\tau)-u_{\text{bp}}=\CO\left((\tau-\tau_{\text{bp}})^{n_{\tau_{\text{bp}}}/n_{\text{bp}}}\right),
\end{equation}
where the  leading coefficient is strictly non-zero. From this we see that the branch point $\tau_{\text{bp}}$ does not necessarily correspond to an $n_{\text{bp}}$-th root, but since the ratio can cancel $\tau_{\text{bp}}$ rather corresponds to a branch point of order 
\begin{equation}\label{BPorder}
\frac{n_{\text{bp}}}{\gcd(n_{\text{bp}},n_{\tau_{\text{bp}}})}.
\end{equation}
It is difficult to compute this integer for a generic branch point, however in all examples discussed below it is equal to $2$, which corresponds to a \emph{square} root.

If the number $\frac{n_{\tau_{\text{bp}}}}{n_{\text{bp}}}\in\mathbb Q\setminus\mathbb Z$ is larger than 1, then it is clear that $\frac{du}{d\tau}(\tau_{\text{bp}})=0$. Conversely, if $\frac{n_{\tau_{\text{bp}}}}{n_{\text{bp}}}<1$ then $\frac{du}{d\tau}(\tau_{\text{bp}})=\infty$. We thus see that any branch point has the property that $\frac{du}{d\tau}$ diverges or vanishes, such that the change of variables from the $u$-plane to the $\tau$-plane is not well-defined.

The branch point locus also allows to find the effective coupling
at the AD points. In the limit where the masses approach the AD locus,
$\bfm\to \bfm_{\text{AD}}$, the AD point $u_{\rm AD}$ is the point where branch
points $u_{\rm bp}$ in the $u$-plane merges with mutually non-local singularities.  
While away from $\bfm_{\text{AD}}$ the effective coupling
$\tau$ of the singularities remain as distinct cusps on the real line, the
branch points move along certain paths inside $\mathbb H$. In an AD
limit $\bfm\to \bfm_{\text{AD}}$, a number of pairs of branch points, $\tau_{\rm bp}$
and $\tau_{\rm bp}'$, coincide at the intersection
of copies of $\CF$, and the branch cut will then disconnect regions
from $\CF_{N_f}$. The effective coupling of the AD
point $\tau_{\rm AD}$ is therefore given by
that of the merged branch points. This is an efficient way to
determine $\tau_{\rm AD}$, which otherwise can only be found by inverting modular
functions. Moreover, if the duality group is a congruence subgroup of $\slz$,
$\tau_{\rm AD}$ corresponds to an elliptic point of the duality group.

\section{The $N_f=1$ curve}\label{sec:nf1}
To make the above discussions more concrete we will now go on to study some specific examples. We will start by including one hypermultiplet. The $N_f=1$ theory has been discussed in some detail in  \cite{AlvarezGaume:1997ek,AlvarezGaume:1997fg,Dolan:2005dw,dolan:2006zc,Huang:2009md}.  

In the massive $N_f=1$ theory, there are three (in general) distinct strong coupling singularities where a hypermultiplet becomes massless. These remain at distinct points in the massless limit, while for special values of the mass two of them can merge into AD points. To analyse the $N_f=1$ theory we will start by restricting to the massless case and then go to an AD mass. Here we can find closed expressions for $u$ in terms of well-known modular forms. Only in the AD case does the theory turn out to be modular. In the end we can use the knowledge gained from these cases to draw some conclusions of the general massive case.

\subsection{The massless theory}\label{sec:nf1masslesscurve}
Let us begin with the massless $N_f=1$ theory. Using the procedure outlined in Sec. \ref{sec:sexticeqn} we find \cite{Nahm:1996di}
\begin{equation}\label{nf1u}
\begin{aligned}
\frac{u(\tau)}{\Lambda_1^2}=&-\frac{3}{2^{\frac73}}\frac{\sqrt{E_4(\tau)}}{\sqrt[3]{E_4(\tau)^{\frac32}-E_6(\tau)}}\\
=&-\frac{1}{16}(q^{-1/3}+104q^{2/3}-7396q^{5/3}+\CO(q^{8/3})),
\end{aligned}
\end{equation}
where we again have made the choice of solution consistent with our convention, such that $u\to -\infty$ for $\tau\to \im\infty$. This function also appears as an order parameter in pure $SU(3)$ SW theory \cite{Aspman2021} as well as in the description of certain elliptically fibred Calabi-Yau spaces \cite{Klemm:2012sx}.
The singularities of the curve are $\frac{u^3}{\Lambda_1^6}=-\frac{3^3}{2^8}$. They are associated with states of charges $(1,0)$, $(1,1)$ and $(1,2)$ becoming massless. The global $\mathbb Z_3$ symmetry acts as $T^{-1}: u(\tau-1)=\omega_3 u(\tau)$, with $\omega_j=e^{\frac{2\pi\im}{j}}$. 

By restricting to the imaginary axis, we can perform the $S$-transformation. For this, let $\tau=\im\beta$ with $\beta>0$. We have that $E_4(\im/\beta)=(\im\beta)^4E_4(\im\beta)=\beta^4E_4(\im\beta)$. Taking the square root is unambiguous since $E_4$ is real on the imaginary axis and $\beta^4$ is positive. This gives $E_4^{\frac12}(\im/\beta)=\beta^2E_4^{\frac12}(\im\beta)$. On the other hand for $E_6$ we have $E_6(\tau)=(\im\beta)^6 E_6(\im\beta)=-\beta^6E_6(\im\beta)$. This implies that the relative sign of $E_6$ flips, and it holds for $ \tau_D\in\im \mathbb R_{>0}
$ that 
\begin{equation}\begin{aligned}\label{nf1stransformation}
\frac{u_D(\tau_D)}{\Lambda_1^2}&=-\frac{3}{2^{\frac73}}\frac{E_4(\tau_D)^{\frac12}}{\left(E_4(\tau_D)^{\frac32}+E_6(\tau_D)\right)^{\frac13}}\\
&= \frac{3}{2^{\frac83}}\left(1+ 144 q - 3456 q^2 + 596160 q^3+\CO(q^4)\right).
\end{aligned}\end{equation}
With the $\mathbb Z_3$ symmetry $u(\tau-1)=\omega_3 u(\tau)$ this confirms the strong coupling singularities given above.

The monodromies on the massless $N_f=1$ $u$-plane are  \cite{Seiberg:1994aj}
\begin{equation}\begin{aligned}\label{nf1monod}
M_1&=\left(\begin{smallmatrix}
 1 & 0 \\
 -1 & 1 \\
\end{smallmatrix}\right)= STS^{-1},\\
  M_2&=\left(
\begin{smallmatrix}
 0 & 1 \\
 -1 & 2 \\
\end{smallmatrix}\right)=(TS)T(TS)^{-1}, \\
 M_3&=\left(
\begin{smallmatrix}
 -1 & 4 \\
 -1 & 3 \\
\end{smallmatrix}\right)=(T^2S)T(T^2S)^{-1}, \\
 M_\infty&=\left(
\begin{smallmatrix}
 -1 & 3 \\
 0 & -1 \\
\end{smallmatrix}\right)=PT^{-3},
\end{aligned}\end{equation}
where $P=S^2=-\mathbbm 1$. Note that these matrices generate the full $\slz$
modular group rather than a (congruence) subgroup. Indeed, as fractional linear
transformations acting on the complex structure 
through their matrix representations, they do not leave $u$
invariant. However, we can consider these
matrices as compositions of paths in the fundamental domain,
and as such they do leave $u$ invariant. To make the connection to the
discussion in \cite{Aspman2021} more direct we can note that by using
another choice of homology basis in the present case we can construct
a different set of monodromies, see for example \cite{Eguchi1999},
which exactly coincides with the ones listed for the $SU(3)$ case of
\cite{Aspman2021}.

Since $E_4$ has a simple zero at $\tau=\omega_3$ (and $\slz$-images), $u(\tau)$ has a branch point at $\tau_{\text{bp}}=\omega_3$. The function $u(\tau)$ does not possess a Taylor series at $\tau_{\text{bp}}$ and is therefore not holomorphic at $\tau_{\text{bp}}$. Since $u(\tau_{\text{bp}})=0$, the branch point of $u(\tau)$ indeed agrees with what is found in \eqref{pbp}. Since $u$ is not holomorphic on $\mathbb H$, it can also not be classically modular. Another reason why $u$ is not modular is the following. If we define $x\coloneqq -16\frac{u}{\Lambda_1^2}=q^{-\frac13}+\CO(q^{\frac23})$, then one can read off from the curve that $\CJ=x^6/(x^3-432)$. This implies that  $u$ should be a Hauptmodul of an index 6 subgroup of $\slz$ with width $h(\infty)=3$ and width decomposition $6=3+1+1+1$ (see \eqref{sumwidthindex}). From the classification of index 6 groups in  Table \ref{tableindex6} we see that such a subgroup of $\slz$ does not exist. In fact, no index 6 subgroup of $\slz$ with 4 cusps exists. This distinguishes massless $N_f=1$ from massless $N_f=0,2,3$, where the duality groups are congruence subgroups isomorphic to $\Gamma^0(4)$ \cite{Nahm:1996di}.

From \eqref{nf1u} one finds 
\begin{equation}
\frac{du}{d\tau}=\frac{\pi\im\Lambda_1^2}{2^{\frac73}}\frac{E_4^{\frac32}+E_6}{E_4^{\frac12}\left(E_4^{\frac32}-E_6\right)^{\frac13}}, \qquad \frac{da}{du}=\frac{\im \left(E_4^{\frac32}-E_6\right)^{\frac16}}{2^{\frac13}\sqrt3\Lambda_1}.
\end{equation}
We can explicitly check that these satisfy Matone's relation,  \eqref{matonepnf}, for massless $N_f=1$,
\begin{equation}\label{nf1matone}
\frac{du}{d\tau}=-\frac{16\pi\im}{3}\frac{\Delta}{u}\left(\frac{da}{du}\right)^2.
\end{equation}
 
\begin{figure}
	\begin{subfigure}{.5\textwidth}
		\centering
		\includegraphics[width=\linewidth]{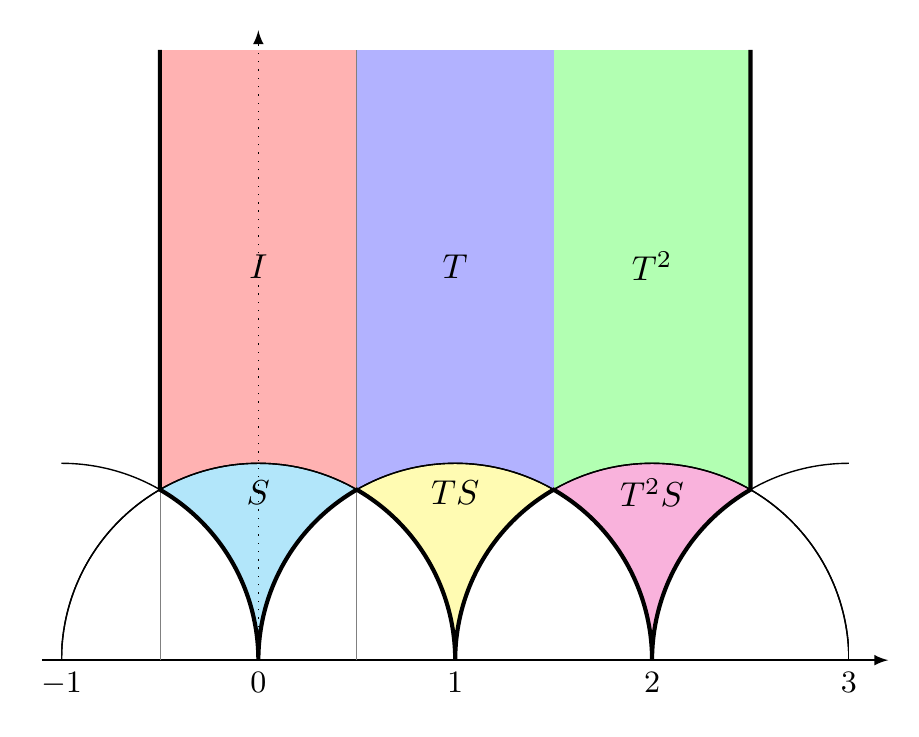}  
	\end{subfigure}
	\begin{subfigure}{.5\textwidth}
		\centering
		\includegraphics[width=.75\linewidth]{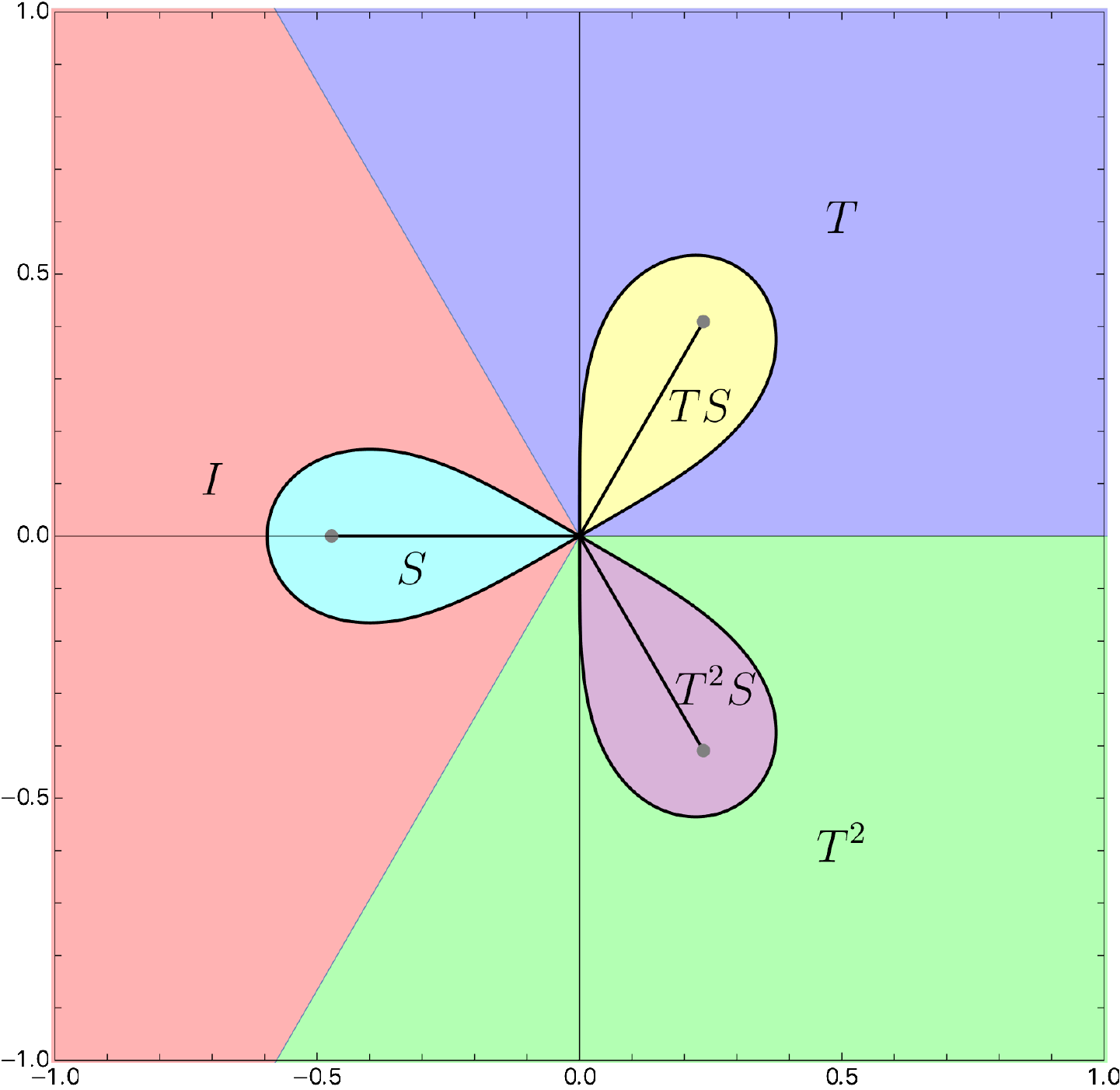}  
	\end{subfigure}
	\caption{Left: Proposal of a fundamental region for massless $N_f=1$. It is clear that it can not be a fundamental domain of any $\slz$ subgroup if we identify the sides, $\tau\sim \tau+3$. This is because the lower boundary is not given by a union of half-disks. If we do \emph{not} identify the sides then the picture is in fact a fundamental domain for $\Gamma^0(4)$. This is because there is an element in $\Gamma^0(4)$ which maps $\im\infty\mapsto 1$. Right: Plot of the massless $N_f=1$ $u$-plane as the union of the images of $u$ under the six $\slz$ images of $\CF$ as in \eqref{funddomainnf1}, in units of $\Lambda_1^2=1$. In particular, the function $u$ is surjective onto $\mathbb C$ on this domain. The singular points sit neatly in the interior of the strong coupling regions $u(S^\ell\CF)$, $\ell=0,1,2$. The origin $u=0$ is $\tau=\omega_3+\mathbb Z$ as discussed above, and bounds all six domains  as is clear from the left figure.}
	\label{fig:nf1domains}
\end{figure}

The fundamental region 
\begin{equation}\label{funddomainnf1}
\CF_{1}(0)=\bigcup_{\ell=0}^{2} T^\ell \CF\,\cup T^\ell S\CF
\end{equation}
as in \eqref{fbfm} was obtained in \cite{Aspman2021}.  It is shown in Fig. \ref{fig:nf1domains}, together with its image under $u$ to the $u$-plane. We stress that \eqref{funddomainnf1} can not be written as $G\backslash \mathbb H$ for any subgroup $G\subseteq \slz$.

 In the massless $N_f=1$ theory, the partitioning \eqref{tmj} is contained in the algebraic plane curve $T_{(0)}(x,y)=0$, where $\frac{u}{\Lambda_1^2}=x+\im y$ and 
\begin{equation}\begin{aligned}\label{t0t00}
T_{(0)}(x,y)&=y (3 x^2 - y^2) (27 x^3 + 128 x^6 - 81 x y^2 + 384 x^4 y^2 + 
   384 x^2 y^4 + 128 y^6).
   \end{aligned}\end{equation}
The first two factors of $T_{(0)}(x,y)$ contains also values which correspond to $j>12^3$ and they need to be sufficiently truncated. The identification of the algebraic curve with the partitioning of $\mathbb H$ is immediate from Fig. \ref{fig:nf1domains}.

\subsection{Type $II$ AD mass}\label{sec:nf1ad}
In Section \ref{sec:ramifications} we saw that the $N_f=1$ theory has
AD fixed points in its moduli space. To study these, we fix the mass
to be one of the AD values, $m\to \mad=\frac{3}{4} \Lambda_1$,
specified by the zero locus of the AD polynomials
\eqref{ADpolys}. Two mutually non-local singularities now collide at the AD
point $u=\uad=\frac{3}{4}\Lambda^2_1$ while the third one simplifies
to $u_0=-\frac{15}{16}\Lambda_1^2$, such that the discriminant reads 
\begin{equation}\label{AD1sing}
	\Delta=(u-\uad)^2(u-u_0).
\end{equation}
From the curve we now find
\begin{equation}\label{uAD1gamma03}
	\frac{u(\tau)}{\Lambda_1^2}=-\frac{f_{3\text{B}}(\tau)+15}{16},
\end{equation}
where
\begin{equation}\begin{aligned}\label{f3b}
		f_{3\text{B}}(\tau)&=\left(\frac{\eta(\tfrac\tau3)}{\eta(\tau)}\right)^{12}=27\left(\frac{b_{3,0}(\tfrac\tau3)}{b_{3,1}(\tfrac\tau3)}\right)^3-27\\
		&=q^{-\frac13}-12+54\, q^{\frac13}-76\, q^{\frac23}-243 \, q+\CO(q^{\frac 43})
\end{aligned}\end{equation}
is the McKay-Thompson series of class 3B for the Monster group
\cite{Conway:1979qga}, and we are again careful to choose the solution
for $u$ with the consistent asymptotics. Substitution of the $q$
series (\ref{f3b}) in (\ref{uAD1gamma03}) reproduces the $q$-series
based on \cite[Eq. (4.93)]{Huang:2009md}.

\begin{figure}
	\begin{subfigure}{.5\textwidth}
		\centering
		\includegraphics[width=\linewidth]{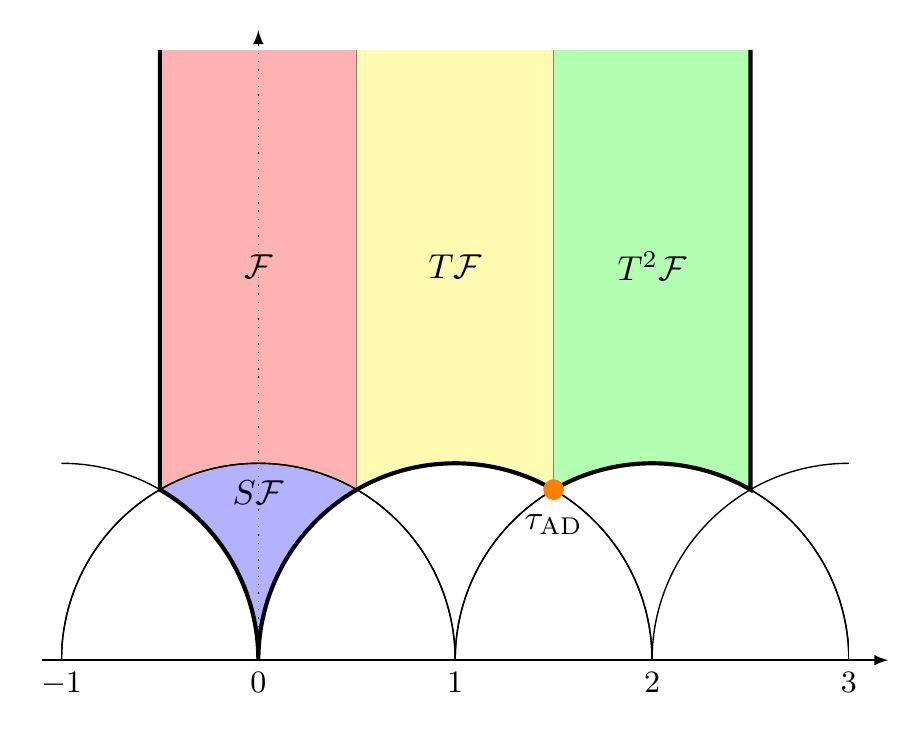}  
	\end{subfigure}
	\begin{subfigure}{.5\textwidth}
		\centering
		\includegraphics[width=.7\linewidth]{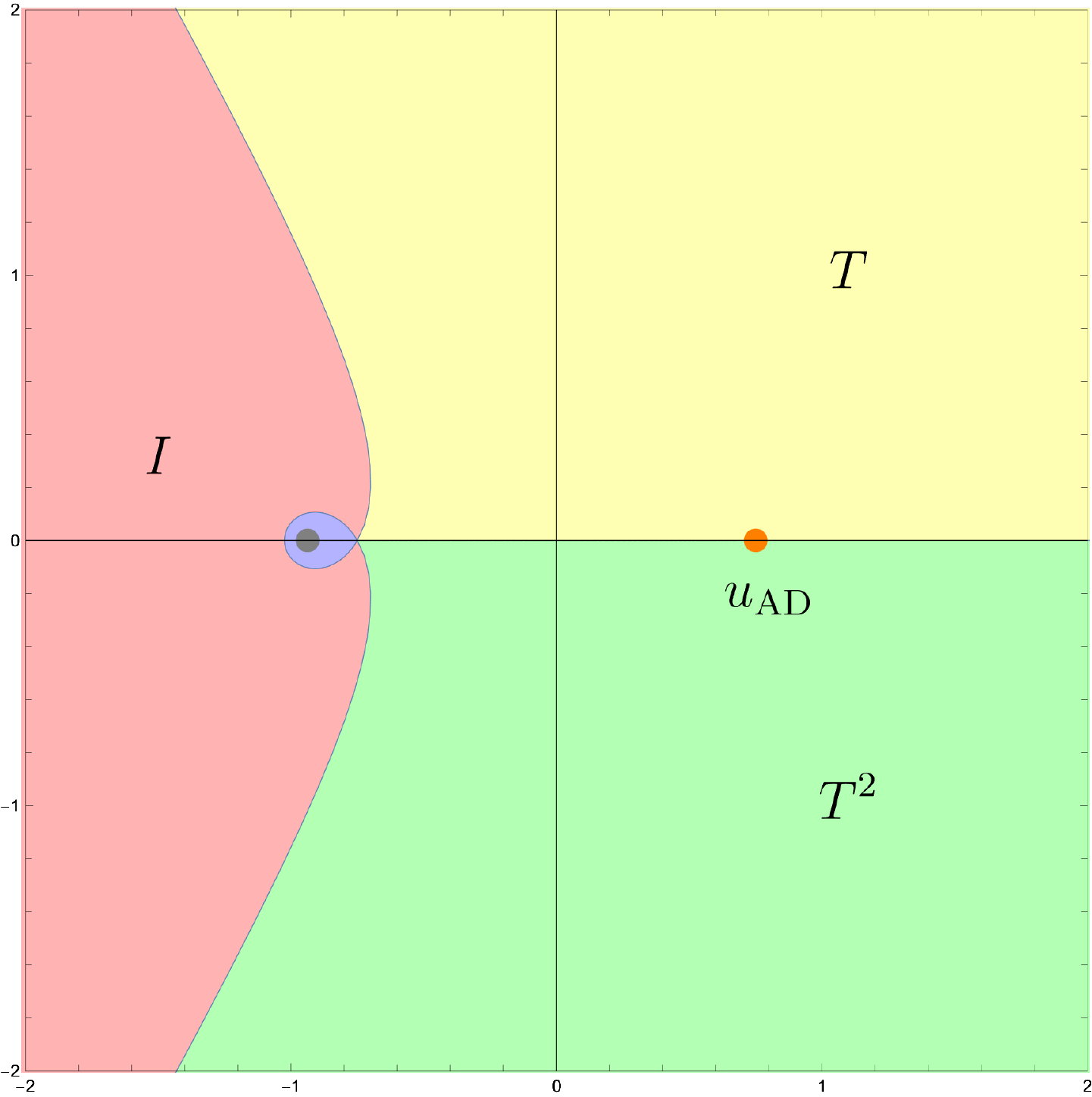}  
	\end{subfigure}
	\caption{Left: Fundamental domain of $\Gamma^0(3)$, the duality group of $N_f=1$ with $m=\mad$. The AD point $\tad=\sqrt3\omega_{12}$ is highlighted. Right: Plot of the $N_f=1$ $u$-plane with AD mass as the union of the images of $u$ under the $6-2=4$  $\slz$ images of $\CF$ forming $\Gamma^0(3)\backslash\mathbb H$, as in the left figure. The complex plane can clearly be covered by 4 triangles. There is only one strong coupling region, which is the circular region. It contains $u_0$ in its interior. The AD point (orange) lies on the boundary of $T^{\pm1}\CF$, as is  clear from the left figure. The areas with the same colours are mapped to each other in the two figures.}
	\label{fig:11domains}
\end{figure}

The functions $\eta$ and
$b_{3,j}$ are defined in Appendix \ref{sec:jacobitheta}. Using Theorem
\ref{thm:etaquotients} in the same Appendix, we find that $\tau\mapsto
f_{3\text{B}}(3\tau)$ is a classical modular function for
$\Gamma_0(3)$ and therefore $u$ is a modular function for
$\Gamma^0(3)$. A fundamental domain of  $\Gamma^0(3)$ is 
\begin{equation}
\CF_{1}(\mad)=\bigcup_{\ell=0}^2T^\ell \CF\cup S\CF.
\end{equation}
This is shown in Fig. \ref{fig:11domains} together with the map to the $u$-plane. The cusps are $\im\infty$ and $0$, with widths 3 and 1, respectively. We take from \cite[Table 4.1]{schultz2015} that $\Gamma^0(3)$ has an elliptic fixed point of order 3.

Using the transformation properties of the $\eta$-function \eqref{etatransformation} it is straightforward to show that the locations of the singularities of \eqref{AD1sing} in the $\tau$-plane are given by ($\omega_j=e^{2\pi\im/j}$)
\begin{equation}
u(\im\infty)=\infty, \qquad u(0)=u_0, \qquad u(\sqrt3\omega_{12})=u_{\text{AD}},
\end{equation}
where the proper limits are understood.
The AD point $\sqrt3\omega_{12}$ is stabilised by the order 3 element
$\left(\begin{smallmatrix} -1&3\\-1&2\end{smallmatrix}\right)\in
\Gamma^0(3)$, and it is therefore the order 3 elliptic fixed point of
$\Gamma^0(3)$. Comparing the locations to the massless case we see
that the regular singularity $u_0$ has stayed on $\tau=0$, while,
contrary to the massless case, the cusps with the two mutually non-local
singularities are disconnected (or cut) from the domain for
massless $N_f=1$, and leaves as remnant the point
$\tau_{\rm AD}$ into the interior of $\mathbb{H}$.\footnote{The
  disconnected region has a physical interpretation as the $u$-plane
  of the AD curve. See the discussion around Eq. (\ref{ADcurve})} This procedure
also reduces the index of the solution: Indeed, from
\eqref{indexcongruence} we compute that $\ind \Gamma^0(3)=4$, where
the $6-4=2$ AD points do not contribute since they are not cusps. This
can also be seen from the fact that  
\begin{equation}\label{f3Bhauptmodul}
j=\frac{(f_{3\text{B}}+3)^3(f_{3\text{B}}+27)}{f_{3\text{B}}}.
\end{equation}
Indeed, since $\CJ=1728 \frac{g_2^3}{\Delta}$, a common factor $(f_{3\text{B}}+27)^2$ of $g_2^3$ and $\Delta$ has cancelled.  The last factor  $f_{3\text{B}}+3=0$ in   \eqref{f3Bhauptmodul} implies $j(\tau)=0$ and therefore $\tau=\omega_3\mod \slz$. In fact, it corresponds to $u(\omega_3)=u(\omega_3+1)=-\frac34\Lambda_1^2$ and it is just a regular point in the $u$-plane. We can also read off from this that $\ord(g_2, g_3, \Delta)=(1,1,2)$ and therefore the AD theory in $N_f=1$ is according to Table \ref{kodaira} a type $II$ singular fibre \cite{Argyres:2015ffa}.

We can also study more characteristic functions of the theory with the AD mass. Using Appendix \ref{sec:jacobitheta}, we can differentiate \eqref{uAD1gamma03} to find 
\begin{equation}
\frac{du}{d\tau}(\tau)=\frac{\pi\im\Lambda_1^2}{24}f_{3\text{B}}(\tau)b_{3,0}(\tfrac\tau3)^2.
\end{equation}
This implies that $\frac{du}{d\tau}$ is a modular form of weight 2 for $\Gamma^0(3)$, without phases.
One can also show that 
\begin{equation}\label{daduAD11}
\frac{da}{du}(\tau)=\frac{\sqrt2\im}{\sqrt{27}\Lambda_1}\sqrt{\frac{b_{3,1}(\tfrac\tau3)^3}{b_{3,0}(\tfrac\tau3)}}.
\end{equation}
An expression for $da/du$ in terms of $_2F_1$ was given in \cite[Eq. (4.13)]{Masuda:1996xj}.

The $q$-expansion of $\frac{da}{du}$ has growing denominators, and therefore $\frac{da}{du}$ is not a modular form of weight $1$ for $\Gamma^0(3)$. However, it is straightforward to check that $\left(\frac{da}{du}\right)^2$ is a modular form of weight $2$ for $\Gamma^0(3)$. We thus find the Matone relation
\begin{equation}
\frac{d u}{d\tau}=-\frac{16\pi\im}{3} \widehat \Delta\left(\frac{da}{d u}\right)^2,
\end{equation}
where $\widehat{\Delta}$ denotes the reduced discriminant. This is consistent with \eqref{matonepnf}.

The monodromies can be found from the ones of the massless theory \eqref{nf1monod},
\begin{equation}\label{11monodromies}
M_1=STS^{-1}=\begin{pmatrix} 
1 & 0 \\
-1 & 1 
\end{pmatrix}, \quad M_{\text{AD}}=M_2M_3=T^2(ST)^{-1} T^{-2}=\begin{pmatrix} -1&3\\-1&2\end{pmatrix}.
\end{equation}
They generate the duality group $\Gamma^0(3)$ and give the large $u$ monodromy  $M_1M_{\text{AD}}=PT^{-3}$. Furthermore, $M_1$ stabilises $\tau=0$ and $M_{\text{AD}}$ stabilises the AD point $\tad=\sqrt3\omega_{12}$. We have that $M_{\text{AD}}^6=\mathbbm 1$ and therefore $\tad$ is indeed an elliptic fixed point. The AD monodromy is conjugate to $(ST)^{-1}$, which fixes $\tau=\omega_3$. Since $\tad=\omega_3+2$, this gives a path in $\tau$-space.

\subsection{Generic real mass}
By turning on a generic real mass, the singularities do not split compared to the massless case since there are already $N_f+2=3$ singularities. Therefore, the fundamental domain of the massive theory should look similar to the massless one of Fig. \ref{fig:nf1domains}, but we now need to consider the presence of branch points and cuts in more detail. We will discuss this and the limits to the pure theory as well as the the theory with the AD mass now. 

For generic mass we have not been able to find a closed expression for $u$ as a function of $\tau$. By expanding $\CJ(u,m,\Lambda_1)$ and inverting  the series we can, however, get an expansion of $u$ for the general massive theory near any cusp. For example, 
the expansion near $\tau=\im\infty$ reads ($\mu=\frac{m}{\Lambda_1}$)
\begin{equation}\label{umassiveNf1exp}
	\frac{u(\tau)}{\Lambda_1^2}=-\frac{1}{16}q^{-\frac13}-\frac13\mu^2+\left(\frac{32}{9}\mu^4-6\mu\right)q^{\frac13}-\left(\frac{5120}{81}\mu^6-\frac{160}{3}\mu^3+\frac{13}{2}\right)q^{\frac23}+\CO(q),
\end{equation}
where we are careful to choose the expansion such that $u\to-\infty$ for consistency with our conventions. It is easy to see that this reproduces the earlier expansions, \eqref{nf1u} and \eqref{uAD1gamma03}, in the respective limits.

\begin{figure}[h]\centering
	\includegraphics[scale=1.2]{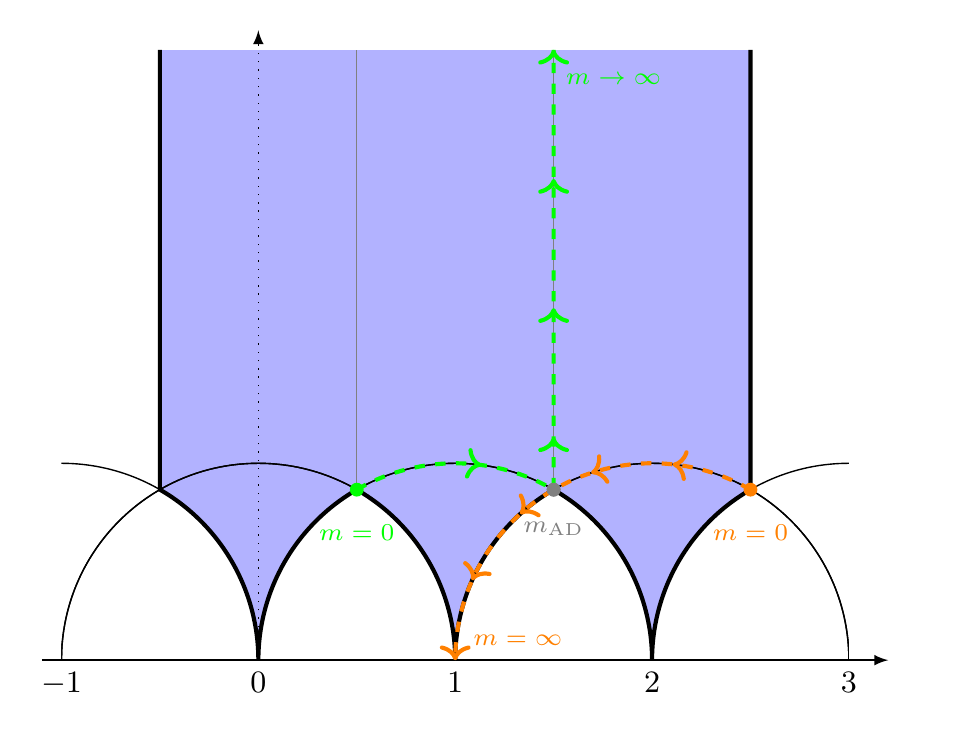}
	\caption{Conjectured paths of the branch points in the fundamental domain of the massive $N_f=1$ theory.}\label{fig:nf1mBP}
\end{figure}

The branch point locus is given by the zero locus of \eqref{pbp}. By calculating $\CJ(u,m,\Lambda_1)$ from the curve and plugging it into the polynomial $D^{\text{bp}}_1$ we find that the zero of the linear polynomial is $u=\ubp=\frac{4}{3}m^2$, and we recognise that this is the polynomial appearing in the denominator of the generalised Matone relation, \eqref{pnf}, such that $\frac{du}{d\tau}$ diverges here. In the massive $N_f=2,3$ theories, where the theories can be studied in detail, we argue that it corresponds to two branch points in the closure of the fundamental domain, which are connected by a branch cut. Motivated by these analyses, we can draw the two branch point loci for positive mass. It is given in Fig. \ref{fig:nf1mBP}. For $m=0$, the branch point is located at the origin $u=0$. At the AD point, they collide, the branch cut vanishes and the order parameter becomes holomorphic, and even modular. In the $m\to\infty$ limit, the branch points also move to infinity.

We can also confirm this from the analysis in Section \ref{sec:branchpoints}. By expanding $\CJ(u)-\CJ(u_{\text{bp}})$ around $u_{\text{bp}}$ for generic mass $m$, the linear coefficient is zero. The $(u-u_{\text{bp}})^2$ coefficient vanishes if and only if either $m=0$ or $m=m_{\im}\coloneqq \frac{3}{4\sqrt[3]{2}}$. For $m=0$ we have $\tau_{\text{bp}}\sim \omega_3$, such that $n_{\omega_3}=3$ in the notation of Section \ref{sec:branchpoints}. Furthermore, $n_{\text{bp}}=6$, such that the order of the branch point \eqref{BPorder} is the denominator of the reduced fraction $\frac{3}{6}$, namely $2$. Since $E_4$ has a simple zero at $\tau_{\text{bp}}$, this agrees with \eqref{nf1u} having a \emph{square} root. 

From Fig. \ref{fig:nf1mBP} we see that the branch point loci pass through $\tau_{\text{bp}}=1+\im$ where $m=m_{\im}$, such that $n_{\im}=2$. Furthermore we find $n_{\text{bp}}=4$, and thus the order of the branch point is $2$. Since $\frac24=\frac12$, it is indeed again the branch point of a square root. 

For any other mass $m\in \mathbb R_{\geq 0}\setminus\{0,m_{\im},\mad\}$ we have $n_{\text{bp}}=2$ while $n_{\tau_{\text{bp}}}=1$, such that the branch point is again of order $2$. This demonstrates that the loci in Fig. \ref{fig:nf1mBP} are complete: there is a single branch point on the Coulomb branch $\CB_1$, and for any mass there are two branch points of a square root in $\mathbb H$, which are connected by a single branch cut. It also implies that if an expression such as \eqref{nf1u} existed for generic mass, while it could contain higher roots of modular forms, they can never have zeros in $\mathbb H$ (as is the case also for $m=0$).

We can study the partition of the $u$-plane provided by   \eqref{tmj} in detail. For $m\neq \mad$ the $u$-plane is partitioned into six regions, whose union of boundary pieces is included in the algebraic curve given by the zero locus  \eqref{tm0} of $T_1$, where $T_1=y \tilde T_1$ and 
\begin{equation}\footnotesize\begin{aligned}
\tilde T_1&=972 \mu ^4+8192 \mu ^2 x^7-21504 \mu ^3 x^5-12096 \mu ^2 x^4+18432 \mu ^4 x^3+1296 \mu ^3 x^2\\&+16128 \mu  x^6 +1944 \mu  x^3+8192 \mu ^2 x^5 y^2+22528 \mu ^3 x^3
   y^2-8192 \mu ^2 x^3 y^4-3456 \mu ^2 x^2 y^2\\&+2304 \mu  x^4 y^2-6912 \mu  x^2 y^4-16384 x^6 y^2-12288 x^4 y^4+4320 x^3 y^2-6144 x^8\\&-1296 x^5-5184 \mu ^5 x-729
   \mu ^2 x-18432 \mu ^4 x y^2-21504 \mu ^3 x y^4-8192 \mu ^2 x y^6\\&-1944 \mu  x y^2-1296 x y^4+4752 \mu ^3 y^2+8640 \mu ^2 y^4+6912 \mu  y^6+2048 y^8.
\end{aligned}\end{equation}
Since the AD point $m=\mad$ corresponds to a phase transition, we have to study the two cases $m<\mad$ and $m>\mad$ separately.

\subsubsection*{The case $m<\mad$}

\begin{figure}
	\begin{subfigure}{.49\textwidth}
		\centering
		\includegraphics[width=\linewidth]{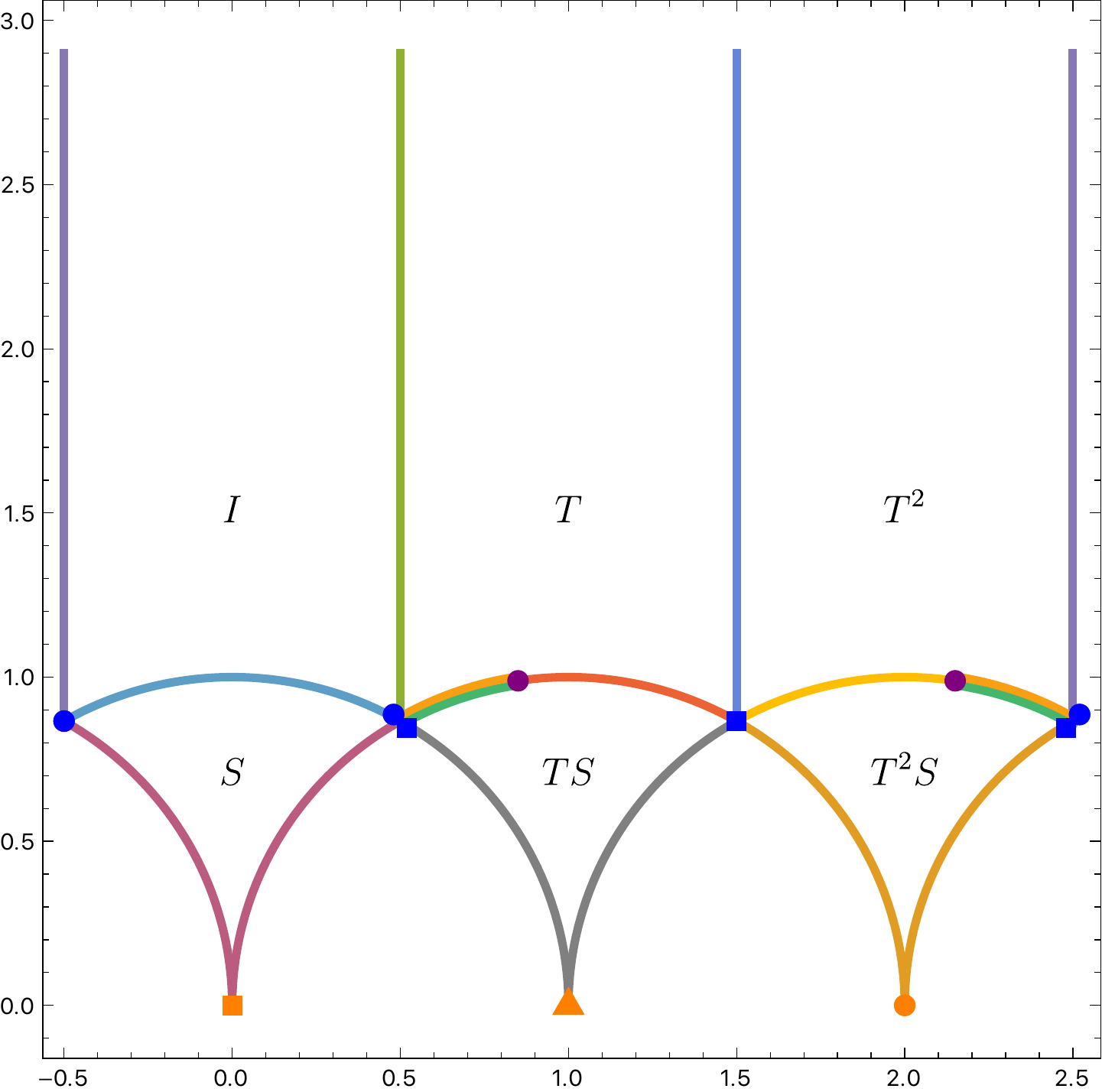}  
			\end{subfigure}
	\begin{subfigure}{.49\textwidth}
		\centering
		\includegraphics[width=\linewidth]{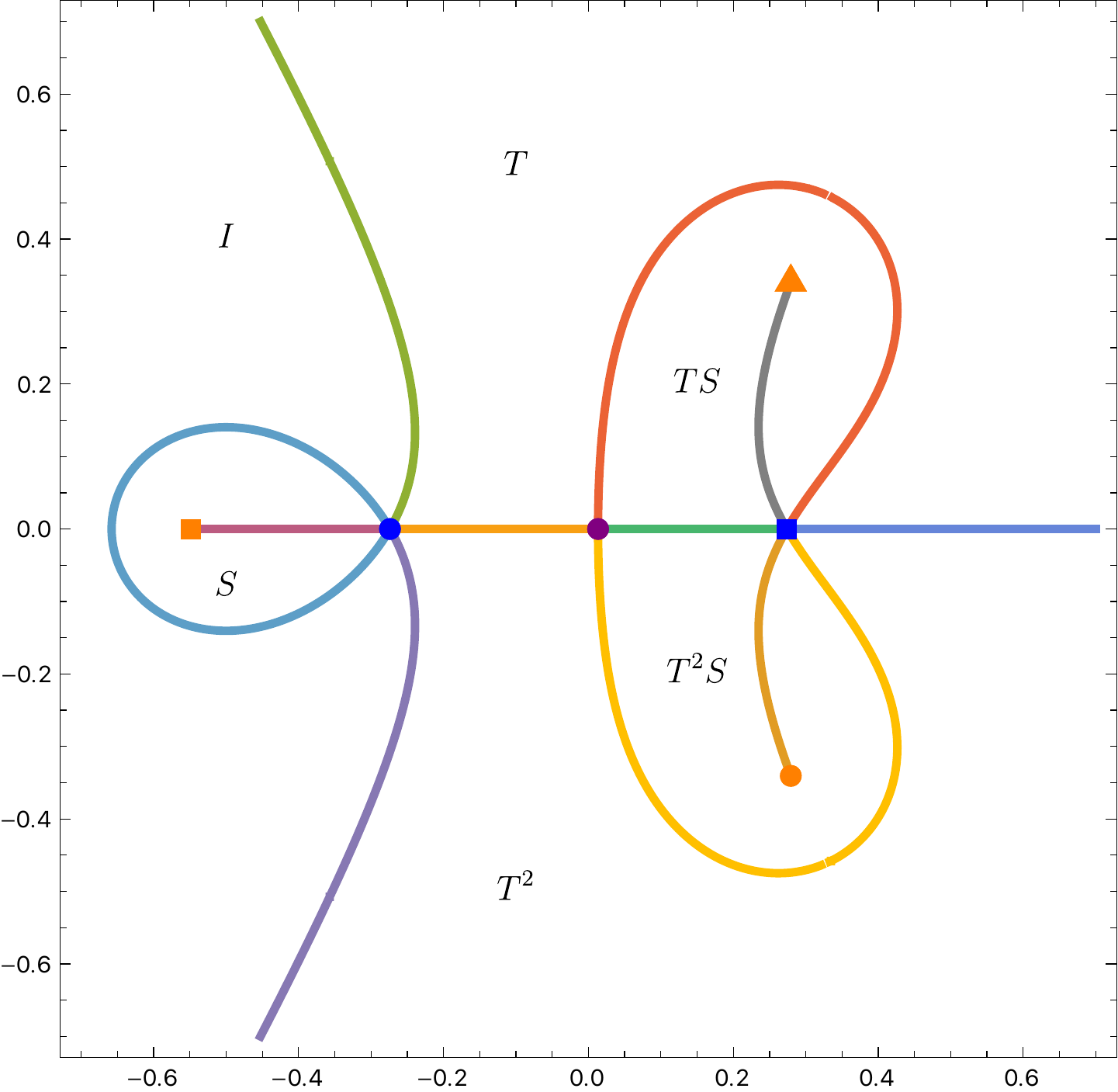}  
\end{subfigure}
	\caption{Identification of the components of the partitioning $\CT_{m}$ in $N_f=1$ for $\mu<\mu_{\text{AD}}$, here for the choice $\mu=\frac{1}{10}$. The $u$-plane $\CB_1$ is partitioned into 6 regions $u(\alpha\CF)$, with the $\alpha\in\slz$ given in both pictures.  The branch point (purple) identifies two points on $\partial\CF_1(m)$. The green/orange arcs are separated by a branch cut, and the opposite ends are glued with the corresponding identical colour on the other side. The blue points correspond to $j=\CJ=0$ and form the intersection points of $\CT_1$.}
	\label{fig:nf1partition}
\end{figure}

From Fig. \ref{fig:nf1mBP} we can take the location of the branch points. There is one singularity $u_1$ on the negative real line, and the other two are complex conjugates (as $\Delta_1$ is a real polynomial).  Using the definition \eqref{tmj}, it is straightforward to show that not all of $y=0$ lies in $\CT_1$, but rather only the real half-line with $u\geq u_1$. Furthermore, the lines truncate at the singularities. On the upper-half plane, the branch points can be viewed as endpoints of  branch cuts coming from $\tau=\frac12+\frac{\sqrt3}{2}\im$  and $\tau=\frac52+\frac{\sqrt3}{2}\im$. See Fig. \ref{fig:nf1partition}. From this it is straightforward to see how taking the massless limit gives back Fig. \ref{fig:nf1domains}.

\subsubsection*{The case $m>\mad$}
\begin{figure}
	\begin{subfigure}{.49\textwidth}
		\centering
		\includegraphics[width=\linewidth]{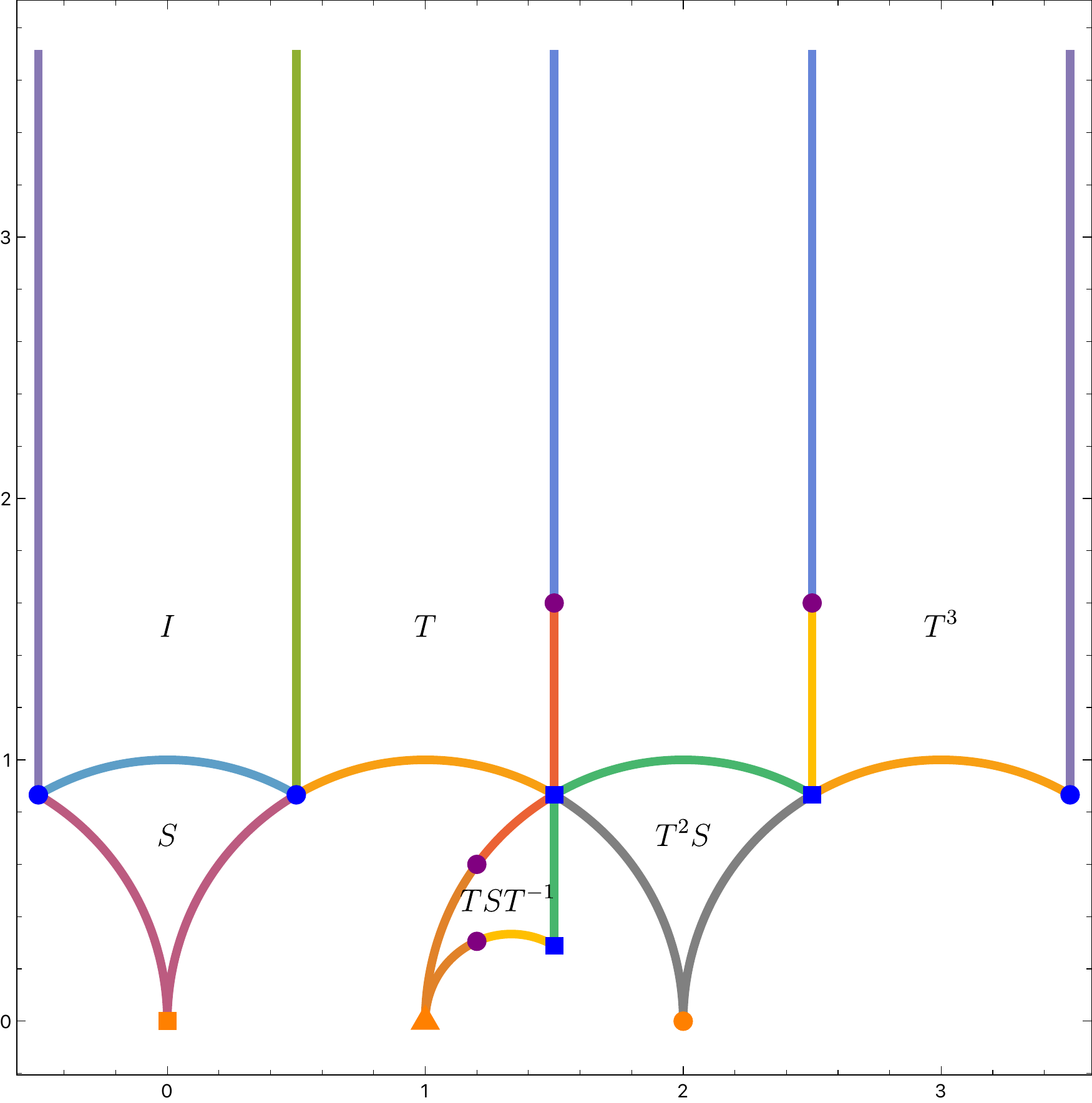}  
			\end{subfigure}
	\begin{subfigure}{.49\textwidth}
		\centering
		\includegraphics[width=\linewidth]{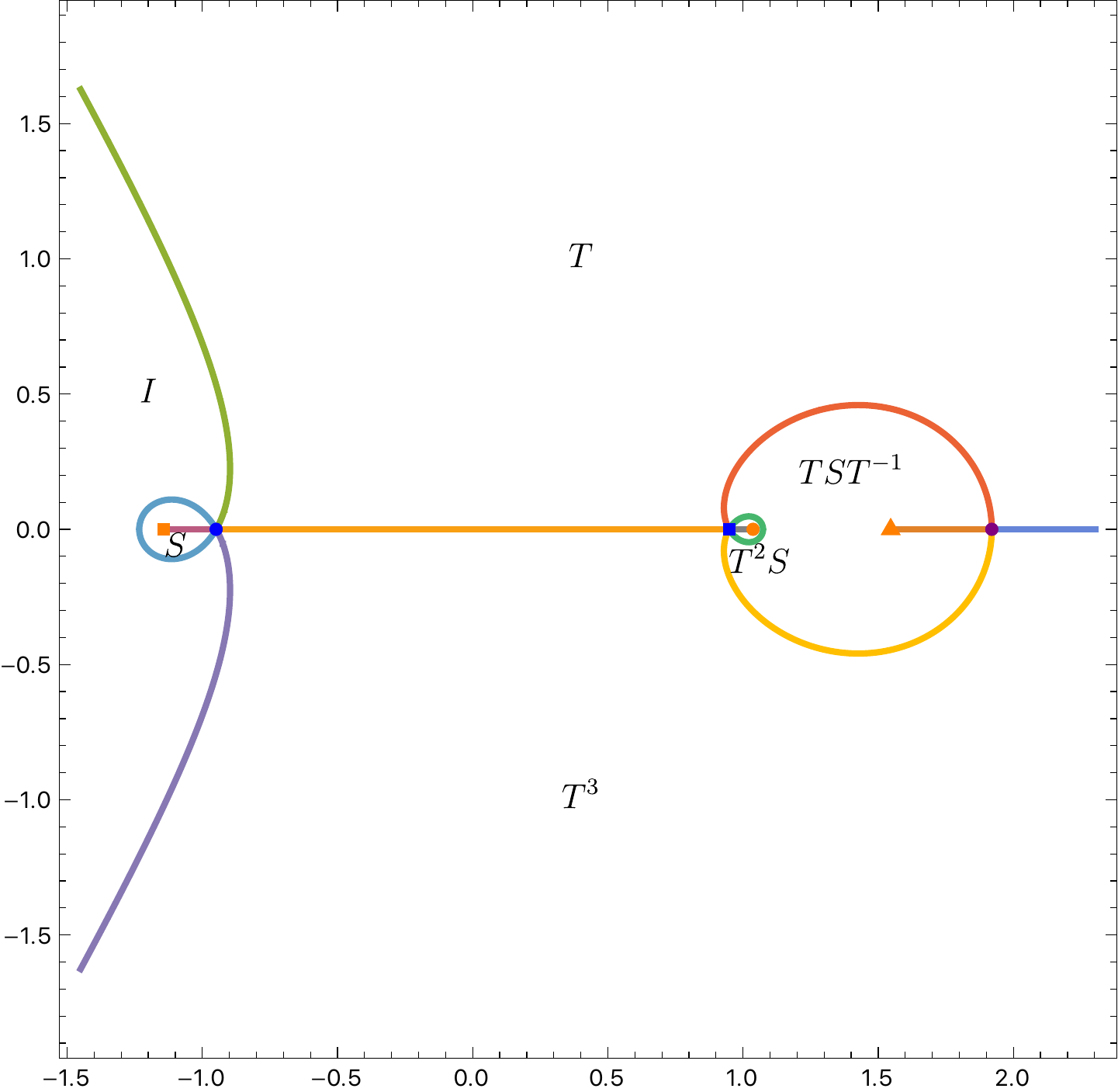}  
\end{subfigure}
	\caption{Identification of the components of the partitioning $\CT_{m}$ in $N_f=1$ for $\mu>\mu_{\text{AD}}$, here for the choice $\mu=\frac65$. The $u$-plane $\CB_1$ is partitioned into 6 regions $u(\alpha_j\CF)$, with the $\alpha_j\in\slz$ given in both pictures.  The branch point (purple) identifies four points on $\partial\CF_1(m)$.}
	\label{fig:nf1partition_m>mAD}
	\end{figure}

At $m=\mad$ two singularities collide, and $\Delta_1$ has a  double root. Since $\Delta_1$ is a real polynomial and depends smoothly on $m$, the two singularities which are complex for $m<\mad$ are real for $m>\mad$. There is no meaningful identification of the singular points when going through $m=\mad$, however for large $m$ there is a distinguished singularity $u_*$ that diverges. We can make the choice of $\CF_m$ suitable for the limit $m\to \infty$, where we should obtain Fig. \ref{fig:fundgamma0(4)}. By studying the dependence of the partition of the $u$-plane on the mass, one finds that $u_*$ is bounded by a region whose area grows as $m\to \infty$. It squeezes into $T\CF$ and $T^3\CF$ and becomes $T^2\CF$ in the limit $m\to\infty$. However, as we want to put the singularities on the real line we need it to touch this axis for finite $m>\mad$. In order to find the corresponding fundamental domain, we can glue parts of the boundary $\partial\CF_1$, such that it not only agrees with the geometry of the partition of the $u$-plane, but also the decoupling procedure is inherent. See Fig. \ref{fig:nf1partition_m>mAD}.

\begin{figure}[h]
	\begin{subfigure}{.5\textwidth}
		\centering
		\includegraphics[width=\linewidth]{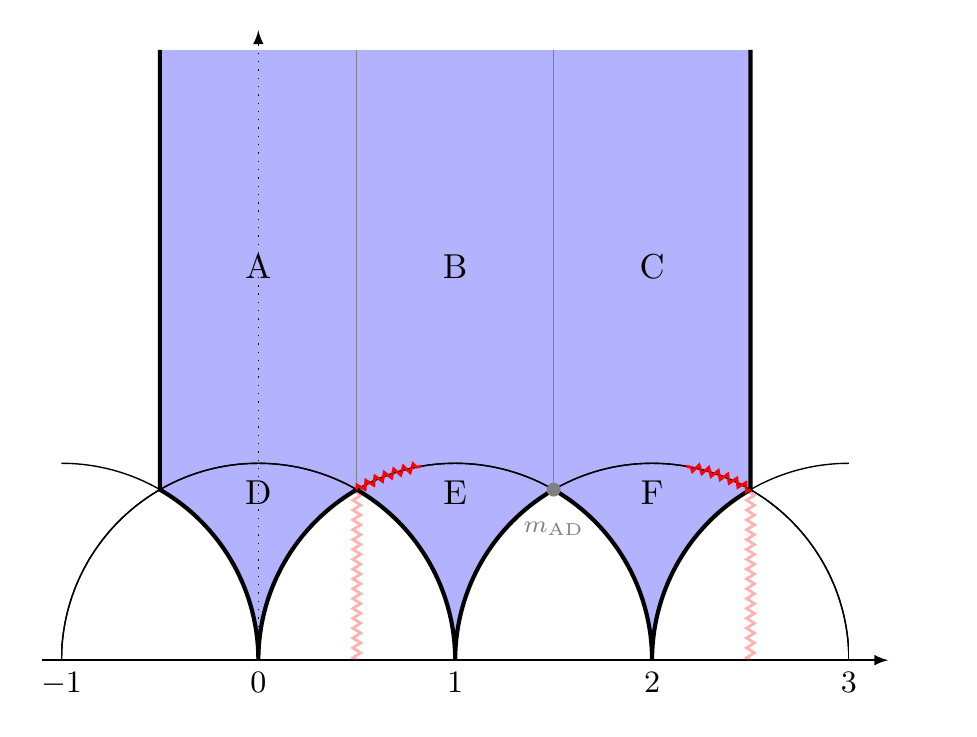}  
		\caption{}
	\end{subfigure}
	\begin{subfigure}{.5\textwidth}
		\centering
		\includegraphics[width=\linewidth]{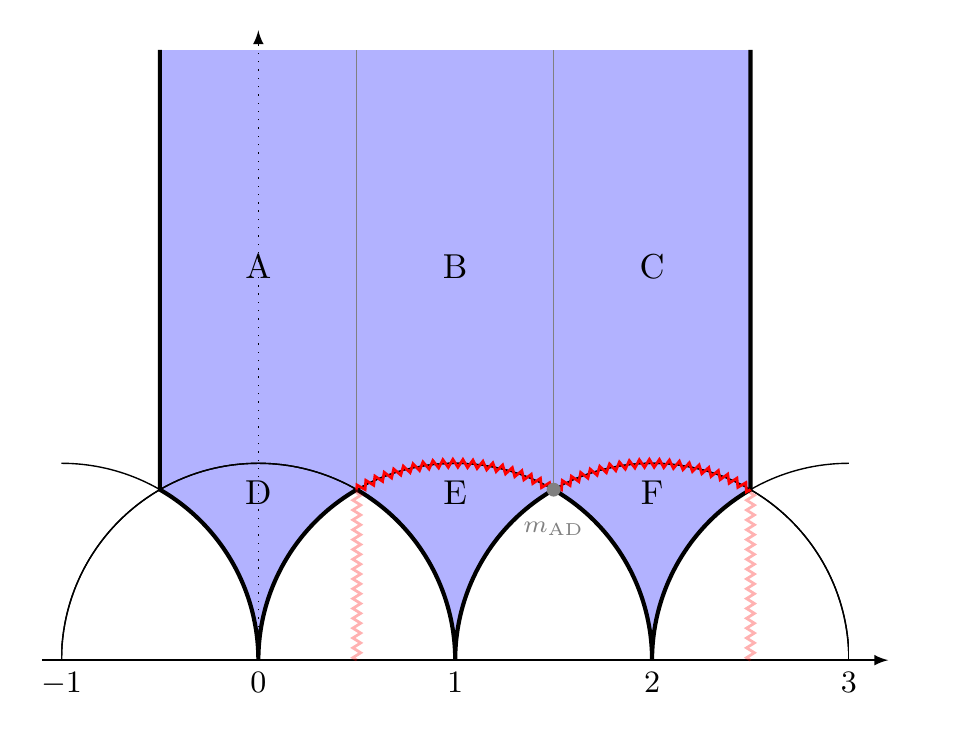} 
		\caption{} 
	\end{subfigure}
\newline
	\begin{subfigure}{.5\textwidth}
		\centering
		\includegraphics[width=\linewidth]{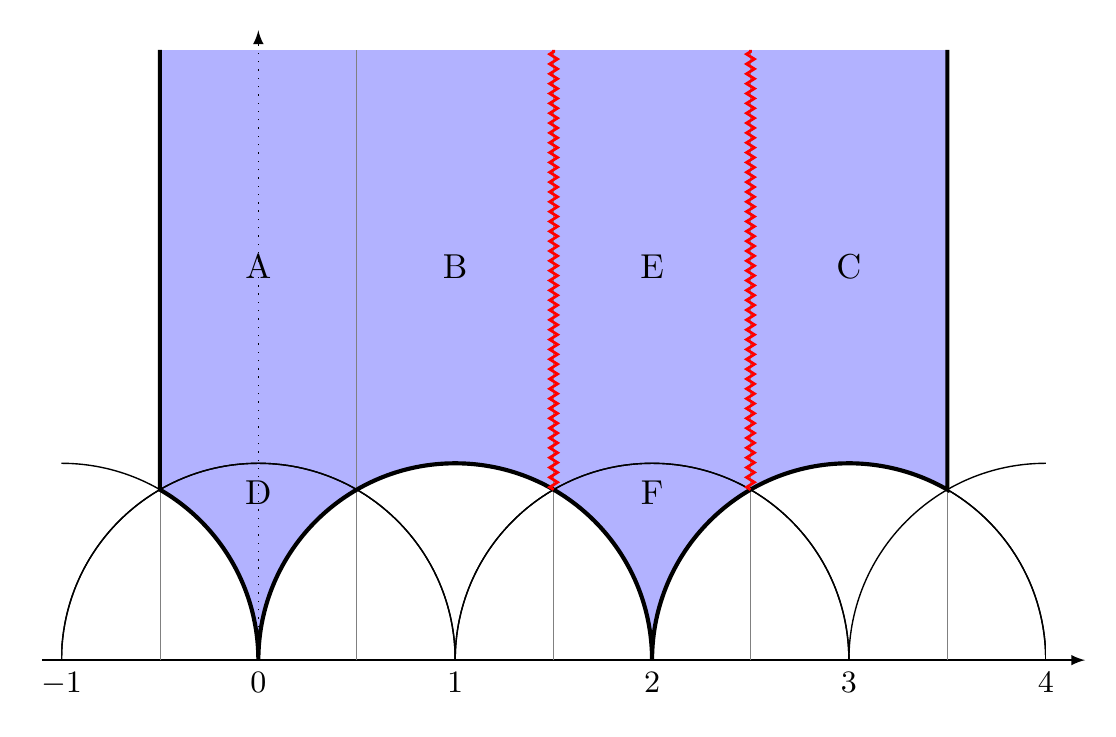}  
		\caption{}
	\end{subfigure}
\begin{subfigure}{.5\textwidth}
	\includegraphics[width=\linewidth]{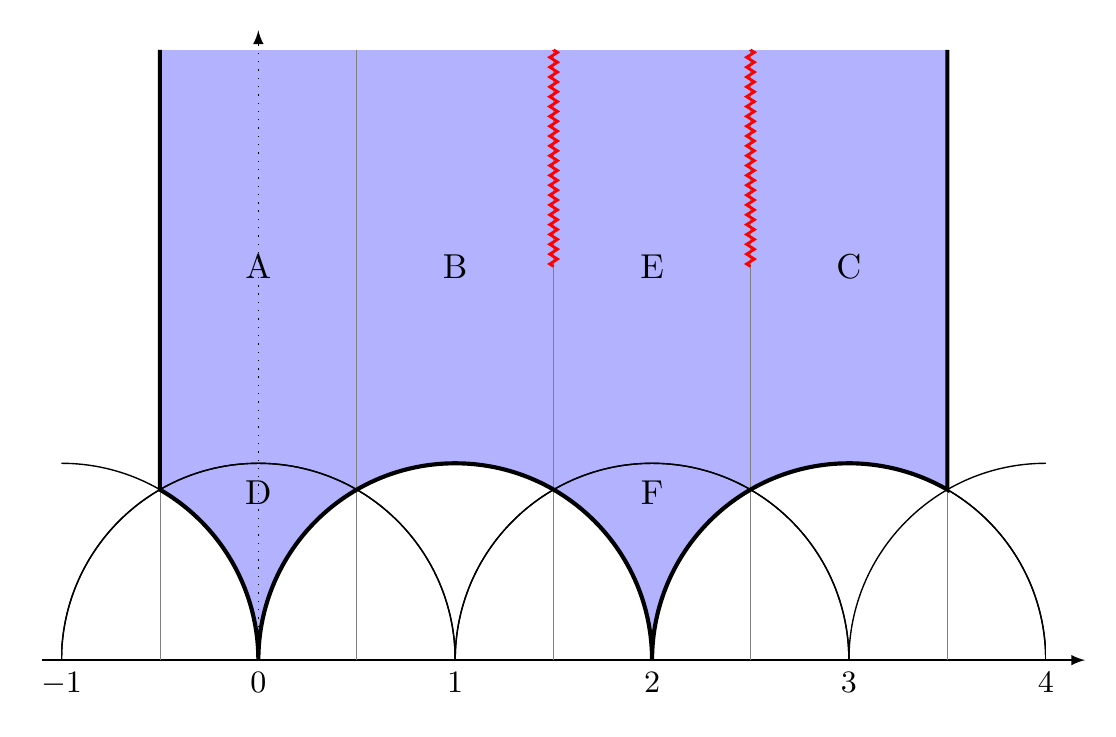}
	\caption{}
\end{subfigure}
	\caption{Choice of branch cuts (red zigzag lines) for varying mass in $N_f=1$. Starting with a small mass in Figure (a) we cut along the paths shown in Fig. \ref{fig:nf1mBP}. At the AD mass we can use the identifications of the different boundaries to reorganise the domain in Figure (b) to the one of Figure (c). In (b) we identify the upper part of the cut in region B with the upper part of the cut in region C, and similarly in E and F. In Figure (c) we instead identify the sides of the cuts such that the boundaries of regions B and C are identified and the two boundaries of E are identified. When we increase the mass further the cuts of Figure (c) move upwards as in Figure (d) eventually reaching infinity and disappearing, leaving us with the domain of the pure theory.}
	\label{fig:nf1CutsAlt}
\end{figure}

An alternative way of depicting how the cuts change the fundamental domain is given in Fig. \ref{fig:nf1CutsAlt}. Here, we lift the restriction that we want to have all the singularities situated on the real axis once we go to mass larger than the AD value. The Figures \ref{fig:nf1CutsAlt} are equivalent to Figs. \ref{fig:nf1partition} and \ref{fig:nf1partition_m>mAD}, as is easily seen by following the identification of the various boundaries. From this description it is direct to see the change in the domain for the different special limits of the mass.

\subsubsection*{The case $m=\mad$}
Finally, let us return to the case $m=\mad$ discussed in detail in
Section \ref{sec:nf1ad}.
As explained above, a region is disconnected from $\CF_{N_f}$ for
 this special value of the mass. The disconnected domain is labelled by $TS$ and $T^2S$ in Figure \ref{fig:nf1partition}, 
$TST^{-1}$ and $T^2S$ in Figure \ref{fig:nf1partition_m>mAD}, and by $E+F$ in
Figure \ref{fig:nf1CutsAlt}. It is also isomorphic to the domain in Figure \ref{fig:IVdomains}.
This region has a physical meaning,
namely as the fundamental domain for the order parameter $\tilde u$ of
the AD theory, obtained after taking the scaling limit to the conformal
field theory. To see this, recall that the scaling limit brings the $N_f=1$ curve
(\ref{eq:curves}) to the AD curve \cite{Argyres:1995jj, Argyres:1995xn}
\be
\label{ADcurve}
y^2=x^3-\frac{1}{4}\Lambda_1^3 \tilde mx-\frac{1}{16}\Lambda_1^4\tilde u, 
\ee
which gives for $\tilde u$ in terms of the effective coupling $\tau$ \cite{Moore:2017cmm},
\be
\tilde u(\tau)= \frac{4}{\sqrt{27}}\Lambda_1^{1/2}\tilde m^{3/2}\frac{E_6(\tau)}{E_4(\tau)^{3/2}},
\ee
for which the disconnected domain is indeed a fundamental domain. This
splitting of the fundamental domain $\CF_{1}$ at the AD point appears to fit well with
the $u$-plane integral for $N_f=1$ discussed in \cite{Moore:2017cmm}.

\subsection{Generic complex mass}\label{sec:nf1complexm}
We can also consider a generic complex mass. The locus of AD masses \eqref{ADpolys} is then real codimension 2. In fact, it is just $\omega_3\, \mad$, with $\omega_3$ a cube root of unity. If $m$ is not any of these three values, the corresponding Coulomb branch has three distinct singularities. 

\begin{figure}
	\begin{subfigure}{.5\textwidth}
		\centering
		\includegraphics[width=\linewidth]{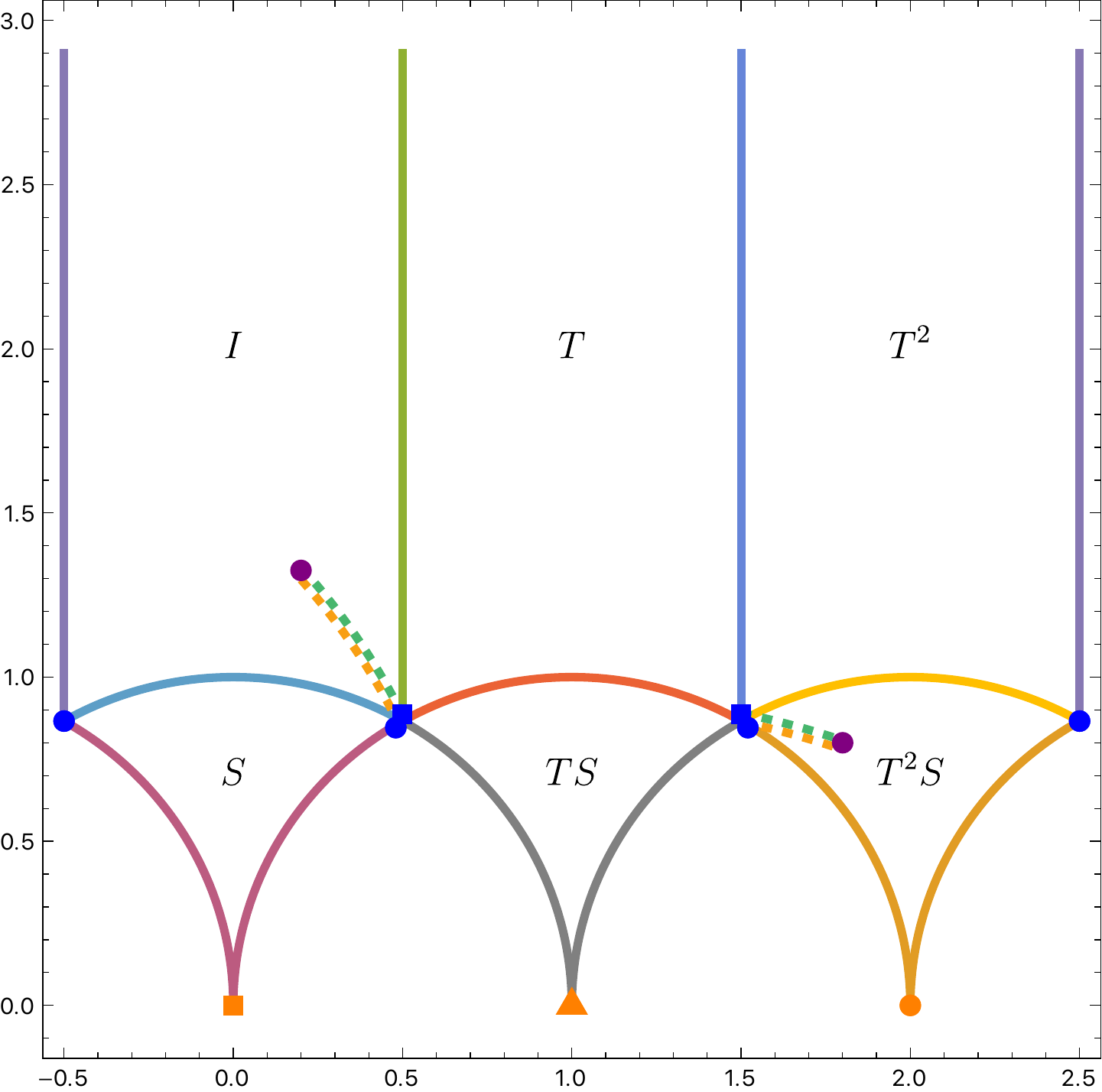}  
			\end{subfigure}
	\begin{subfigure}{.5\textwidth}
		\centering
		\includegraphics[width=\linewidth]{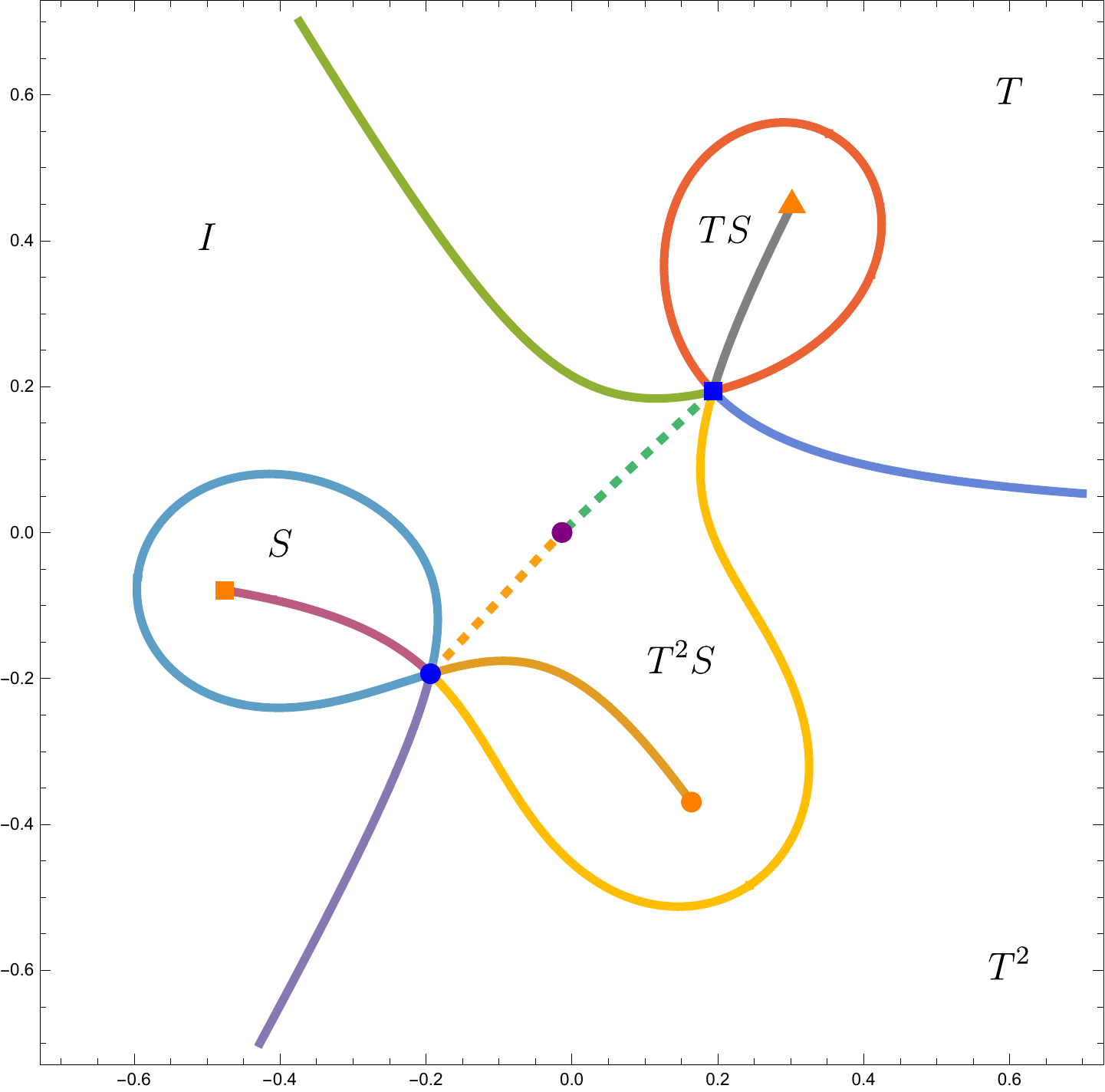}  
\end{subfigure}
	\caption{Identification of the components of the partitioning $\CT_{m}$ in $N_f=1$ for a complex mass $\mu\in\mathbb C$, here for the choice $\mu=\frac{\im}{10}$. The $u$-plane $\CB_1$ is partitioned into $5$ regions, which is due to the fact that $u(\CF)$ and $u(T^2S\CF)$ are glued at the branch cuts (dashed). The purple dot is the branch point in $N_f=1$.}
	\label{fig:nf1complexm}
	\end{figure}

We can decompose $m=a+\im b$, and $T_m$ is then a polynomial in $a$, $b$, $x$ and $y$. From \eqref{pbp} we see that if $m$ is complex, then $j(\tau_{\text{bp}})=\CJ(\ubp)$ is also complex, such that $\tau_{\text{bp}}$ is generically an interior point of $\CF$ or an $\slz$ copy thereof. The branch cuts most conveniently run from such branch points to the intersection points of the curves, where $\CJ(u)=j(\tau)=0$. From \eqref{jinvg2g3} it is clear that they correspond to the two solutions of $g_2(u)=0$.
We plot the partitioning of the $u$-plane with the branch cuts for an imaginary mass in Fig. \ref{fig:nf1complexm}.

Due to the fact that $\ubp\not\in\CT_1$, the branch cuts run to the interior of the $\alpha_j\CF$. The four sides of the two cuts are pairwise identified, which makes points on the branch cut smooth points on the Riemann surface. This identification  glues different regions $u(\alpha_j\CF)$ together, in this case $u(\CF)$ and $u(T^2S\CF)$. This is clearly visible in Fig. \ref{fig:nf1complexm}, where the dashed lines in the $u$-planes are the images of the branch cuts in $\CF_m$, and they do \emph{not} belong to the partitioning $\CT_1$. As a result, the $u$-plane is  partitioned into five and not six components. This is not in contradiction with Section \ref{sec:mod_appr} because the fundamental domain $\CF_1(m)$ is still a union of six copies of $\CF$: the cutting and glueing along the branch cuts is an additional feature of the domain.

\section{The $N_f=2$ curve}\label{sec:nf2}
Let us now move on to discuss the theory with two hypermultiplets. This theory has four strong coupling singularities where massless hypermultiplets appear. For general masses they are distinct points while for special mass configurations one or more singularities can collide. We will begin by restricting to the case of equal masses, $m_1=m_2=m$, where we can find explicit expressions for $u$ as a function of $\tau$. Then we briefly discuss the case of two distinct masses before moving on to discuss what happens in the simpler cases of massless hypermultiplets and when fixing the mass to an AD value. 

\subsection{Equal masses}
Let us consider first the equal mass case, $\bfm=(m,m)$, where $m\neq0$ and $m\neq \mad=\frac12\Lambda_2$. It is discussed in detail in \cite{Bilal:1997st}. In this case, the discriminant factors as
\begin{equation}\label{discrnf2equal}
\Delta=(u-u_*)^2(u-u_+)(u-u_-),
\end{equation}
where $u_*=m^2+\frac{\Lambda_2^2}{8}$ and $u_\pm=-\frac{\Lambda_2^2}{8}\pm m\Lambda_2$. It is easy to check that $\{u_*,u_+,u_-\}$ never collide other than in the two cases mentioned above.
Using the modular lambda function, $\lambda=\frac{\jt_2^4}{\jt_3^4}$, as a generator of the intermediate field $\Gamma(2)$, the sextic equation factors into three quadratic polynomials over $\Gamma(2)$. These equations can now be solved exactly. In $N_f=2$, two solutions have the property that $|u(\tau)|\to\infty$ when $\tau\to\im\infty$. Following our (usual) convention, we choose
\begin{equation}\label{unf2mm}
	\begin{aligned}
		\frac{u}{\Lambda_2^2}=&-\frac{\vartheta_4^8+\vartheta_2^4\vartheta_3^4+(\vartheta_2^4+\vartheta_3^4)\sqrt{16\frac{m^2}{\Lambda_2^2}\vartheta_2^4\vartheta_3^4+\vartheta_4^8}}{8\vartheta_2^4\vartheta_3^4}\\
		=&-\frac{1}{64}q^{-1/2}-\frac{m^2}{\Lambda_2^2}+\left(64\frac{m^4}{\Lambda_2^4}-32\frac{m^2}{\Lambda_2^2}-\frac{5}{16}\right)q^{1/2}+\CO(q).
	\end{aligned}
\end{equation}
Due to the appearance of the square root in \eqref{unf2mm} $u$ is not holomorphic, and similarly to the $N_f=1$ case there will be branch points in the fundamental domain. From Section \ref{sec:ramifications} we expect them to be given by
\begin{equation}\label{nf2bpjm}
	j^{\text{bp}}(m)=16\frac{(16m^2-\Lambda_2^2)^3}{m^2}. 
\end{equation}
By plugging in the solution for $\CJ(u,m,\Lambda_2)$ we find that this corresponds to $u=\ubp=2m^2-\frac{\Lambda_2^2}{8}$. We recognise this as the root of the polynomial $P^M_2$ of the generalised Matone relation. By using standard relations between the $j$-invariant and Jacobi theta function we can also check that this coincides with the zeros of the square root.

Defining $f_2(\tau)\coloneqq 16\tfrac{m^2}{\Lambda_2^2}\jt_2(\tau)^4\jt_3(\tau)^4+\jt_4(\tau)^8$, we see that the branch point of the square root is $f_2(\tau_0)=0$. Near $\tau_0$, the expansion of $f_2$ reads $f_2(\tau)=(\tau-\tau_0)h(\tau)$, where $h(\tau)$ is holomorphic near $\tau_0$ and $h(\tau_0)\neq0$. Then one branch of the square root reads $\sqrt{f_2(\tau)}=\sqrt{\tau-\tau_0}\sqrt{h(\tau)}$. Now since $h(\tau_0)\neq0$, we have that $\tau\mapsto \sqrt{h(\tau)}$ is nonzero and in fact holomorphic in a neighbourhood of $\tau_0$. However, $\tau\mapsto \sqrt{\tau-\tau_0}$ is strictly non-holomorphic at $\tau_0$. This proves that $u$ is not holomorphic at $\tau_0$. 

From \eqref{unf2mm} we can also calculate the other interesting quantities,
\begin{equation}\begin{aligned}\label{massivenf2quantities}
		\frac{da}{du}&=-\frac{\im}{\Lambda_2}\frac{\jt_2^2\jt_3^2}{\sqrt{\jt_2^4+\jt_3^4+\sqrt{f_2}}}, \\
		\frac{du}{d\tau}&=\pi\im\Lambda_2^2\jt_4^8\frac{2(4\frac{m^2}{\Lambda_2^2}+1)\jt_2^4\jt_3^4+\jt_4^8+(\jt_2^4+\jt_3^4)\sqrt{f_2}}{8\jt_2^4\jt_3^4\sqrt{f_2}}.
\end{aligned}\end{equation}
We can again explicitly check that they satisfy Matone's relation, \eqref{matonepnf},
\begin{equation}\label{Nf2massiveMatone}
	\frac{du}{d\tau}=-\frac{16\pi i}{2}\frac{\widehat\Delta}{u-u_{\text{bp}}}\left(\frac{da}{du}\right)^2.
\end{equation}
On the rhs, the double singularity $u_*$ has cancelled, while, as discussed in Section \ref{sec:matone}, the branch point $u_{\text{bp}}=2m^2-\frac18\Lambda_2^2$ remains in the denominator.

\subsubsection*{Fundamental domain}
A fundamental domain can be found in the following way. The six roots of the sextic equation gives the six \emph{cusp expansions}. In order to simplify the expressions, let us  momentarily set $\Lambda_2=1$ and $(a,b,c)\coloneqq (\jt_2^4,\jt_3^4,\jt_4^4)$. All six expressions can be brought to a canonical form, see Table \ref{cuspexpnf2m}. 

\begin{table}[h]\begin{center}
		\begin{tabular}{|>{$\displaystyle}l<{$}| >{$\displaystyle}l<{$}|  >{$\displaystyle}l<{$}  |}
		\hline
\alpha_j & \alpha_j(\im\infty) & \alpha_j u  \\ \hline
\text{id}&\im\infty & -\frac{c^2+ab+(a+b)\sqrt{c^2+16m^2ab}}{8ab} \\
T& \im\infty& -\frac{b^2-ac+(-a+c)\sqrt{b^2-16m^2ac}}{-8ac} \\
S& 0& -\frac{a^2 + b c + (b + c) \sqrt{a^2+ 16  m^2b c}}{8bc} \\
TS& 1& -\frac{b^2 - a c + (a - c) \sqrt{b^2 - 16  m^2a c}}{-8 a c}  \\
TST^{-1}& 1& -\frac{ c^2+a b  + (-a - b) \sqrt{c^2 + 16 m^2a b }}{8 a b}  \\
T^2ST & 2 & -\frac{a^2 + b c + (-b - c) \sqrt{a^2 + 16 m^2b c }}{8 b c} 
\\ \hline 
		\end{tabular}
		\caption{Cusp expansions and associated coset representatives $\alpha_j$ of the solution for $N_f=2$ with mass $\bfm=(m,m)$. } \label{cuspexpnf2m}\end{center}
\end{table} 

The overall sign can be fixed from the purely quadratic term in the numerator. Using the Jacobi identity $a+c=b$, such a representation is unique and the expressions cannot be further simplified. Then instead of studying which transformations give the right values at the cusps, we can take the cusp expansions and try to find maps $\alpha_j\in \slz$ that takes $u(\tau)$ to the  functions under study. Due to the square root, this is very subtle.  For instance, for $T^2ST$ the Jacobi theta functions transform as $(a,b,c)\mapsto (e^{2\pi\im}a,b,c)\mapsto (e^{2\pi\im}c,b,a)\mapsto (e^{2\pi\im}b,c,e^{\pi\im}a)$. We ignore the weight factors since numerator and denominator are homogeneous in the modular weight. This implies that 
\begin{equation}
\sqrt{c^2+16m^2ab}\mapsto \sqrt{e^{2\pi\im}a^2+16m^2 e^{2\pi\im}bc}=-\sqrt{a^2+16m^2bc},
\end{equation}
and gives precisely the last row in Table \ref{cuspexpnf2m}. The other transformations can also be proven directly. Such identifications are valid as long as $m$ is generic, and in particular such that the square root does not resolve. This obviously excludes the cases $m=0$ and $m=\pm \mad$, and it is conceivable that these are the only cases. We continue by assuming that it is true.

As argued above, there will also be branch points in the fundamental domain due to the square roots appearing in the solution for $u$. For generic complex mass these points will lie inside the fundamental domain. If we restrict to positive masses we see from \eqref{nf2bpjm} that $\lim_{m\searrow 0}j^{\text{bp}}(m)=-\infty$, while $j^{\text{bp}}(\tfrac{\Lambda_2}{4})=0$, $j^{\text{bp}}(\mad)=12^3\Lambda_2^4$ with $\mad=\tfrac12\Lambda_2$ and $\lim_{m\to \infty}j^{\text{bp}}(m)=+\infty$. Furthermore, one finds that $j^{\text{bp}}:(0,\infty) \to \mathbb R$ is monotonically increasing, and $\mathbb R$ is partitioned into $j^{\text{bp}}((0,\tfrac{\Lambda_2}{4}])=(-\infty,0)$, $j^{\text{bp}}([\tfrac{\Lambda_2}{4},\tfrac{\Lambda_2}{2}])=[0,12^3\Lambda_2^4]$ and $j^{\text{bp}}([\tfrac{\Lambda_2}{2},\infty))=[12^3\Lambda_2^4,\infty)$. We aim to find a curve in $\tau$-space with these properties. 

The branch point is located at $u=u_{\text{bp}}=2m^2-\tfrac{\Lambda_2^2}{8}$. In the case $m=0$, $u_{\text{bp}}=u_+=u_-$ collide. For $m=\tfrac{\Lambda_2}{4}$, the branch point $u_{\text{bp}}=0$ is at the origin. At the AD mass $m=\tfrac{\Lambda_2}{2}$, the branch point collides with $u_*$ and $u_+$ at $\tad=1+\im$ (see Fig. \ref{fig:fund_dom_gamma_02F}). We can use this knowledge to conjecture the branch point paths in $\tau$-space. 

\begin{figure}[h]\centering
	\includegraphics[scale=1]{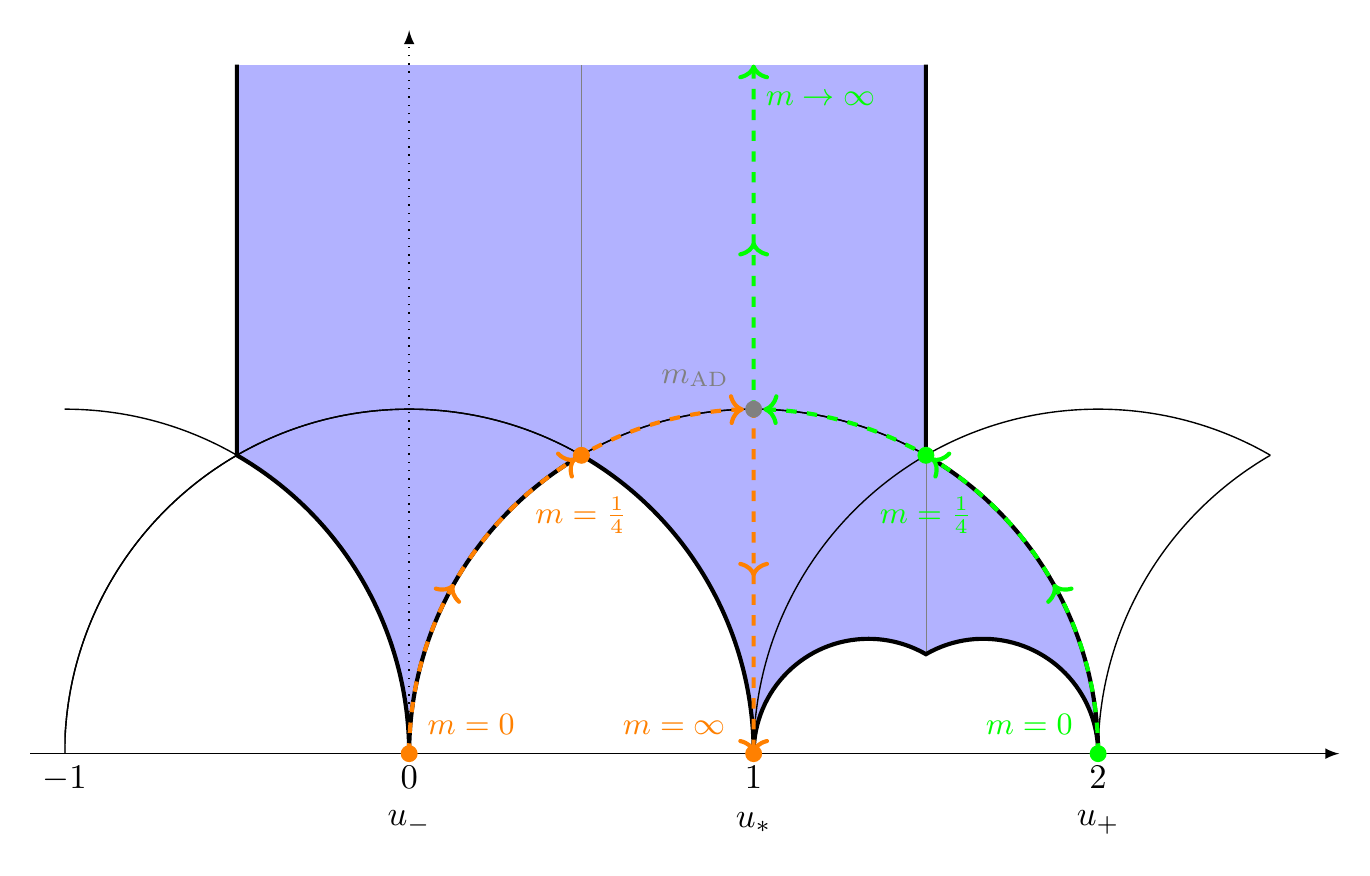}
	\caption{Fundamental domain $\CF_2(m,m)$ of massive
          $\bfm=(m,m)$ $N_f=2$ theory. The dashed lines correspond to
          the conjectured paths of the branch points from zero to
          infinite mass. For given positive mass $m$, the two branch
          points are identified under $TST^{-1}$, such that there is
          only one branch point
          $u_{\text{bp}}=2m^2-\tfrac{\Lambda_2^2}{8}$ on the
          $u$-plane. At $m=\mad$ the two branch points meet, the
          square root in $u(\tau)$ resolves, and $u(\tau)$ becomes
          holomorphic and modular.}\label{fig:fundu2m} 
\end{figure}

The cosets that we found above allow to construct a fundamental domain 
\begin{equation}\label{fmmnf2}
	\CF_{2}(m,m)=\CF\cup T\CF\cup S\CF\cup TS\CF\cup TST^{-1}\CF\cup T^2ST\CF.
\end{equation}
where we take the union of the elements in Table \ref{cuspexpnf2m}. This is drawn in Fig. \ref{fig:fundu2m} together with the conjectured paths of the branch points. Since the $\alpha_j$ generate the whole $\slz$, it is clear that this domain is not a fundamental domain of any congruence subgroup of $\slz$. By computing the $q$-series of all the cusp expansions, one can match the singularities with the cusps,
\begin{equation}
	u(0)=u_-, \quad u(1)=u_*, \quad u(2)=u_+.
\end{equation}
The generic mass case $\bfm=(m_1,m_2)$ splits the singularity $u_*$ further and removes either $TS\CF$ or $TST^{-1}\CF$ away from $\tau=1$.

Let us give some further evidence for the paths of the branch
points. The points for $m=0$, $m=\mad$ and $m=\infty$ are fixed from
the fact that in all three limits the duality group of the theory
becomes a congruence subgroup. The
branch points approach either $\tau=1$ or $\im\infty$ in the
decoupling limit, since these are identified under $\Gamma^0(4)$. This
agrees with the fact that $u_{\text{bp}}\to \infty$ for
$m\to\infty$. We can also check it against the solution
\eqref{unf2mm}. The branch point satisfies $f_2=0$, for which $u$
simplifies, 
\begin{equation}\label{ubpnf2mfunction}
	\frac{u(\tau_{\text{bp}})}{\Lambda_2^2}=-\frac{f_{2\text{B}}(\tfrac{\tau_{\text{bp}}}{2})+16}{128},\qquad f_{2\text{B}}(\tau)=\left(\frac{\eta(\tau)}{\eta(2\tau)}\right)^{24}=256\frac{\jt_3(\tau)^4\jt_4(\tau)^4}{\jt_2(\tau)^8},
\end{equation}
where $f_{2\text{B}}$ is a Hauptmodul of the congruence subgroup $\Gamma_0(2)\subseteq \SL$. One can plot this $u$ over the curves given in Fig.  \ref{fig:fundu2m}, and not only find that it is real everywhere, but it behaves as $u_{\text{bp}}=2m^2-\tfrac{\Lambda_2^2}{8}$ as a function of $m$. In particular, it is monotonically increasing and has the correct intermediate and limiting points $m\in\{0,\tfrac{\Lambda_2}{4},\tfrac{\Lambda_2}{2},\infty\}$. Therefore, the curves in Fig. \ref{fig:fundu2m} are parametrisations of \eqref{nf2bpjm} compatible with our solution for $u$. 

For any mass, the pair of branch points is identified under $u$. In order to see this, note that the value of $u$ at a branch point is given by \eqref{ubpnf2mfunction}. Since it is a modular function for $\Gamma^0(2)$, it is invariant under $TST^{-1}$. This relates the two loci in Fig. \ref{fig:fundu2m} at both smooth components of each curve,
\begin{equation}\begin{aligned}\label{tstibp}
		TST^{-1}:\begin{cases}1+e^{\varphi\im}\!\!\!\!&\longmapsto 1+e^{(\pi-\varphi)\im},\\
			1+\im \delta&\longmapsto 1+\frac1\delta\im.\end{cases}
\end{aligned}\end{equation}
The pair of two such points are the branch points of the square root, and the branch cut can be any path connecting the two branch points \cite{gamelin2003complex}. For $m> \mad$ for instance, on can take it to be the complex interval $\CI_\delta=1+[\tfrac1\delta,\delta]\im$. This can also be seen from the fact that when $\tau$ traverses a small circle around one branch point, the expression $u(\tau)$ receives a minus sign in front of the square root. According to Table \ref{cuspexpnf2m} this interchanges the cusp expansions in the regions $T\CF$ and $TS\CF$, and the transition map is precisely $(TS)T^{-1}$ as in \eqref{tstibp}. For $m<\mad$ the branch points sit on the boundaries of $S\CF$ and $T^2ST\CF$, and the transition map $S(T^2ST)^{-1}=TST^{-1}$ is identical.  In order to achieve single-valuedness, any path encircling one branch point must also encircle the other. On a \emph{dogbone} contour around the interval $\CI_\delta$ the function $u(\tau)$ returns to the original value, as it picks up twice the phase factor $-1$.
The function $u(\tau)$ is then a continuous single-valued function on the slit plane $\CF(m,m)\backslash \CI_\delta$, which one may interpret as a Riemann surface.

\subsubsection*{Limits to zero, AD and infinite mass}
The limits to other theories are given as follows. For $m\to 0$, the singularities $u_+$ and $u_-$ merge at $-\frac{\Lambda_2^2}{2^8}$, which we located at $\tau=0$. This agrees with the fact that for $m=0$ the order parameter is modular for $\Gamma(2)$ and in particular invariant under $T^2$. More precisely, we can use $\Gamma(2)$ to move the copies $TST^{-1}\CF$ and $ T^2 ST\CF$ in order to obtain a more canonical form of $\Gamma(2)\backslash\slz$. For this, note that we can identify $ST^{-1}\CF$ and $T^2ST\CF$, since 
\begin{equation}
ST^{-1}(T^2ST)^{-1}=\begin{pmatrix}1&-2\\2&-3\end{pmatrix}\in \Gamma(2): 2\longmapsto 0.
\end{equation}
Similarly, we can identify $TST^{-1}\CF$ with $TST\CF$, as the transition function is also in $\Gamma(2)$. This gives precisely Fig. \ref{fig:masslessNf2domains}.  In fact, since these transition functions are in $\Gamma(2)$, Fig. \ref{fig:fundu2m} gives an \emph{equivalent} fundamental domain for $\Gamma(2)$. It is however not the preferred choice for two reasons.  First, not all copies of $\CF$ are in a strip of width $2$. Second, not all possibilities for cusp identifications have been taken, and it is preferable to only show inequivalent cusps. It is noteworthy that the (horizontal) width of the domain for fixed imaginary part never exceeds $2$, and that it is possible to draw the fundamental domain as an actual \emph{domain}, i.e. a connected open subset of $\mathbb H$.

The decoupling limit $m\to \infty$ to $N_f=0$ is also interesting. The triangle $TST^{-1}\CF$ can be identified with $T^2\CF$ since 
\begin{equation}
(T^2)^{-1}TST^{-1}=\begin{pmatrix}-1&0\\1&-1\end{pmatrix}\in \Gamma^0(4): 1\longmapsto \im\infty. 
\end{equation}
Similarly, we can identify $TS\CF$ with $T^3\CF$ as the transition map is  in $\Gamma^0(4)$ and also maps $1\mapsto \im\infty$. Lastly, the triangle $T^2ST\CF$ around $\tau=0$ can be identified with $T^2S\CF$. This demonstrates that not only do we get the domain $\Gamma^0(4)$ as in Fig. \ref{fig:fundgamma0(4)}, but in fact  the domain in Fig. \ref{fig:fundu2m} is also a fundamental domain for $\Gamma^0(4)$. Aside from the disclaimer of the above paragraph, it is not a sound modular domain for $\Gamma^0(4)$ as the lines with constant real parts are not identified. The flow to the low energy effective theory with no hypermultiplets can be understood from the modular curve perspective as identifying the cusp $\tau=1$ of width $2$ with the cusp $\im\infty$, such that the number of rational cusps decreases by $2$, while the width of the cusp $\im\infty$ increases by $2$.

In the AD limit $m\to \pm\mad$, the mutually non-local singularities
$u_*$ and $u_+$ collide and become elliptic points of the curve. This
eliminates all the triangles near these cusps: In this case the
regions $TS\CF$, $TST^{-1}\CF$ and $T^2ST\CF$ are removed and the
domain of the theory with this mass, see
Fig. \ref{fig:fund_dom_gamma_02F}, remains. The AD point $\tau_{\rm
  AD}$ lies in the interior of $\mathbb H$, and is an elliptic point
of the duality group $\Gamma^0(2)$. 

\subsubsection*{$u$-plane of AD theory}
Similarly to the AD point for $N_f=2$, the disconnected cusps
corresponding to the non-local singularities form the fundamental
domain for order parameter of the AD curve. The disconnected cusps
form a fundamental domain for $\Gamma_0(2)$, which is incidentally congruent
to the duality group of the asymptotically free theory at the AD point.
To demonstrate this, recall that the AD curve reads
\cite{Argyres:1995xn},  
\be
y^2=x^3-\frac{\Lambda_2^2}{4} \tilde u x-\frac{\Lambda_2^3}{12}m\tilde u +\frac{\Lambda_2^3}{27}m^3. 
\ee
This gives for the order parameter
\be
\tilde u(\tau)=\frac{4}{3}m^2-\frac{m^2\,f_{2B}(\tau)}{64+f_{2B}(\tau)},
\ee
with $f_{2B}(\tau)$ as in (\ref{ubpnf2mfunction}). $f_{2B}$ is a
Hauptmodul for $\Gamma_0(2)$, such that the disconnected domain is
indeed a fundamental domain for $\tilde u$.

\subsubsection*{Partitioning of the $u$-plane}
Finally, we can study the partitioning that the domain \eqref{fmmnf2} induces on the $u$-plane under the map \eqref{unf2mm}. As studied in Section \ref{sec:uplanetiling}, the partitioning is contained in a real algebraic plane  curve, which is given by the equation $\text{Im}\,\CJ(u,\bfm,\Lambda_2)=0$. For generic $\mu=\frac{m}{\Lambda_2}$, we can compute it as the zero locus of the polynomial
\begin{equation}\footnotesize\begin{aligned}\label{mmpolynomial}
T_{(m,m)}=&y \left(-128 \mu^2 x+48 \mu^2+64 x^2-16 x+64 y^2-3\right) \\
&\times \big(-720896 \mu^4 x^2 y^2+262144 \mu^2 x^2 y^4-262144 \mu^2 x^4 y^2-262144 \mu^2 x^3
   y^2\\
   &-303104 \mu^2 x^2 y^2-589824 \mu^6 x^2+688128 \mu^4 x^4+737280 \mu^4 x^3+27648 \mu^4 x^2\\&
   -262144 \mu^2 x^6-786432 \mu^2 x^5-86016 \mu^2 x^4-36864 \mu^2
   x^3-1728 \mu^2 x^2\\&
   -49152 \mu^4 x y^2+524288 \mu^2 x y^4+24576 \mu^2 x y^2-221184 \mu^6 x+20736 \mu^4 x\\&+589824 \mu^6 y^2+688128 \mu^4 y^4+9216 \mu^4
   y^2+262144 \mu^2 y^6+12288 \mu^2 y^4\\&
  -2880 \mu^2 y^2+165888 \mu^8-1944 \mu^4+81 \mu^2+786432 x^3 y^4+65536 x^2 y^4\\&+786432 x^5 y^2+131072 x^4
   y^2+40960 x^3 y^2+10240 x^2 y^2+262144 x^7\\&+65536 x^6+12288 x^5+6144 x^4-576 x^3+144 x^2+262144 x y^6+28672 x y^4\\&-1344 x y^2-27 x\big).
\end{aligned}\end{equation}
The second factor on the rhs gives a circle on the $x+\im y=\tfrac{u}{\Lambda_2^2}$-plane with radius $|\mu^2-\frac14|$ and centre $(x,y)=(\mu^2+\frac18,0)$. By tuning the mass $\mu$ from $0$ to $\infty$, one passes through the AD point $\mu=\frac12$ where the radius of the circle shrinks to $0$. For this mass, three regions defined through $T_{(m,m)}=0$ collapse to a point $x+\im y=\tfrac{\uad}{\Lambda_2^2} =\frac38$,  which is the only root over $\mathbb R^2$ of the quadratic polynomial.  This gives further evidence that the domain 
\eqref{fmmnf2} is in fact correct for all $\mu\in\mathbb (0,\infty)\backslash \{\frac12\}$.

\begin{figure}
	\begin{subfigure}{.49\textwidth}
		\centering
		\includegraphics[width=\linewidth]{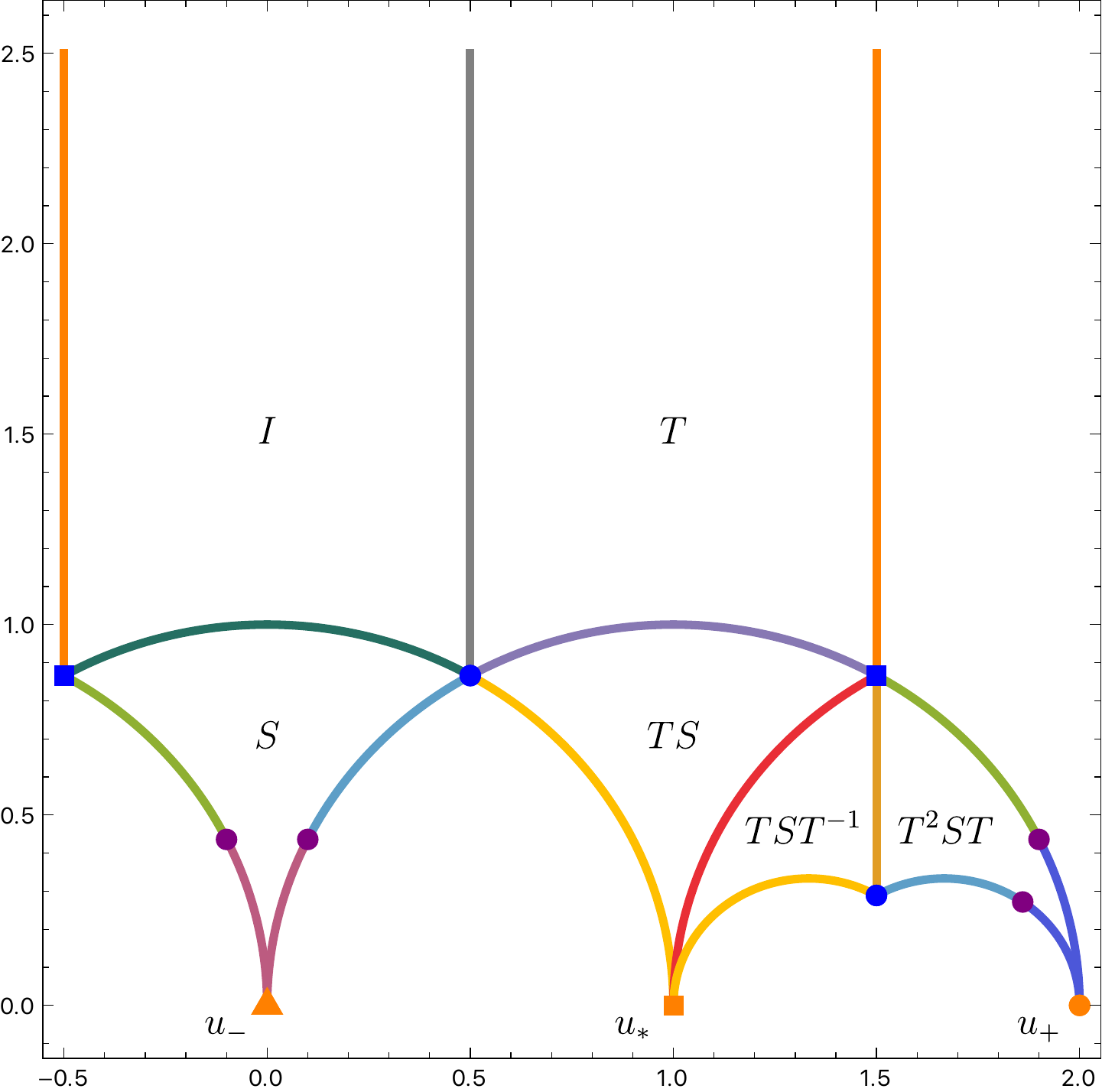}  
			\end{subfigure}
	\begin{subfigure}{.49\textwidth}
		\centering
		\includegraphics[width=\linewidth]{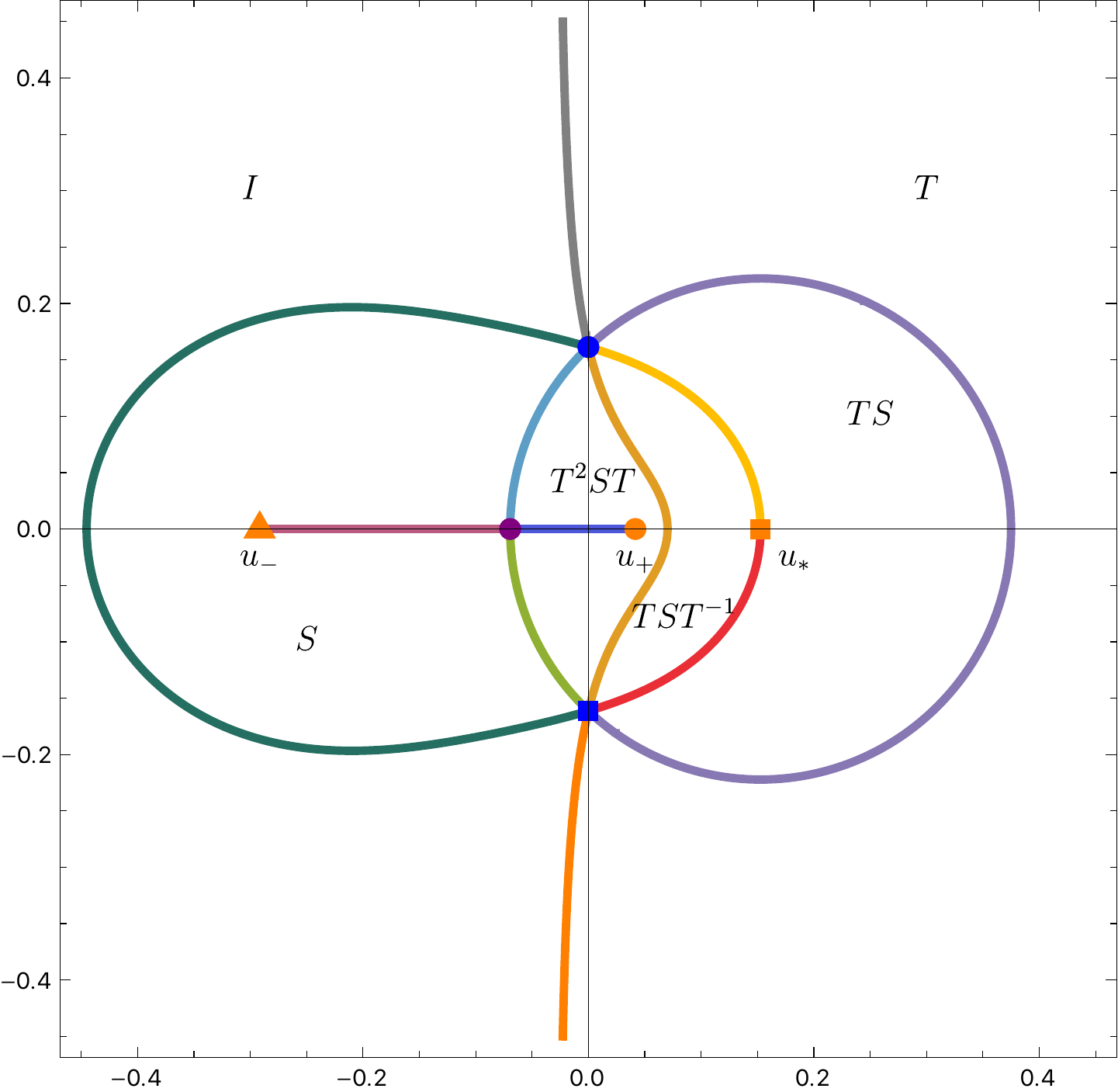}  
\end{subfigure}
	\caption{Identification of the components of the partitioning $\CT_{(m,m)}$ in $N_f=2$ for the particular choice $m=\frac{\Lambda_2}{6}$. The $u$-plane $\CB_2$ is partitioned into 6 regions $u(\alpha\CF)$, with the $\alpha\in\slz$ given in both pictures.  The branch point (purple) identifies four points on $\partial\CF(m,m)$. A natural choice of branch cut is on the circle around $\tau=1$ with radius $1$, as suggestive in Fig. \ref{fig:fundu2m} (we omit it in this Figure for readability). The singularities $u_\pm$ correspond to a single massless particle each and thus lie in the interior of a $u(\alpha\CF)$. The singularity $u_*$ is double and thus lies on the boundary of two such regions. The boundary pieces of $\CF(m,m)$ are pairwise identified, which can be found by comparing $\CF(m,m)$ with the curve $T_{(m,m)}=0$. Glueing the corresponding boundary pieces results in a Riemann surface of genus 0 with punctures. }
	\label{fig:nf2partition}
\end{figure}

We can find the truncations of the  
zero locus of \eqref{mmpolynomial} that gives the partitioning \eqref{tmj} in the following way. The locus $y=0$ cannot be contained fully in $\CT_\bfm$, since otherwise the partition of $\CB_2$ would be into more than 6 parts. By direct computation one can show that for $0<\frac{m}{\Lambda_2}<\frac14$ we have $\CJ(u,\bfm)\leq 12^3$ for $u_-<u<u_+$ (recall that $\CJ(u,\bfm)$ diverges for all $u$ approaching a singularity). This proves that the line from $u_-$ to $u_+$ is contained in $\CT_\bfm$. It allows to identify the boundary pieces $\alpha_j\partial\CF$ on $\mathbb H$ with the boundary pieces $\partial (u(\alpha_j\CF))$ on $\CB_2$, which is depicted in Fig. \ref{fig:nf2partition}.

\subsection{Two distinct masses}
\begin{figure}
	\begin{subfigure}{.49\textwidth}
		\centering
		\includegraphics[width=\linewidth]{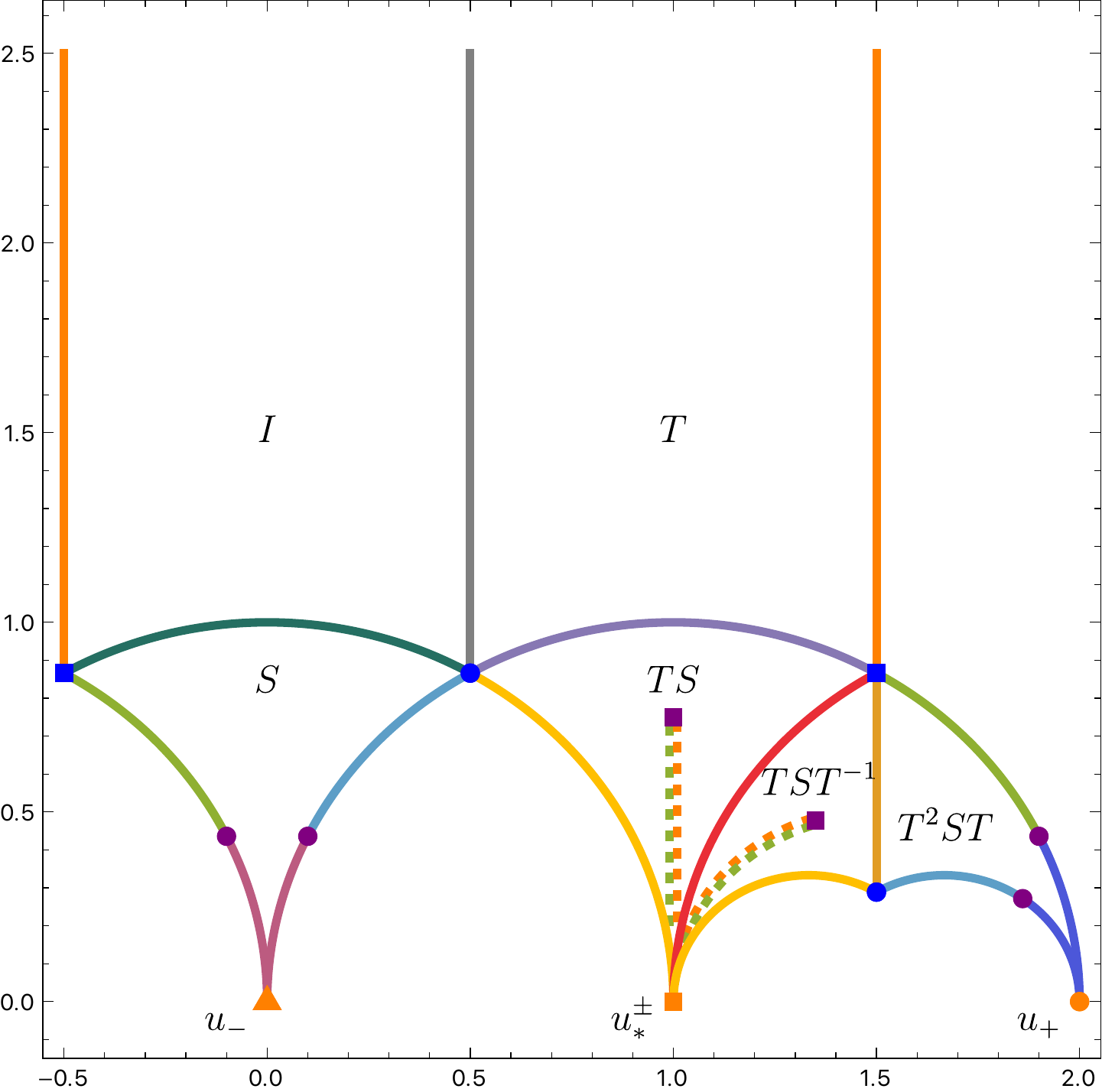}  
			\end{subfigure}
	\begin{subfigure}{.49\textwidth}
		\centering
		\includegraphics[width=\linewidth]{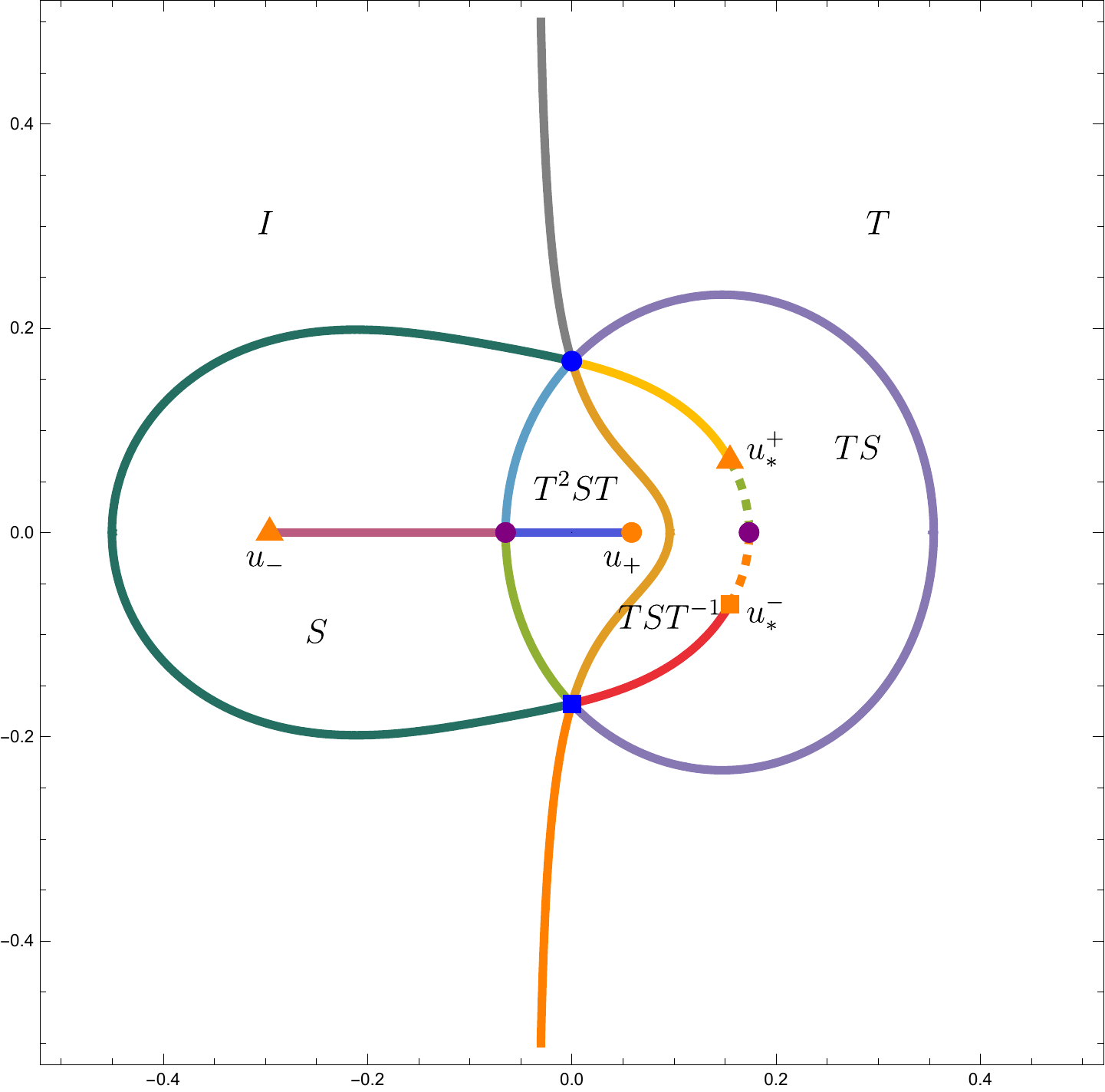}  
\end{subfigure}
	\caption{Identification of the components of the partitioning $\CT_{(m_1,m_2)}$ in $N_f=2$ for the particular choice $\mu_1=\frac{1}{10}$ and $\mu_2=\frac14$. The $u$-plane $\CB_2$ is naively partitioned into six regions $u(\alpha\CF)$, with the $\alpha\in\slz$ given in both pictures. Two regions $u(TS\CF)$ and $u(TST^{-1}\CF)$ are however glued along the pairs of branch cuts (dotted), running from the two singular points $u_*^\pm$ (orange, square) to the branch point $\tau_{\text{bp},2}$ (purple, square). They do not  belong to the partitioning $\CT_\bfm$. A natural choice for the branch cut is along the lines where $j(\tau)$ is real.}
	\label{fig:nf2m1m2partition}
\end{figure}

In the generic case, the two masses are distinct. As in $N_f=1$, we can expand and invert the $\CJ$-invariant for large $u$ to find the series ($\mu_i=\frac{m_i}{\Lambda_2}$)
\begin{equation}\label{uNf2genMass}
	\begin{aligned}
		\frac{u(\tau)}{\Lambda_2^2}=-&\frac{1}{64}q^{-\frac12} - 
		\frac12 (\mu_1^2 + \mu_2^2) + \left(24 (\mu_1^4 + \mu_2^4) + 
		16 \mu_1^2 \mu_2^2 - 32 \mu_1 \mu_2 - \frac{5}{16}\right) q^{\frac12} \\
		-& 
		128 \left(\mu_1^2 + \mu_2^2\right) \left(16 (\mu_1^4 + \mu_2^4) - 
		14 \mu_1 \mu_2 + 1\right) \,q+\CO(q^{\frac32}).\end{aligned}
\end{equation}
The double singularity $u_*$ in the equal mass case now splits into two distinct singularities, $u_*^\pm$. Due to the locus of masses giving rise to $u$-planes with AD points, it is difficult to give a fundamental domain $\CF_2(\bfm)$ for any choice of $\bfm=(m_1,m_2)$. From \eqref{pbp} it is clear that there are two distinct branch points in $\CB_2$. When both $m_1$ and $m_2$ are real and small, i.e. have not made a phase transition compared to $\bfm=0$, one branch point $u_{\text{bp},1}$ belongs to $\CT_\bfm$, while the other $u_{\text{bp},2}$ does not. However, $\CJ(u_{\text{bp},2})=j(\tau_{\text{bp},2})\in\mathbb R$ is also real but larger than $12^3$. A natural choice of branch cuts is along the tessellation $\{\tau\in\mathbb H\,|\, j(\tau)\in\mathbb R\}$, which aside from \eqref{slztessellation} contains the $\slz$ images of the positive imaginary axis.  The plot of the partitioning $\CT_\bfm$ shows a feature found already in $N_f=1$ with a complex mass (see Section \ref{sec:nf1complexm}): The $u$-plane is partitioned into only $5$ regions, which is due to two regions $u(\alpha_j\CF)$ being glued along pairs of branch cuts (see Fig. \ref{fig:nf2m1m2partition}). The splitting of $u_*$ into two distinct singularities in this case does not require the two regions $TS\CF$ and $TST^{-1}\CF$ to taper to distinct cusps, as we have that both  $TS, TST^{-1}:\im\infty\mapsto 1$. The two singularities are rather split due to the branch cut, and the limit of $u(\tau)$ as $\tau\to 1$ depends on the path from which $\tau=1$ is approached. This is different from $u_+\neq u_-$, where the boundary pieces near the cusps are not identified.

\subsection{The massless theory}\label{sec:masslessnf2}
When we go to the massless theory we now find
\begin{equation}\label{masslessnf2u}
\begin{aligned}
\frac{u(\tau)}{\Lambda_2^2}=&-\frac{1}{8}\frac{\vartheta_3(\tau)^4+\vartheta_4(\tau)^4}{\vartheta_2(\tau)^4}=-\frac18-\frac{1}{64}\left(\frac{\eta(\frac\tau2)}{\eta(2\tau)}\right)^8\\
=&-\frac{1}{64}(q^{-1/2}+20q^{1/2}-62q^{3/2}+216q^{5/2}+\CO(q^{7/2})).
\end{aligned}
\end{equation}
This function is the completely replicable function of class 4C and is a Hauptmodul for $\Gamma(2)$ \cite{Alexander:1992,ford1994,Ferenbaugh1993}. The physical discriminant becomes $\Delta=(u+\tfrac{\Lambda_2^2}{8})^2(u-\tfrac{\Lambda_2^2}{8})^2$. 
The two cusps correspond to  $u(0)=-\frac{\Lambda^2_2}{8}$ and $u(1)=+\frac{\Lambda_2^2}{8}$. They are associated with the particles of charges $(1,0)$ and $(1,1)$ becoming massless.

\begin{figure}
	\begin{subfigure}{.5\textwidth}
		\centering
		\includegraphics[width=\linewidth]{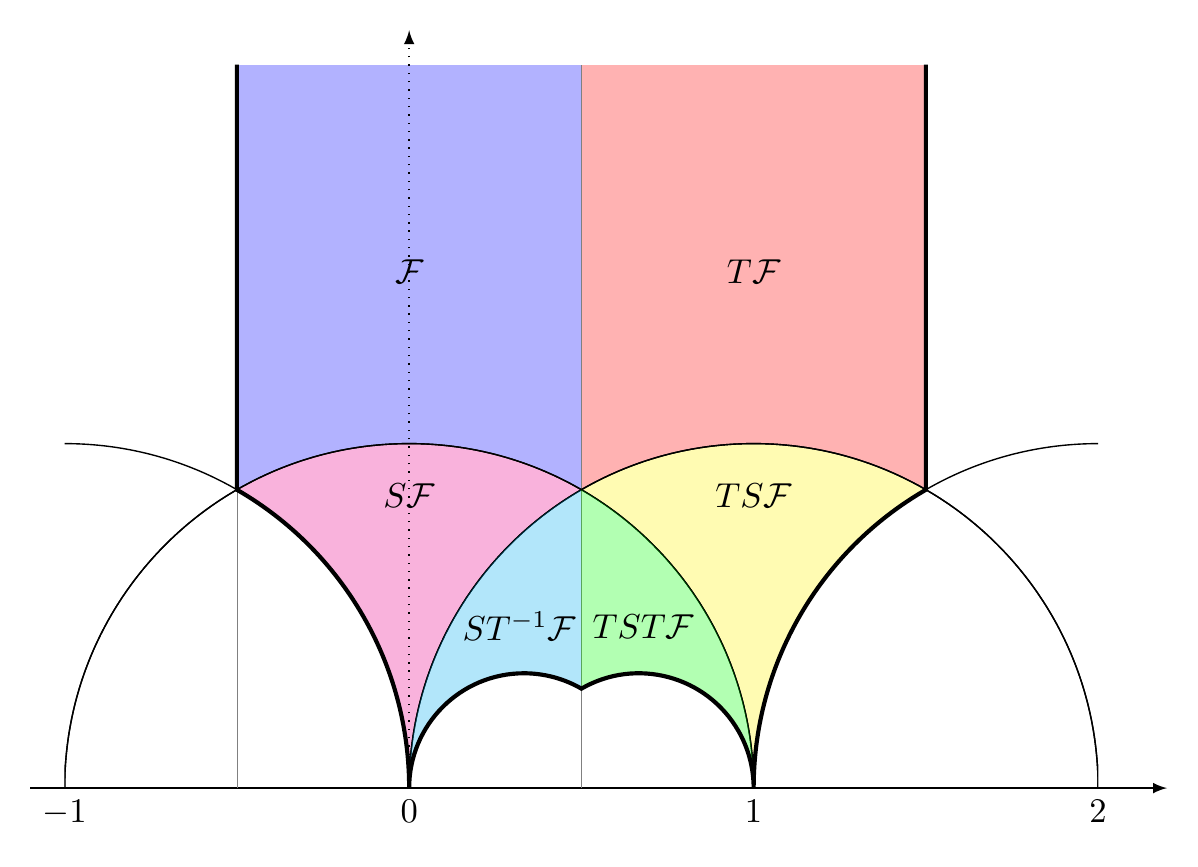}  
	\end{subfigure}
	\begin{subfigure}{.5\textwidth}
		\centering
		\includegraphics[width=.65\linewidth]{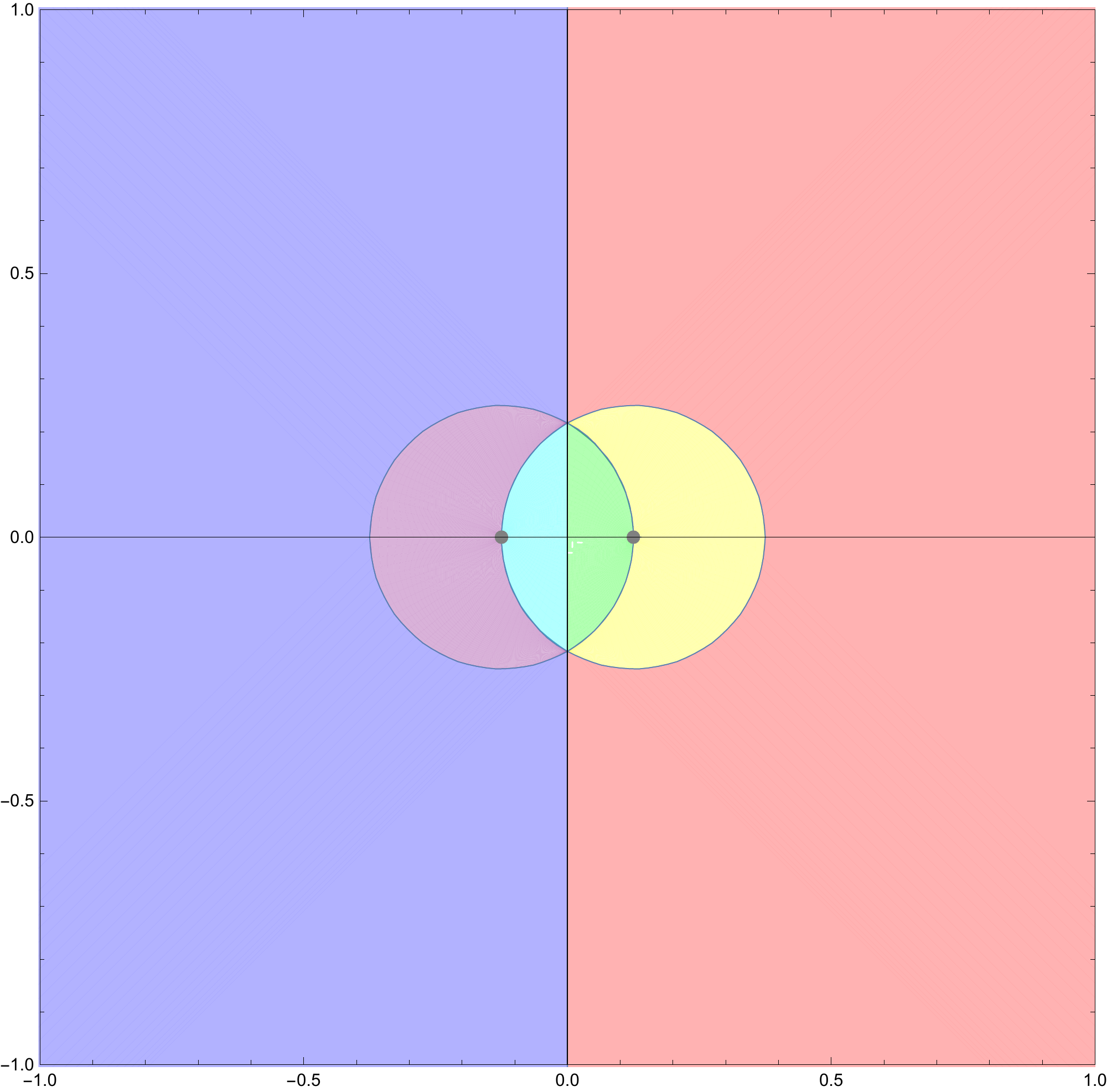}  
	\end{subfigure}
	\caption{Left: Fundamental domain of $\Gamma(2)$. This is the duality group of massless $N_f=2$. All three cusps $\{\im\infty,0,1\}$ have width 2. Right: Plot of the massless $N_f=2$ $u$-plane as the union of the images of $u$ under the $\ind\Gamma(2)=6$ $\slz$ images of $\CF$. Here, we use the decomposition $\Gamma(2)\fd=\bigcup_{k,\ell=0}^1T^\ell S^k \CF \cup ST^{-1}\CF\cup TST\CF$. There is a $\mathbb Z_2$ symmetry which acts by $u\mapsto -u$. The singularities $\tau=0, 1$ are both touched by two triangles each. }
	\label{fig:masslessNf2domains}
\end{figure}

A fundamental domain for $\Gamma(2)$ is given by
\begin{equation}
\CF_2(0,0)=\CF\cup T\CF\cup S\CF\cup TS\CF\cup ST^{-1}\CF\cup TST\CF
\end{equation}
and is plotted in Fig. \ref{fig:masslessNf2domains} together with the map to the $u$-plane. This picture gives rise to the \emph{dessin d'enfant} of the $j$-invariant \cite[Fig. 6]{juanzacarias2020}, as $u$ is a linear function of the modular $\lambda$-invariant, which has critical points $\lambda=0,1,\infty$.

\subsection{Type $III$ AD mass}\label{sec:adnf2}

\begin{figure}[h]\centering
	\includegraphics[scale=0.7]{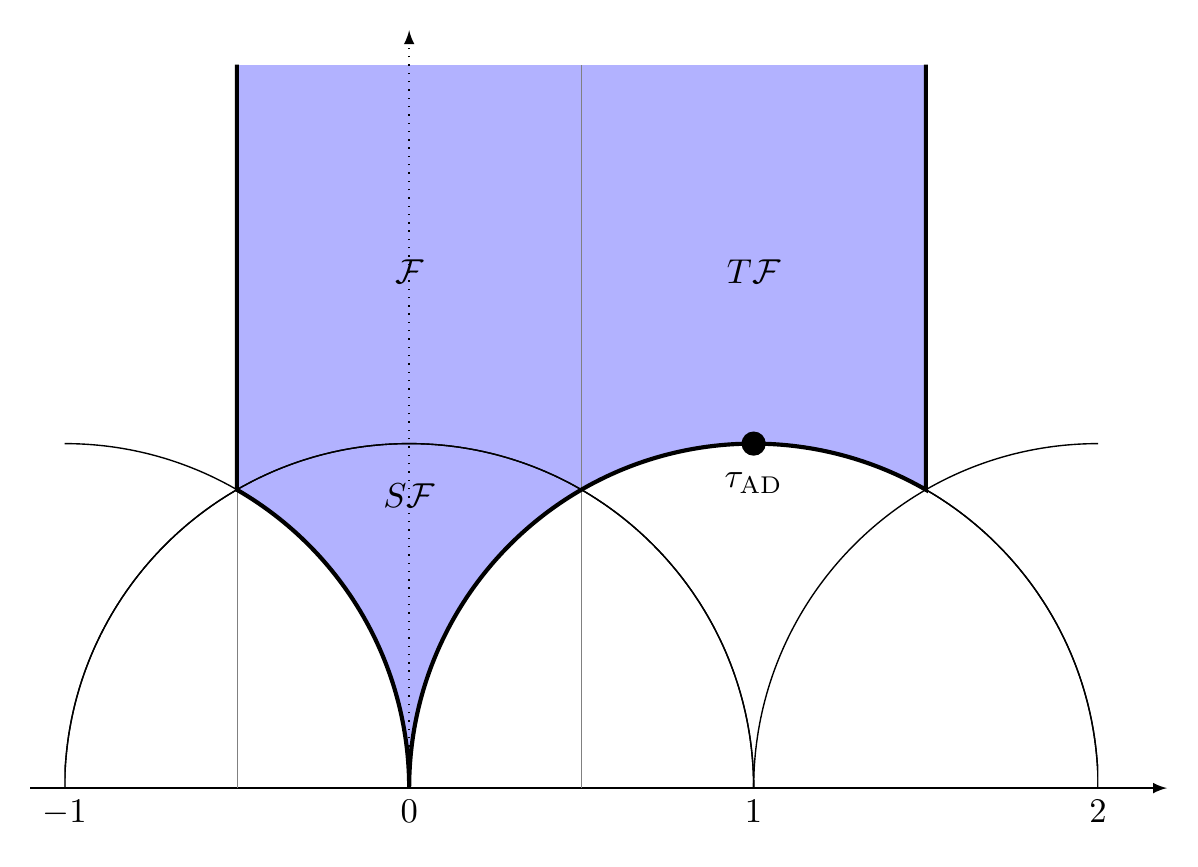}
	\caption{Fundamental domain of $\Gamma^0(2)$. This is the duality group of $N_f=2$ with masses $\bfm=\tfrac12(\Lambda_2,\Lambda_2)$. The AD point corresponds to the elliptic fixed point $\tad=1+\im$.}\label{fig:fund_dom_gamma_02F}
\end{figure}

If we choose $m_1=m_2=\mad=\tfrac12 \Lambda_2$, we find a $u$-plane with an AD theory of type $III$ located at $u=\uad=\tfrac{3}{8}\Lambda_2^2$ \cite{Argyres:1995xn}. Three singularities collide in this point, while one remains at $u_0=-\tfrac58 \Lambda_2^2$. The discriminant now takes the form
\begin{equation}\label{nf2ADdisc}
\Delta=(u-u_{\text{AD}})^3(u-u_0).
\end{equation}
Using $\Gamma(2)$ as an intermediate field of the sextic equation, we can show that
\begin{equation}\label{uIIInf2minus}
\frac{u(\tau)}{\Lambda_2^2}=-\frac{f_{\text{2B}}\left(\tfrac\tau2\right)+40}{64}=-\frac{1}{64}\left(q^{-1/2}+16+276q^{1/2}-2048q+\CO(q^{3/2})\right),
\end{equation}
where $f_{\text{2B}}$ was defined in \eqref{ubpnf2mfunction}, and it is the McKay-Thompson series of class 2B \cite{Alexander:1992,ford1994,Ferenbaugh1993}. It is a Hauptmodul for $\Gamma_0(2)$. Therefore, $u$ is a modular function for $\Gamma^0(2)$. A fundamental domain of $\Gamma^0(2)$ is 
\begin{equation}
\CF_2(\bfm_{\text{AD}})=\CF\cup T\CF\cup S\CF
\end{equation}
and is shown in  Fig. \ref{fig:fund_dom_gamma_02F}. It has index 3 in $\psl$, since three mutually non-local singularities have collided. This can also be seen from the fact that the curve reads
\begin{equation}\label{jf2b}
j(\tau)=\frac{(f_{2\text{B}}(\tfrac\tau2)+16)^3}{f_{2\text{B}}(\tfrac\tau2)}.
\end{equation}

 One has that $u(\tad)=\uad$ whenever $f_{2\text{B}}(\tfrac\tad2)=-64$, whose solution locus intersects with our choice of $\Gamma^0(2)\backslash\mathbb H$ in  $\tad=1+\im$. This can be proven from the $S$-transformation of the Dedekind $\eta$ function. It is also easy to check that $u(0)=u_0$. Taking the proper limits in \eqref{massivenf2quantities} we directly find $\frac{du}{d\tau}$ as well as $\frac{da}{du}$ and we can check that they satisfy the Matone relation
\begin{equation}\label{21Matone}
\frac{d u}{d\tau}=-\frac{16\pi\im}{2} \widehat \Delta\left(\frac{da}{d u}\right)^2,
\end{equation}
consistent with \eqref{matonepnf}. Both branch points of the $N_f=2$ theory have collided along with the singularities where mutually non-local states become massless. The monodromies are
\begin{equation}\label{gamma^02mono}
M_0=STS^{-1}=\begin{pmatrix}1&0\\ -1&1\end{pmatrix},\quad M_{\text{AD}}=TS^{-1}T^{-1}=\begin{pmatrix}-1&2\\-1&1\end{pmatrix},
\end{equation}
and they satisfy $M_0 M_{\text{AD}}=M_\infty$ with $M_\infty=PT^{-2}$. Furthermore, $M_{\text{AD}}^4=\mathbbm 1$, such that $\tad$ indeed is an elliptic fixed point of $\Gamma^0(2)$. The AD monodromy is conjugate to $S^{-1}$, which fixes $\tau=\im$. Since $\tad=\im+1$, this gives a path in $\tau$-space.

\section{The $N_f=3$ curve}\label{sec:nf3}
We will start by discussing the $N_f=3$ theory with one non-zero mass, $\bfm=(m,0,0)$, where we can find an explicit expression for $u$ in terms of Jacobi theta functions. After this we discuss the generic mass case, the  massless theory and a number of theories with specific AD masses.

\subsection{One non-zero mass}
For the general theory it turns out to be complicated to find closed expressions for $u$, but if we only keep one non-zero mass, $\bfm=(m,0,0)$, we can make more progress. Four of the strong coupling singularities now merge in pairs of two and the physical discriminant becomes
\begin{equation}
	\Delta = (u-u_+)^2(u-u_-)^2(u-u_*),
\end{equation}
with 
\begin{equation}\label{nf3m00sing}
	u_\pm= \pm\frac{m\Lambda_3}{8}, \qquad u_*=\frac{\Lambda_3^2}{2^8}+m^2.
\end{equation}
There are two AD points at $m=\mad=\pm\tfrac{1}{16}\Lambda_3$ and $u=\uad=\tfrac{1}{128}\Lambda_3^2$ where either $u_+$ or $u_-$ merges with $u_*$ to give a type $III$ singular fibre. We now find that the sextic equation for $u$ again splits over the intermediate field $\Gamma(2)$. In this case there is only one solution that has $|u|\to\infty$ for $\tau\to\im\infty$, and as has been mentioned before this has $u\to -\infty$. This is then the reason why we have persistently chosen this convention in all other cases, to make the decoupling limits from $N_f=3$ consistent. We find that
\begin{equation}\begin{aligned}\label{uMassiveNf3}
	\frac{u}{\Lambda_3^2}&=-\frac{2\vartheta_3^4\vartheta_4^4+(\vartheta_3^4+\vartheta_4^4)\sqrt{f_3}}{64\vartheta_2^8}\\
	&=-\frac{1}{2^{12}}\left(\frac 1q+ (-8 + 4096 \mu^2) + 4 (5 + 32768 \mu^2 - 4194304 \mu^4) q+\CO(q^{2})\right),
\end{aligned}\end{equation}
where we have defined $f_3=\frac{64m^2}{\Lambda_3^2}\vartheta_2^8+\vartheta_3^4\vartheta_4^4$ and $\mu=\frac {m}{\Lambda_3}$. It is straightforward to calculate the other interesting quantities explicitly from \eqref{uMassiveNf3} and to check that the generalised Matone relation is satisfied also in this case. 

Similarly to what we saw in the equal mass $N_f=2$ case, the square roots will introduce branch points in the moduli space. They are given by
\begin{equation}\label{jbpnf3}
	j^{\text{bp}}(m)=\frac{(\Lambda_3-8m)^3(\Lambda_3+8m)^3}{16m^4\Lambda_3^2}.
\end{equation}
By plugging in the expression for $\CJ$ in terms of $u$ we find that the branch point lies at $u_{\text{bp}}=2m^2$ in the $u$-plane, as is also found by studying the Matone polynomial \eqref{pnf}. We can also use known relations between the $j$-invariant and theta functions to check that \eqref{jbpnf3} coincides with $f_3=0$, such that the branch point of $u$ is that of the square root in \eqref{uMassiveNf3}.

\subsubsection*{Fundamental domain}
\begin{figure}[h]\centering
	\includegraphics[scale=0.8]{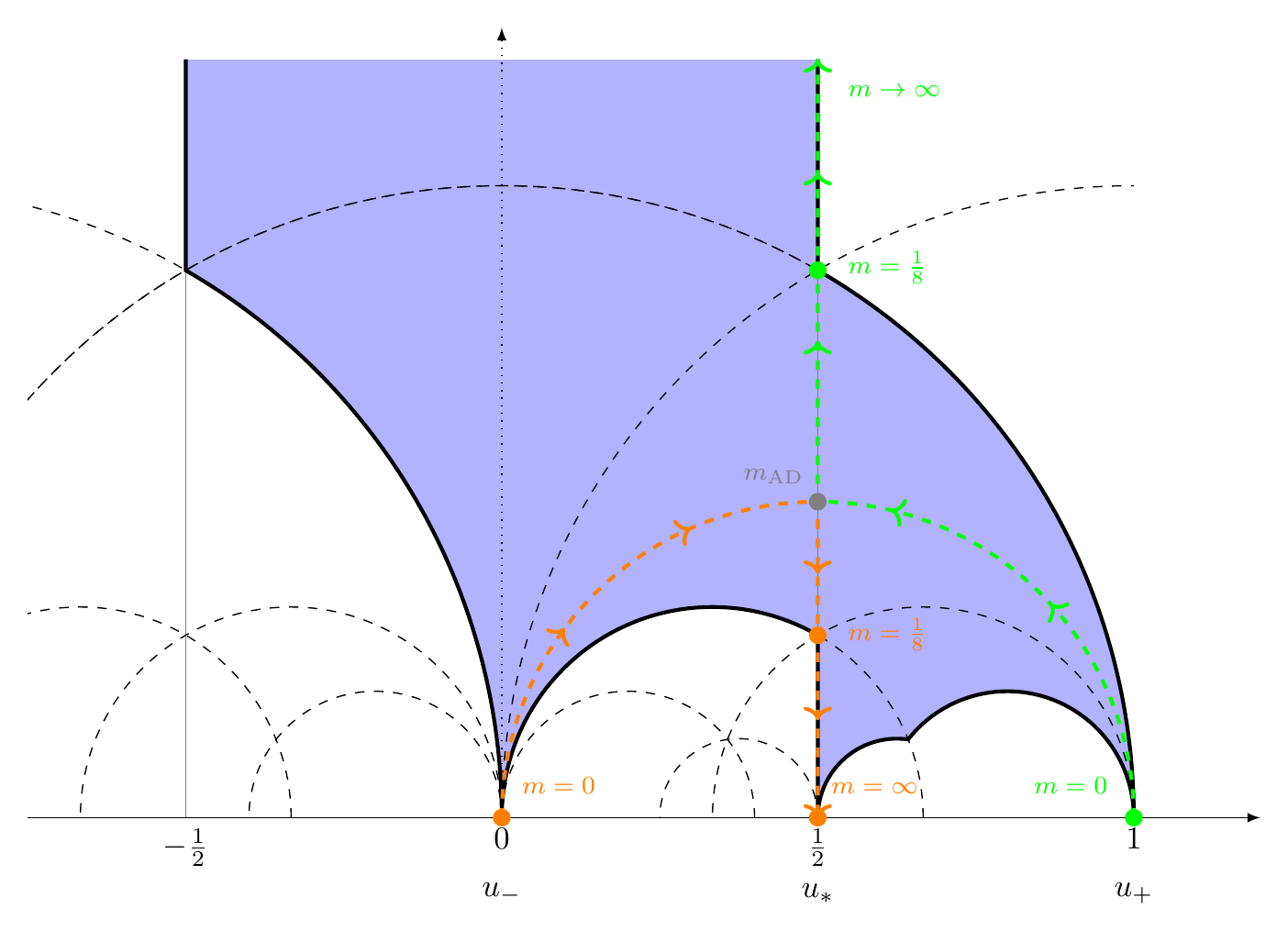}
	\caption{Fundamental domain for $N_f=3$ with $\bfm=(m,0,0)$. The dashed lines corresponds to the path of the branch points on the positive ray from massless to infinite mass.}\label{fig:fundu3m}
\end{figure}

We can repeat the method developed in $N_f=2$ with $\bfm=(m,m)$ and write down all the cusp expansions. They can be canonically normalised to match the form of the expansion at $\infty$. This allows to find the maps $\alpha_j\in\slz$, which give the fundamental domain 
\begin{equation}
\CF_{3}(m,0,0)=\CF\cup S\CF\cup ST^{-1}\CF\cup TST\CF\cup TST^2\CF\cup TST^2S\CF,
\end{equation}
shown in Fig. \ref{fig:fundu3m}. It is valid for all masses $m$ that do not allow the square root to resolve. We prove below that this does not happen unless $m=0$ or $m=\mad=\frac{1}{16}\Lambda_3$.

Let us also study the paths of branch points in the fundamental domain. Similarly as in massive $N_f=1,2$, we analyse the critical values of \eqref{jbpnf3}. We have that $\lim_{m\searrow 0}j^{\text{bp}}(m)=+\infty$, $j^{\text{bp}}(\mad)=12^3$, $j^{\text{bp}}(\tfrac{\Lambda_3}{8})=0$ and $\lim_{m\to\infty}j^{\text{bp}}(m)=-\infty$. It is easy to show that $j^{\text{bp}}: (0,\infty)\to \mathbb R$ is monotonically decreasing and therefore injective. 

Since $u_{\text{bp}}=2m^2$, we have that at $m=0$ the branch points coincides with $u_+$ and $u_-$. At the AD point, $m=\tfrac{\Lambda_3}{16}$, it collides along with $u_*$ and $u_+$. Finally, for $m\to \infty$ it diverges, just as $u_*$ does. This fixes the points $\tau=0$ and $\tau=1$ for $m=0$, $\tau=\tad=\tfrac12+\tfrac\im2$ for $m=\mad$ and $\tau=\tfrac12$ or $\im\infty$ for $m= \infty$. The simplest curves connecting these three points are quarter-circles with radius $\tfrac12$ around $\tau=\tfrac12$ starting from either $\tau=0$ or $\tau=1$, followed by a vertical path from $\tfrac12+\tfrac\im2$ to either $\tfrac12$ or $\im\infty$. 

The fundamental domain together with the path of the branch points found from the above considerations is shown in Fig. \ref{fig:fundu3m}.

The various checks of the branch points paths are analogous to $N_f=1,2$. We can plot $j(\tau)$ along these curves and find that it has the same global properties and critical points as \eqref{jbpnf3}. The intermediate value $j(\tau)=0$ corresponds to $m=\tfrac{\Lambda_3}{8}$ and $\tau=\tfrac12+\tfrac{1}{2\sqrt2}\im$, which is in the $\slz$-orbit of $\omega_3$. Along the branch point locus, $u$ simplifies to 
\begin{equation}
	u(\tau_{\text{bp}})=-\frac{f_{2\text{B}}(\tau_{\text{bp}})}{2^{13}},
\end{equation}
with $f_{2\text{B}}$ given by \eqref{ubpnf2mfunction}. On the paths in Fig. 
\ref{fig:fundu3m} this function behaves precisely as $u_{\text{bp}}(m)=2m^2$.

\subsubsection*{Limits to zero, AD and infinite mass}
As in the $\bfm=(m,m)$ $N_f=2$ theory, there are three interesting limits: $m\to 0$, $m\to \infty$ and $m\to \mad$. In the massless limit, we aim to recover $\Gamma_0(4)\backslash\mathbb H$. This is not difficult to see: Under $\Gamma_0(4)$, we can identify $TST^2S\CF$ with $ST^{-2}S\CF$, and similarly $TST^2\CF$ with $ST^{-2}\CF$, since the transition maps are in $\Gamma_0(4)$. This gives precisely Fig. \ref{fig:fund_dom_gamma_04}.

By decoupling the massive hypermultiplet, the theory flows to massless $N_f=2$. We find that $u_*\to \infty$, while $u_\pm\to \pm\frac{\Lambda_2^2}{8}$. From Section \ref{sec:masslessnf2} it is clear that the singularities $u_\pm$ do not move in $\tau$-space. The cusp region $TST^2S\CF$ is identified with $T\CF$ under the duality group $\Gamma(2)$ of massless $N_f=2$. 
Moreover, the remaining differing triangle $TST^2\CF$ can be mapped to $TS\CF$ using $\Gamma(2)$. This then gives precisely $\Gamma(2)\backslash\mathbb H$ as in Fig. \ref{fig:masslessNf2domains}. 

Finally, in the limit $m\to \mad=\tfrac{1}{16}\Lambda_3$ the singularities $u_*$ and $u_+$ collide. Since they drop out of the curve, we should remove all regions near those cusps. In Fig. \ref{fig:fundu3m} we can remove the triangles $TST\CF$, $TST^2\CF$ and $TST^2S\CF$, after which the index 3 group $\Gamma_0(2)$ remains. This is precisely what is found as the duality group of the $\bfm=(\mad,0,0)$ theory, as shown in Fig. \ref{fig:fund_dom_gamma_02F}. The pre-image of the merged non-local singularities $\uad$ is the point $\tad$, which lies in the interior of $\BH$ and corresponds to the point where the branch points have collided.

\subsection{Generic masses} 
For generic masses $\bfm=(m_1,m_2,m_3)$, the order parameter reads
\begin{equation}\label{genericunf3}
\begin{aligned}
\frac{u}{\Lambda_3^2}&=-\frac{1}{2^{12}}\bigg(\frac 1q +(-8 + 4096  M_2) \\&+ 
 4 \left(5 + 32768  M_2 + 3670016 M_3 - 4194304  M_4 - 4194304  M_4'\right) q+\CO(q^2)\bigg),
	\end{aligned}
\end{equation}
where the coefficients $M_i$ are the symmetric polynomials defined in \eqref{nf3masspoly} for the variables $\frac{m_i}{\Lambda_3}$. There are five generally distinct singular points. 

Due to the $N_f=3$ distinct branch points on the $u$-plane, the fundamental domain for a given mass $\bfm$ has an intricate web of branch cuts. Furthermore, the fundamental domains $\CF_3(\bfm)$ change as $\bfm$ passes through $\CL_3^{\text{AD}}$ (see Fig. \ref{fig:ADlocusNf3mmu}). A fundamental domain $\CF_3(\bfm)$  can also change when $\bfm$ is varied such that $\Delta_3$ has any double root, and when branch points in $\mathbb H$ pass through the $\slz$ tessellation $\CT_{\mathbb H}$ \eqref{slztessellation}. 

For any given mass $\bfm$ one easily computes $T_3$ from \eqref{imj}, and truncates the plot of the level set to the region where $\CJ(u,\bfm,\Lambda_3)\leq 12^3$. The branch points $u_{\text{bp}}$ are the zeros of $P_3^{\text{M}}$ \eqref{pnf}. On the upper half-plane $\mathbb H$, a branch point $\tau_{\text{bp}}$ is any of the $\slz$-images of $(j|_{\CF})^{-1}(\CJ(u_{\text{bp}}))$. When $\CJ(u_{\text{bp}})\leq 12^3$, then obviously $\tau_{\text{bp}}\in \mathbb \CT_{\mathbb H}$. If $\CJ(u_{\text{bp}})> 12^3$, then $\tau_{\text{bp}}\in \slz \cdot \im \mathbb R_{>0}$. Lastly, if  $\CJ(u_{\text{bp}})\not\in\mathbb R$, then $\tau_{\text{bp}}$ is an interior point of an $\slz$ copy of $\CF$.

\begin{figure}
	\begin{subfigure}{.49\textwidth}
		\centering
		\includegraphics[width=\linewidth]{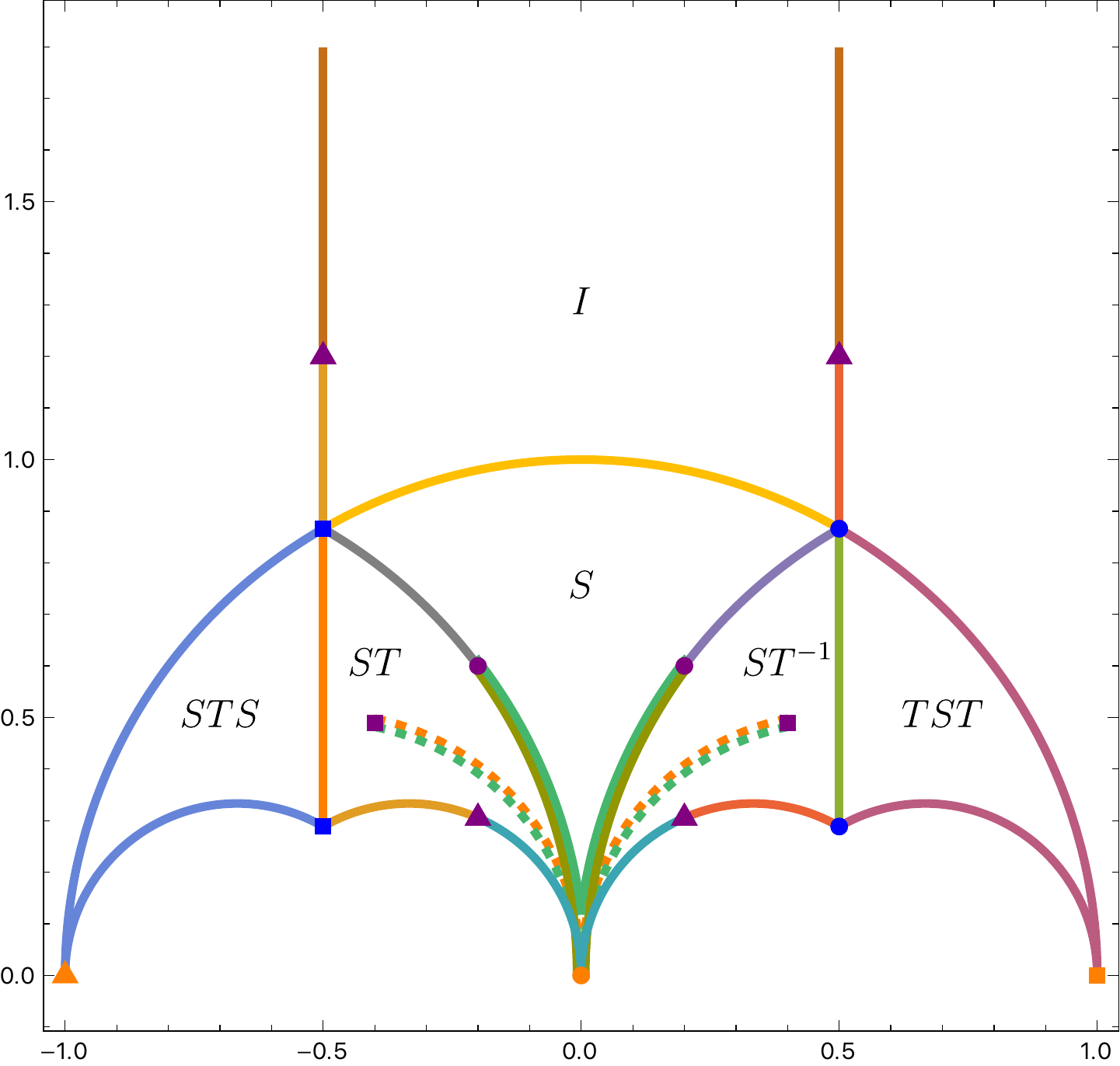}  
			\end{subfigure}
	\begin{subfigure}{.49\textwidth}
		\centering
		\includegraphics[width=\linewidth]{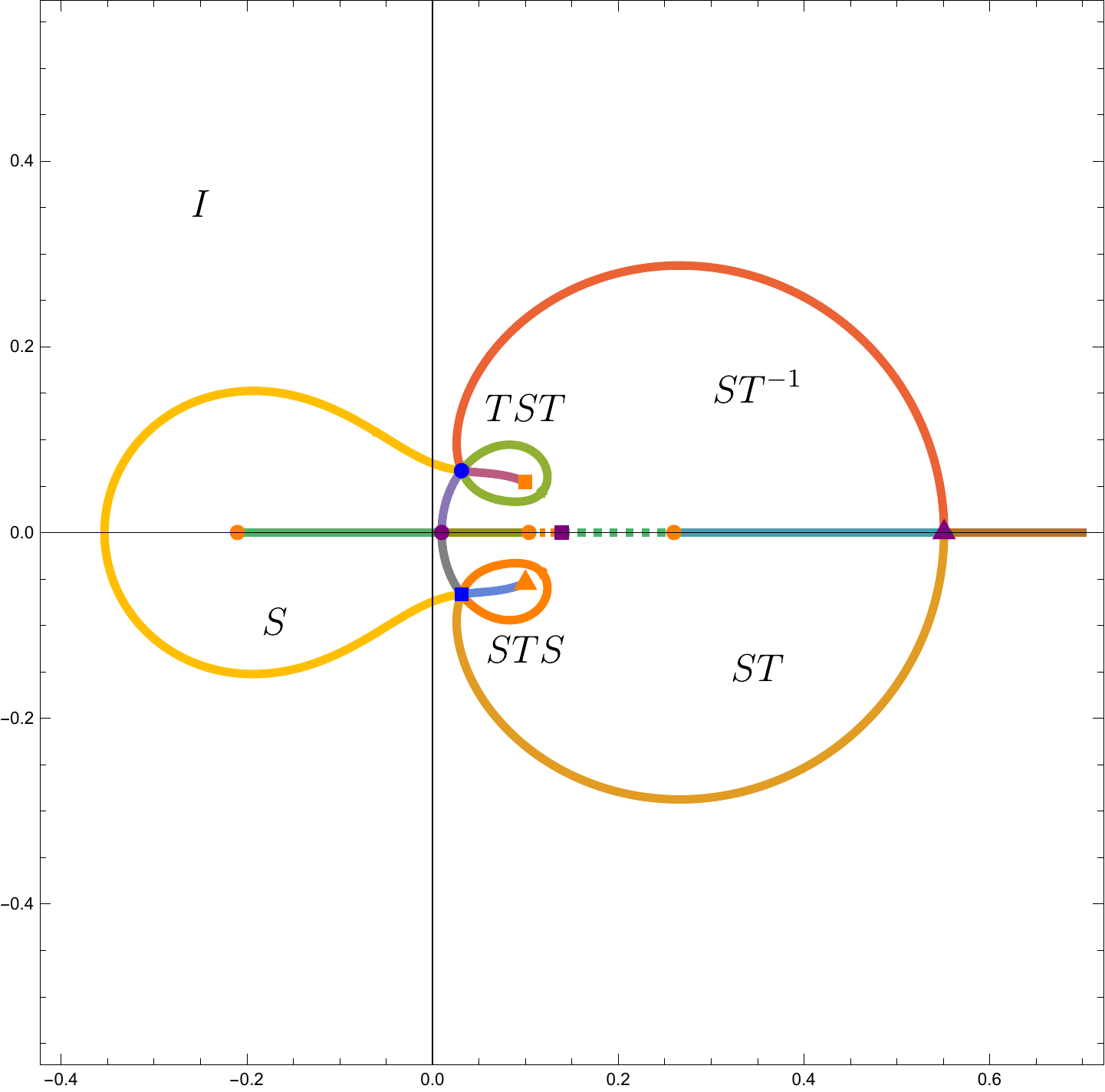}  
\end{subfigure}
	\caption{Identification of the components of the partitioning $\CT_{(m_1,m_2,m_3)}$ in $N_f=3$ for the particular choice $\mu_1=\frac{1}{10}$, $\mu_2=\frac{3}{10}$ and $\mu_3=\frac{5}{10}$. The $u$-plane $\CB_3$ is  partitioned into five regions $u(\alpha\CF)$, as those for $ST^{-1}$ and $ST$ are glued by two branch cuts. The fundamental domain is given by six copies of $\CF$, with three pairs of branch points (purple). Two branch points (triangle and disk) lie on $\CT_3$. The third (square) lies in the interior and glues the copies $ST\CF$ and $ST^{-1}\CF$. A natural choice for the branch cuts not lying in $\CT_3$ (dashed) is on the real axis in the $u$-plane, for which the path in $\mathbb H$ is along the tessellation $\slz\cdot \im\mathbb R_{>0}$. The singularity at $\tau=0$ does not have width $3$ as apparent, since due to the branch cuts $u(\tau)$ assumes three different values depending on which side of the branch cut $\tau=0$ is approached from.}
	\label{fig:nf3m1m2m3partition}
\end{figure}

In Fig. \ref{fig:nf3m1m2m3partition} we plot the $u$-plane and corresponding fundamental domain for three distinct masses. The five distinct singular points are partitioned into five regions $u(\alpha\CF)$, where two of them are glued by branch cuts.

\subsection{The massless theory}\label{sec:nf3massless}
When sending $m\to0$ from above we find
\begin{equation}\label{nf3masslesssolution}
\begin{aligned}
\frac{u(\tau)}{\Lambda^2}=&-\frac{1}{64}\frac{\vartheta_3(\tau)^2\vartheta_4(\tau)^2}{(\vartheta_3(\tau)^2-\vartheta_4(\tau)^2)^2}=-\frac{1}{2^{12}}\left(\frac{\eta(\tau)}{\eta(4\tau)}\right)^8 \\
=&-\frac{1}{2^{12}}(q^{-1}-8+20q-62q^3+\CO(q^{9/2})).
\end{aligned}
\end{equation}
It is the completely replicable function of class 4C and a Hauptmodul for $\Gamma_0(4)$ \cite{Alexander:1992,ford1994,Ferenbaugh1993}.
The physical discriminant is $\Delta=u^4(u-\frac{\Lambda_3^2}{2^8})$, and one finds that the singularities are located at $u(0)=0$ and $u(\tfrac12)=\frac{\Lambda_3^2}{2^8}$. 
At $\tau=0$ a dyon with charge $(1,0)$ becomes massless, while at $\tau=\frac12$ one finds instead that a dyon with charge $(2,1)$ becomes massless. The massless $N_f=3$ $u$-plane has no global symmetries. 

\begin{figure}[h]\centering
	\includegraphics[scale=0.8]{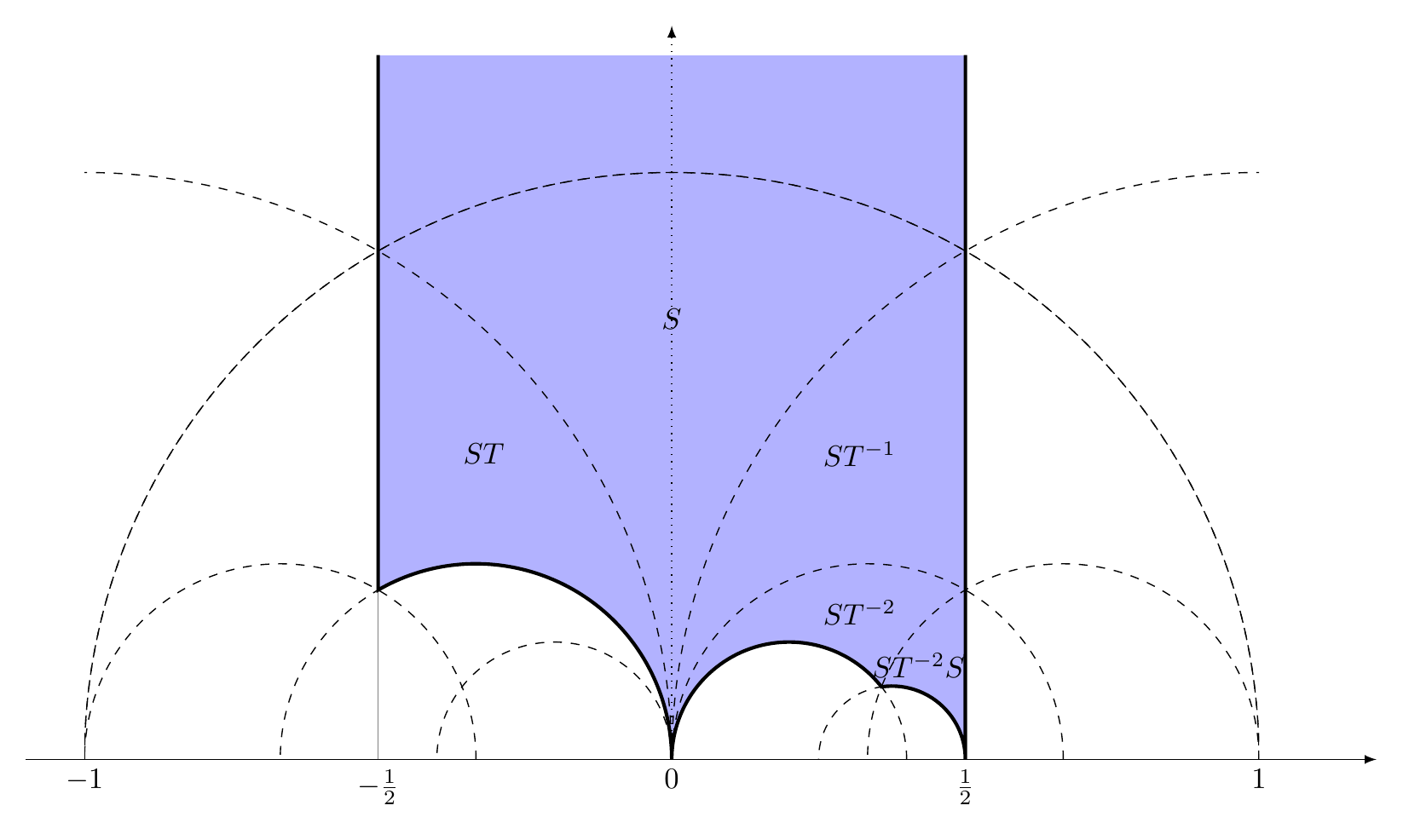}
	\caption{Fundamental domain of $\Gamma_0(4)$, the duality group of massless $N_f=3$. The cusp at $\tau=0$ has width four, while the cusp at $\tau=\tfrac{1}{2}$ has width one. }\label{fig:fund_dom_gamma_04}
\end{figure}

A choice of fundamental domain for $\Gamma_0(4)$ is
\begin{equation}
	\CF_{3}(0,0,0)=\CF\cup S\CF\cup ST\CF\cup ST^{-1}\CF\cup ST^{-2}\CF\cup ST^{-2}S\CF,
\end{equation}
and is shown in Fig. \ref{fig:fund_dom_gamma_04}.

\subsection{Type $IV$ AD mass}\label{31ADtheroy}
As illustrated in Fig. \ref{fig:ADlocusNf3mmu}, on the generic mass $N_f=3$ $u$-plane, there is not only the $IV$ AD point but also a variety of $III$ and $II$ points. We will give a few explicit examples of the $u$-plane of the theories with masses tuned to these specific values, starting with the most symmetric case. 

For $\bfm=(m,m,m)$ and $m=\frac18\Lambda_3$, four mutually non-local singularities collide in $u_{\text{AD}}=\frac{1}{32}\Lambda_3^2$. The remaining singularity is $u_0=-\frac{19}{2^8}\Lambda_3^2$ and never collides with the other four. The physical discriminant is $
\Delta=(u-u_0)(u-u_{\text{AD}})^4$. One finds that 
\begin{equation}\label{unf3AD}
\frac{u}{\Lambda_3^2}=-\frac{j^*+304}{2^{12}},
\end{equation}
where 
\begin{equation}\begin{aligned}\label{ramanujansato}
j^*&=432\frac{\sqrt j+\sqrt{j-1728}}{\sqrt j-\sqrt{j-1728}}=432\frac{E_4^{\frac32}+E_6}{E_4^{\frac32}-E_6}\\
&=\frac{1}{q}-120+10260 q-901120 q^2+91676610 q^3+O\left(q^{4}\right)
\end{aligned}\end{equation}
is the Ramanujan-Sato series of level 1 \cite{Almkvist2013,chan2004,chan2011}.\footnote{The Ramanujan-Sato series generalise Ramanujan's formula for $\tfrac1\pi$ as a series of quotients of modular forms. They exist for level 1 up to 11 and beyond. The level 1 series is the only one whose generating function can not be expressed by an $\eta$-quotient \cite{chan2012}.}

Inverting \eqref{ramanujansato} we find
\begin{equation}\label{nf3ADcurvej}
j=\frac{(j^*+432)^2}{j^*}.
\end{equation}
Using this and a discussion similar to the massless $N_f=1$ case for the transformations of $j^*$ we find that the singularities are located at ($\omega_j=e^{2\pi \im/j}$)
\begin{equation}
	u(\im\infty)=\infty, \quad u(\omega_3)=u_{\text{AD}}, \quad u(0)=u_0.
\end{equation}
We can read off from \eqref{nf3ADcurvej} that $\ord(g_2,g_3,\Delta)=(2,2,4)$ at the AD point, such that according to Table \ref{kodaira} we indeed have a singular fibre of Kodaira type $IV$ \cite{Argyres:2015ffa}.

\begin{figure}
	\begin{subfigure}{.5\textwidth}
		\centering
		\includegraphics[width=\linewidth]{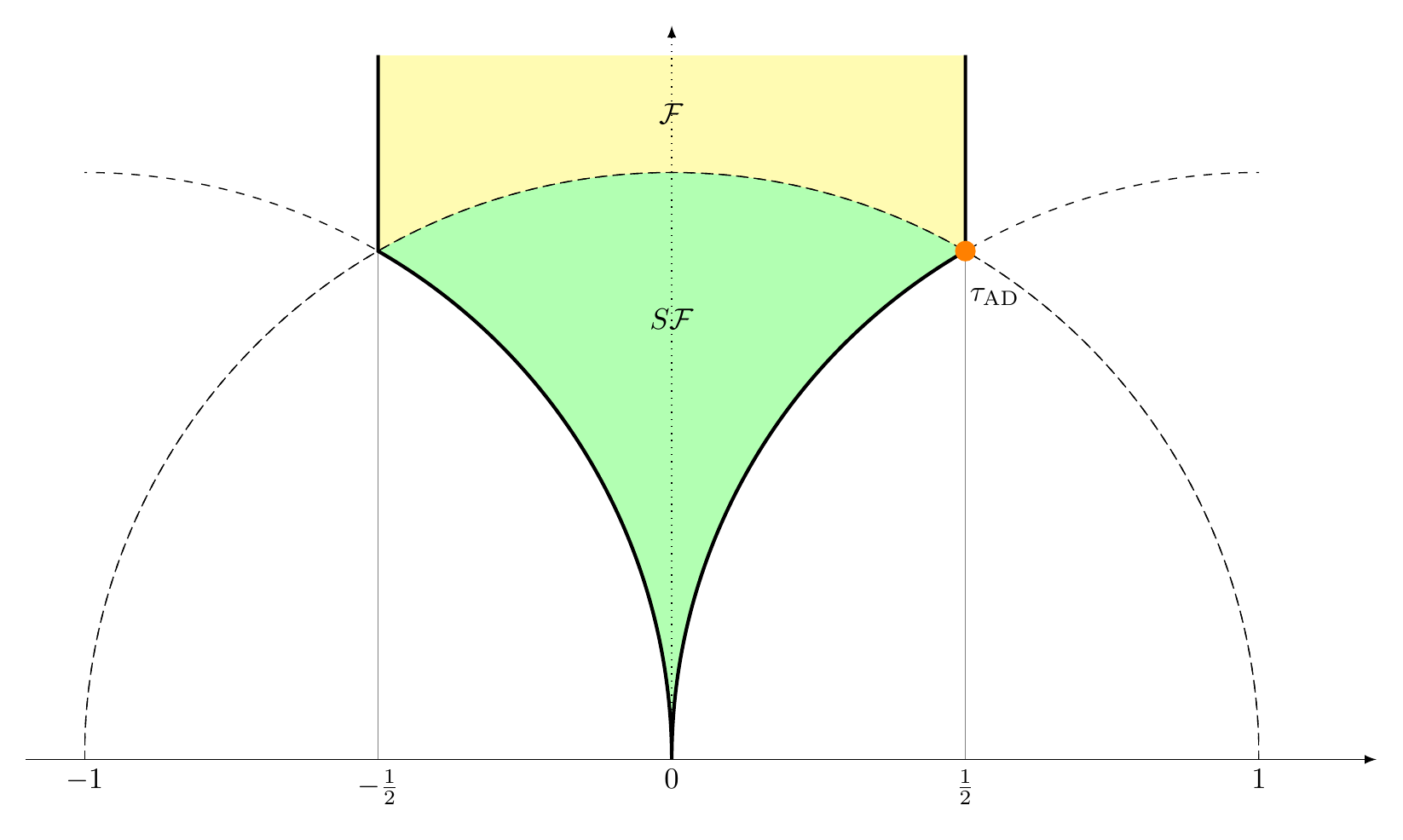}  
	\end{subfigure}
	\begin{subfigure}{.5\textwidth}
		\centering
		\includegraphics[width=.6\linewidth]{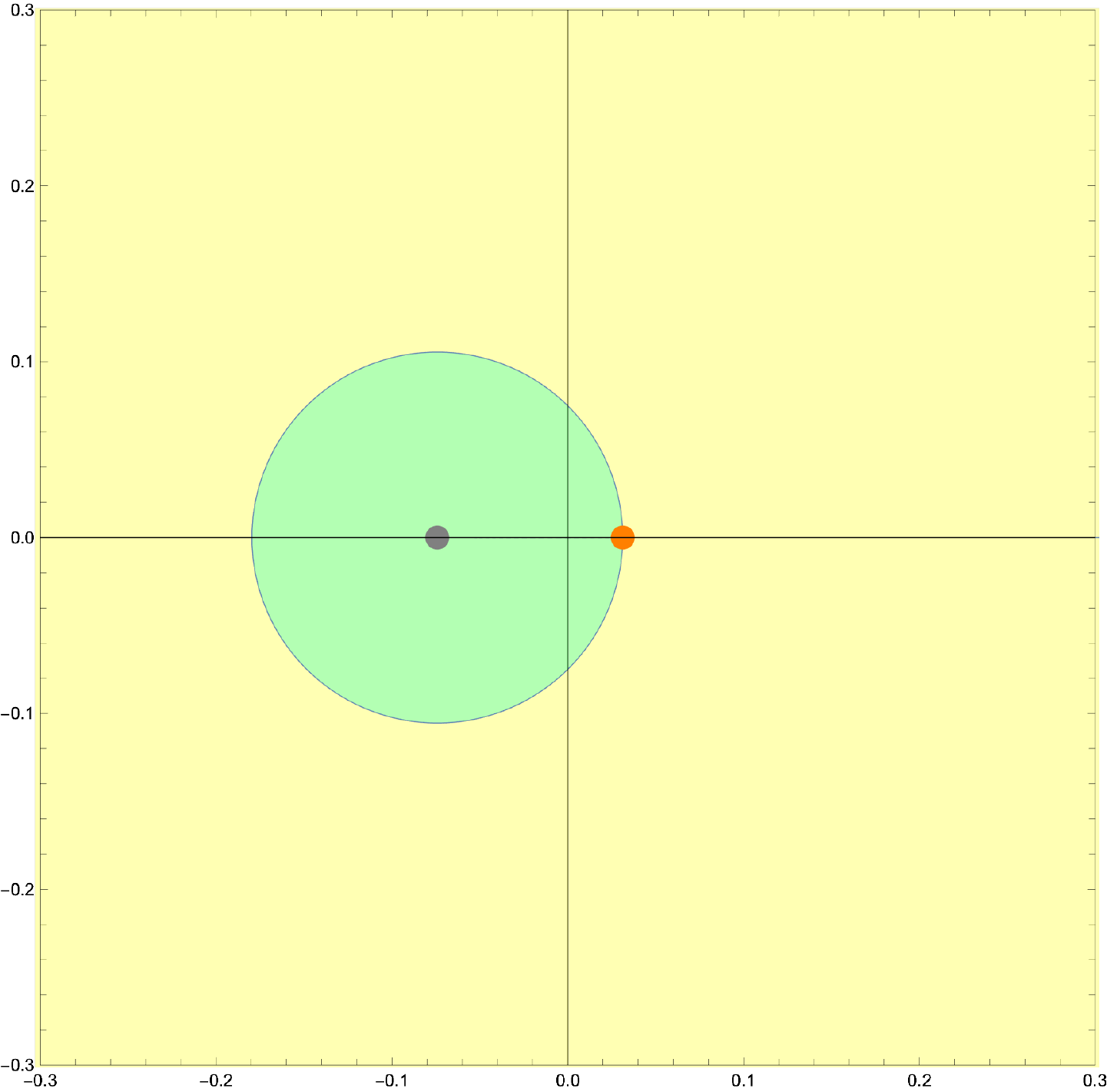}  
	\end{subfigure}
	\caption{Left: Proposed fundamental domain of the duality group $\Gamma_{j^*}$ of the $N_f=3$ $\bfm=\tfrac18(\Lambda_3,\Lambda_3,\Lambda_3)$ theory. It is a region with index $6-4=1+1$, elliptic fixed point $\omega_3$ and cusp at $0$. This uniquely fixes it. Right: Plot of the $u$-plane for the corresponding theory as the union of the images of $u$ under the $6-4=2$  $\slz$ images of $\CF$ as in the left figure. The complex plane can clearly be covered by 2 triangles. This demonstrates that the proposal of the left figure is indeed a fundamental domain for \eqref{unf3AD}. The weak coupling region is covered by $u(\CF)$. The strong coupling region $u(S\CF)$ contains the singular point $u_0$ in its interior. The AD point (orange) lies on the boundary.}
	\label{fig:IVdomains}
\end{figure}

From \eqref{nf3ADcurvej} we read off that the duality group $\Gamma_{j^*}$ has index $2$, which is consistent with the previous cases in $N_f=1,2$ in that a factor of $(u-u_{\text{AD}})^4$ has cancelled from $g_2^3$ and $\Delta$, and therefore does not contribute to the index. The fundamental region of $u$ is therefore of index 2 with $0$, $\omega_3$ and $\im\infty$ on its boundary. However, there is no index 2 subgroup of $\slz$ with two distinct cusps \cite[Table 4.1]{schultz2015}.\footnote{In fact, there is exactly one index 2 subgroup of $\slz$ and it has only one cusp of width two. This group is sometimes referred to as $\Gamma_0(1)^*$ and is generated by $TS$ and $T^2$, and the Hauptmodul is given by $\sqrt{j-1728}=8 \tfrac{(\jt_2^4+\jt_3^4)(\jt_3^4+\jt_4^4)(\jt_4^4-\jt_2^4)}{\jt_2^4\jt_3^4\jt_4^4}$.} This agrees with the fact that \eqref{unf3AD} is not a classical modular form and the monodromy group does not promote to a modular group since its action on $u$ is not associative (see Section \ref{sec:nf1masslesscurve}). We can nevertheless propose a fundamental region
\begin{equation}
\CF_{3}(\bfm_{\text{AD}})=\CF\cup S\CF,
\end{equation}
 see Figure \ref{fig:IVdomains}.

The monodromies are found by consistency,
\begin{equation}
m_0=STS^{-1}=\begin{pmatrix} 1&0\\-1&1\end{pmatrix}, \quad
m_{\text{AD}}=TS^{-1}=\begin{pmatrix}-1&1\\-1&0\end{pmatrix},
\end{equation}
and are unique in $\slz$.\footnote{The overall signs are fixed in the following way. The large $u$ monodromy is $PT^{-1}$. The monodromy at $m_0$ is oriented such that it conjugates  to $T$. This fixes the sign of $m_{\text{AD}}$ from the below relation.}
They fix $\tau=0$ and $\tad=e^{\pi\im/3}=\frac12+\frac{\sqrt3}{2}\im$, respectively, and produce the large $u$ monodromy $m_0 m_{\text{AD}}=PT^{-1}$. Just as in the massless $N_f=1$ case, we note that as matrices they do not form a congruence subgroup but instead generate the whole of $\slz$, since $m_\infty=PT^{-1}$ and $Tm_0T=S$. However, $u$ is not invariant under $S$.

\subsection{Type $III$ AD mass}\label{sec:nf3III}
\begin{figure}[h]\centering
	\includegraphics[scale=0.8]{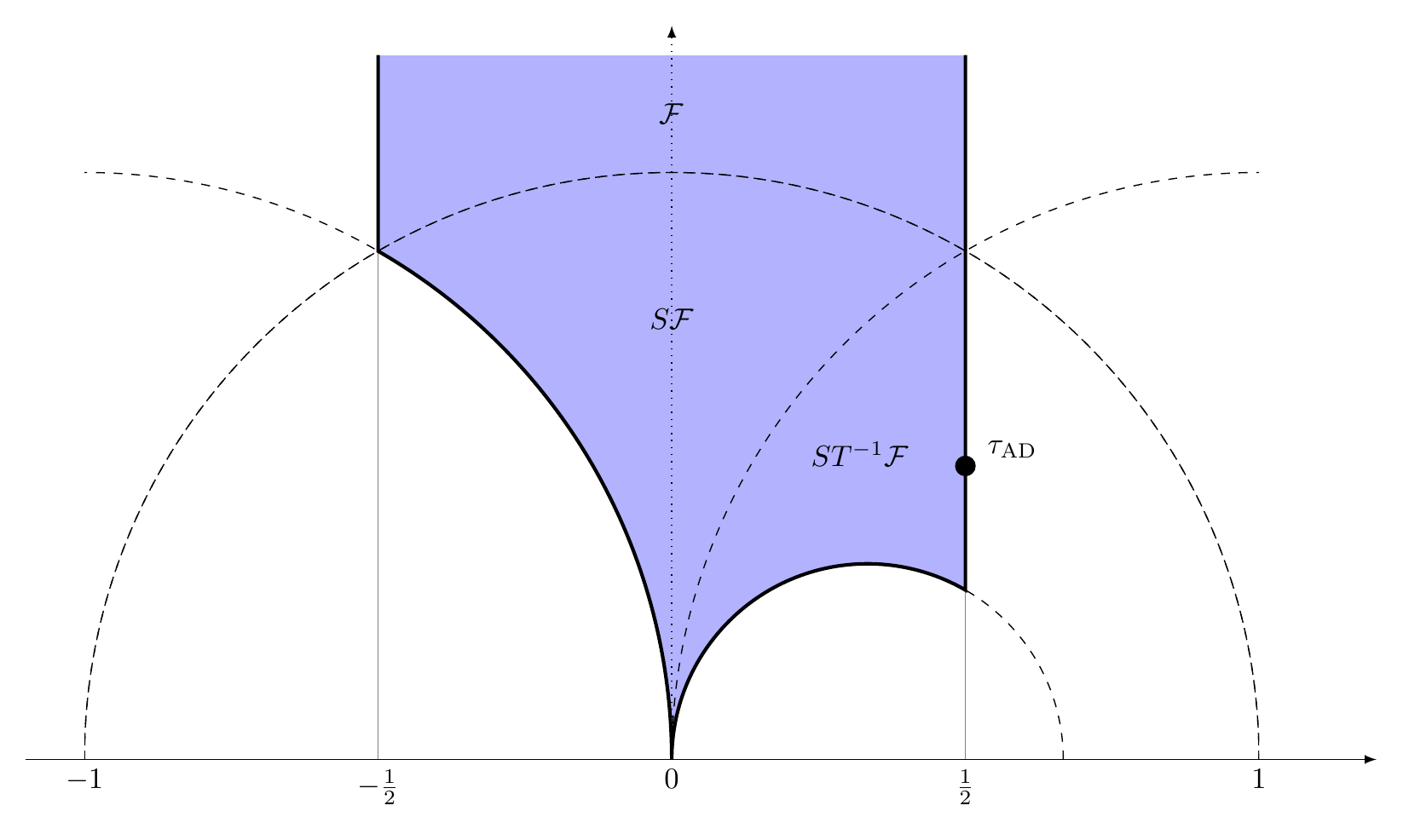}
	\caption{Fundamental domain of $\Gamma_0(2)$. This is the duality group of $N_f=3$ with mass $\bfm=\tfrac{1}{16}(\Lambda_3,0,0)$. The AD point is the elliptic fixed point of the domain, located at $\tad=\frac12+\frac\im2$. }\label{fig:fund_dom_gamma_02}
\end{figure}

In the single mass case $\bfm=(m,0,0)$ with $\mad=\frac{1}{16}\Lambda_3$, the $N_f=3$ curve has an AD point at $\uad=\frac{1}{128}\Lambda_3^2$.  The physical discriminant is $\Delta=(u-\uad)^3(u-u_0)^2$ with $u_0=-\frac{1}{128}\Lambda_3^2$, which is $-\uad$ by coincidence.
One easily finds 
\begin{equation}\label{uIIInf3}
\frac{u(\tau)}{\Lambda_3^2}=-\frac{f_{2\text{B}}(\tau)+32}{2^{12}},
\end{equation}
with $f_{2\text{B}}$ defined in \eqref{ubpnf2mfunction}. This fits nicely into the description as $\ind \Gamma_0(2)=3$ is equal to $6-3=3$, being the number of mutually non-local singularities collided at $\uad$. We find that $u(\tau)=u_0$ is equivalent to $f_{2\text{B}}(\tau)=0$, and one can easily show that $f_{2\text{B}}$ vanishes at the cusp $\tau=0$. Using the $S$-transformation of $\eta$, we can show that $\tad=\frac12+\frac\im2$. In terms of the Hauptmodul, the curve reads $
j=\frac{(f_{2\text{B}}+256)^3}{f_{2\text{B}}^2}$.
This shows that the AD point is indeed a type $III$ singularity. It also follows that $j(\tad)=12^3$ and that $\tad$ is in the $\slz$ orbit of $\im$. The duality group $\Gamma_0(2)$ is generated by $g_1=T$ and $g_2=ST^2S$. The AD point is stabilised by $g_1g_2\in \Gamma_0(2)$, which makes it an elliptic fixed point. A fundamental domain for $u$ is given in Figure \ref{fig:fund_dom_gamma_02}. The effective coupling at the AD point is explained through the fact that the $N_f=3$ branch point collides along with the three mutually non-local singularities in $\uad$, and the two branch points on the upper half-plane as drawn in Fig. \ref{fig:fundu3m} collide at $\tad$ for $m=\mad$. 

The monodromies are
\begin{equation}\label{gamma_02mono}
M_0=ST^2S^{-1}=\begin{pmatrix}1&0\\-2&1\end{pmatrix}, \quad M_{\text{AD}}=(TS)^{-1} S^{-1} TS=\begin{pmatrix}-1&1\\-2&1\end{pmatrix}.
\end{equation}
The first one describes the path around the cusp $\tau=0$, which has width $2$.  The AD monodromy is conjugate to $S^{-1}$, which fixes $\tau=\im$. The path is then given by the map $(TS)^{-1}:\im\mapsto \tad$.
The matrices \eqref{gamma_02mono}  satisfy $M_0M_{\text{AD}}=M_\infty$ with $M_\infty=PT^{-1}$ and $M_{\text{AD}}^2=\mathbbm 1$, such that $\tad$ is indeed an elliptic fixed point for $\Gamma_0(2)$ of order $2$. They are also related to \eqref{gamma^02mono} by conjugation with $\text{diag}(2,1)$, which induces the isomorphism between the $\Gamma_0(2)$ and $\Gamma^0(2)$ curves.

\subsection{Type $II$ AD mass}\label{sec:nf3II}
\begin{figure}[h]\centering
	\includegraphics[scale=0.8]{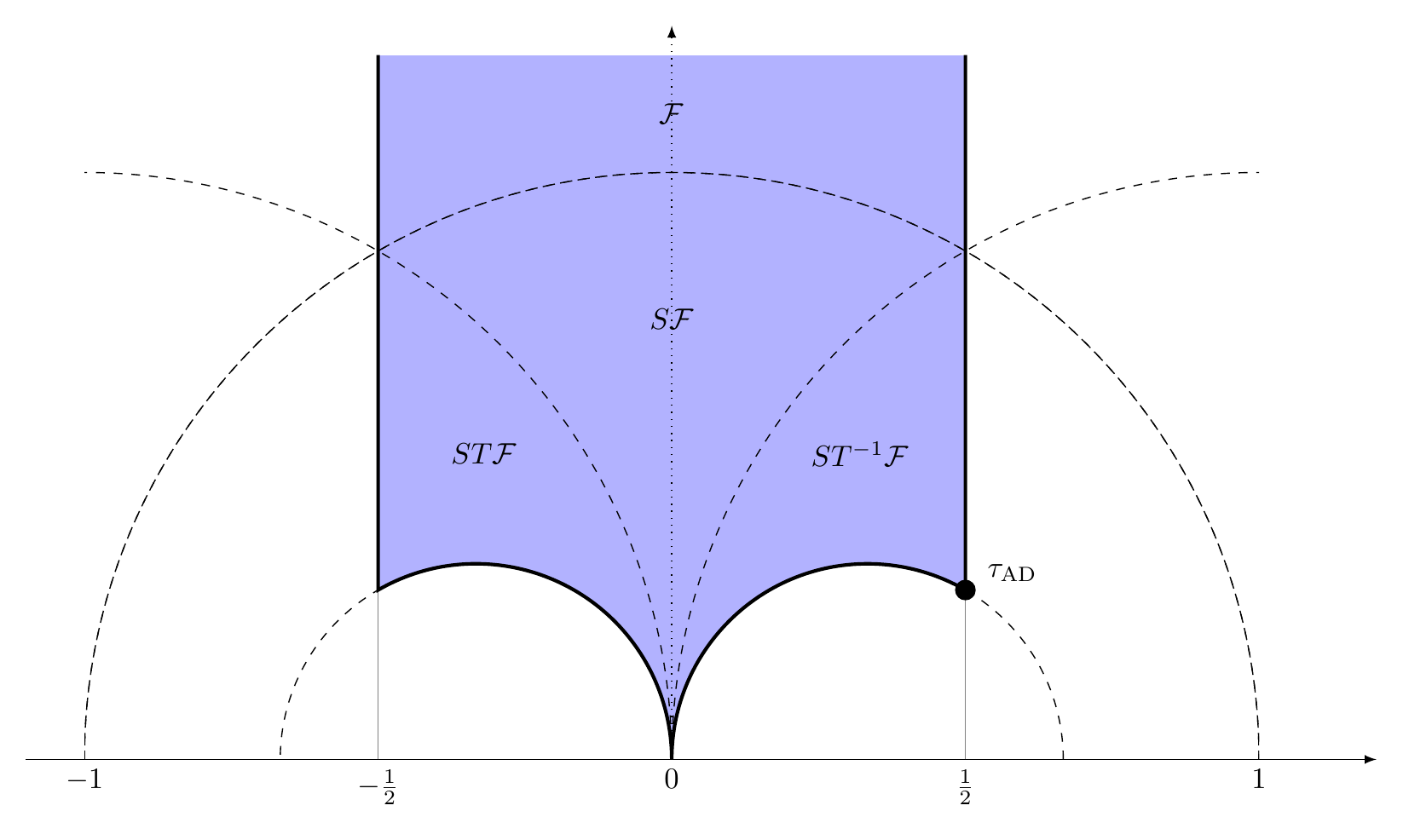}
	\caption{Fundamental domain of $\Gamma_0(3)$, the duality group of $N_f=3$ with mass $\bfm=-\tfrac{1}{64}(\Lambda_3,\Lambda_3,\Lambda_3)$. It has index 4 in $\psl$ and a type $II$ AD point located at the elliptic fixed point $\tad=\frac{\omega_{12}}{\sqrt3}=\tfrac12+\tfrac{1}{2\sqrt3}\im$. The width of the cusp $\tau=0$ is $3$.}\label{fig:fund_dom_gamma_03}
\end{figure}

On the equal mass $\bfm=(m,m,m)$ curve we can also tune the mass to $m=\mad=-\frac{1}{64}\Lambda_3$ to find a type II AD theory at $\uad=\frac{5}{2^{10}}\Lambda_3^2$. By fixing the mass to this value we find 
\begin{equation}
\frac{u(\tau)}{\Lambda_3^2}=-\frac{f_{3\text{B}}(3\tau)+7}{2^{12}}
\end{equation}
with  $f_{3\text{B}}$ given in \eqref{f3b}. At $\uad$ two mutually non-local singularities collide, while the other three reside at $u_0=-\frac{7}{2^{12}}\Lambda_3^2$. The physical discriminant is therefore $\Delta=(u-\uad)^2(u-u_0)^3$.
We know from Section \ref{sec:nf1ad} that $\tau\mapsto f_{3\text{B}}(3\tau)$ is a Hauptmodul for $\Gamma_0(3)$, and in fact the fundamental domain is just given by the one for $\Gamma^0(3)$ as in Fig. \ref{fig:11domains}, with every point divided by $3$. It also decomposes into $\slz$ images of $\CF$, see Fig. \ref{fig:fund_dom_gamma_03}.

The AD point $u(\tad)=\uad$ translates to $f_{3\text{B}}(3\tad)=-27$ which has $\tad=\frac{1}{\sqrt3}\omega_{12}$ as a solution (where $\omega_j=e^{2\pi\im/j}$). The other singularity satisfies $f_{3\text{B}}(3\tau)=0$ and therefore $\tau=0$. In terms of the Hauptmodul of $\Gamma_0(3)$ the $j$-invariant of the curve with above given mass $\bfmad$ reads $j=(f_{3\text{B}}+27)(f_{3\text{B}}+243)^3/f_{3\text{B}}^3$. This proves that the AD singularity is Kodaira type $II$ and therefore indeed equivalent to the $II$ theory in $N_f=1$, see Section \ref{sec:nf1ad}. It is interesting that both curves are parametrised by the same Hauptmodul, as the number of singularities on the curves  are different. 

The monodromies are given by 
\begin{equation}
M_0=ST^3S^{-1}=\begin{pmatrix}1&0\\-3&1\end{pmatrix}, \quad M_{\text{AD}}=(TS)^{-1}(ST)^{-1} (TS)=\begin{pmatrix}-1&1\\-3&2\end{pmatrix},
\end{equation}
which are just \eqref{11monodromies} conjugated by $\text{diag}(3,1)$. They furthermore satisfy $M_0 M_{\text{AD}}=M_\infty$ with $M_\infty=PT^{-1}$. Since $M_{\text{AD}}^6=\mathbbm 1$ in $\psl$, the AD point $\tad$ is an elliptic fixed point  in $\Gamma_0(3)$. Its stabiliser $M_{\text{AD}}$ decomposes into the monodromy $(ST)^{-1}$ around $\tau=\omega_3$, and the path $(TS)^{-1}: \omega_3\mapsto \tad$.

\section{Discussion}\label{sec:discussion}
We have studied the Coulomb branches $\CB_{N_f}$ of $SU(2)$ $\CN=2$
Yang-Mills theories with $N_f\leq 3$ massive hypermultiplets in the
fundamental representation. In particular, we have considered the order
parameter $u$ as function of the effective coupling, and derive
domains $\CF_{N_f}$ such that $u:\CF_{N_f}\to \CB_{N_f}$ is 1-to-1.
We find that generically the function $u$ has square roots appearing in the expressions for $u$, such that $\CF_{N_f}$
is not isomorphic to a domain $\Gamma\backslash \mathbb{H}$ for a
congruence subgroup of $\slz$. Nevertheless, exact expressions can be
determined, such as for $N_f=2$ with 2 equal masses, and $N_f=3$ with
one non-vanishing mass. For other special values, branch
points and cuts can be absent and the fundamental domain is that of a
modular curve for a congruence 
subgroups of $\SL$, as also encountered in cases in the literature \cite{Seiberg:1994rs, matone1996, Nahm:1996di,
  Ito:1995ga}.

We described how the order parameters are naturally expressed as roots
of certain degree six polynomials with modular functions as
coefficients. Many interesting aspects of the theories can be read off
from these polynomials:
\begin{itemize}
\item The degree of the polynomial tells us that the
fundamental domains of the order parameters can be described as six
copies of the ordinary $\SL$ domain. 
\item For the modular theories this
further implies that the duality group needs to be at most index six
in $\SL$. 
\item The discriminant of the sextic polynomials includes the
branch points as well as the superconformal AD fixed points of the
theories. 
\item We further discussed how one can explicitly construct
fundamental regions of order parameters as images of fundamental
domains in $\CF_{N_f}$. The
partitioning of the fundamental regions of the order parameters seem
to generalise aspects of the dessins d'enfants
\cite{juanzacarias2020,schuett2009,He:2013eqa,Ashok:2006br,Ashok:2006du,He:2012kw,Tatitscheff:2018aht,He:2014jva}
to the case of non-modular elliptic surfaces. 
\end{itemize}

Physically, the branch points and cuts provide a mechanism for $\CF_{N_f}$ to evolve as function of the mass.
This is most apparent in the limits where a hypermultiplet decouples
or multiple singularities coincide, where branch cuts appear to
``cut'' and ``glue'' regions of $\CF_{N_f}$. In particular near an AD
point, regions with non-local cusps are disconnected from
$\CF_{N_f}$. This makes it manifest that on the $u$-plane, not
only non-local singularities become coincident, but also branch
points, which ceases to be branch points in the limit because also the pre-images in $\CF_{N_f}$ have collided.

We believe that our methods can be adapted to many other
rank one theories, such as those of class $\CS$
\cite{Gaiotto:2009we,Gaiotto:2009hg}. The present analyses could
perhaps also be used to draw lessons about moduli spaces of other
theories, such as $\CN=2$ SYM with gauge group $SU(N)$ for $N>2$ or
Calabi-Yau compactifications in string theory, where in many cases
similar structures should arise. Remnants of which could perhaps be
seen in \cite{Aspman2021, Klemm:2012sx}. Lastly, we hope our methods
find applications in similar geometries such as F-theory
\cite{Hajouji:2019vxs} and 5d SCFTs \cite{Seiberg:1996bd}.  
Moreover, our findings may benefit the evaluation of the $u$-plane integral
\cite{Witten:1994cg,Moore:1997pc,Korpas:2017qdo, Moore:2017cmm,
  Korpas:2019cwg}. We aim to address this integral for massive $\CN=2$ QCD in future work \cite{AFM:future}. 

We would further like to mention to explore potential physical
consequences of the branch points. It is known that the AD points correspond to critical
points of a second order phase transition \cite{Bilal:1997st,
  Russo:2014nka, Russo:2019ipg}. It might then be natural to think of
the branch cuts in $\CF_{N_f}$, as in for example
Fig. \ref{fig:nf1mBP}, as boundaries over which a first order phase
transition takes place. Since branch points and cuts seem to be a
generic feature, it would suggest that similar
points appear in all theories with these kinds of superconformal fixed 
points. It would of course be very interesting to study this further
and we leave that for future work.

Another potential application is the $\CN=2$ QCD beta function.
In \cite{Dolan:2005mb, Dolan:2005dw}, Dolan gave a proposal for
the beta function of the massless $N_f=0,1,2,3$ theories, expanding on
the work of
\cite{Minahan:1996ws,Ritz:1997he,Kanno:1998qj,Carlino:1999tc,Konishi:2000nq}. For
the cases $N_f=0,2,3$ we can use the notation of the present paper to
collect these conjectured expressions as \footnote{compare with
  Eqs.(2.15), (3.11) and (4.21) of \cite{Dolan:2005dw},} 
\begin{equation}\label{betaDolan}
	\beta_{N_f} = -\widehat{\Delta}\frac{1}{u}\frac{d\tau}{du},
\end{equation} 
where $\widehat{\Delta}$ is the reduced discriminant of
\eqref{gcd}. In \cite{Dolan:2005dw} a shift is also made for $u$ in
$N_f=3$. For $N_f=1$ the argument is revised, basically due to the
square roots appearing in $u$ and the spoilage of modularity. We will
disregard these details in the following discussion. Using our
knowledge from the present paper it is now natural to conjecture that
the beta functions of the general massive theories, at least for
$N_f=2$ and 3, is given by  
\begin{equation}
	\beta_{N_f}= -\frac{\Delta}{\pma_{N_f}}\frac{1}{u}\frac{d\tau}{du}=\frac{4-N_f}{16\pi i}\frac{1}{u}\left(\frac{du}{da}\right)^2
\end{equation}
where the Matone relation was used in the second step. This obviously
gives back the expression \eqref{betaDolan} for the massless
$N_f=0,2,3$. It furthermore satisfies many good qualities, such as
being a weight $-2$ modular form in the cases where the theories are
modular. For the theories where we have explicit expressions for the
relevant quantities we can explicitly check that it has the correct
behaviour near the singular points. This is done in Appendix
\ref{sec:beta}. This proposal is speculative but would serve as
an interesting topic for further investigations.  
\vspace{.4cm}\\
{\it Note added}: While completing this paper, the work
\cite{Closset:2021lhd} by Closset and Magureanu appeared on the arXiv with partial overlapping results on modular
fundamental domains within $\CN=2$ QCD.

\acknowledgments
We are happy to thank Ling Long, Gregory Moore, Ken Ono and Edward Witten for
correspondence and discussions. JA is
supported by the Government of Ireland Postgraduate Scholarship
Programme GOIPG/2020/910 of the Irish Research Council. EF is supported by the TCD
Provost's PhD Project Award. JM is
supported by the Laureate Award 15175 “Modularity in Quantum Field 
Theory and Gravity” of the Irish Research Council. This research was supported in part by
the National Science Foundation under Grant No. NSF PHY-1748958 through the KITP program “Modularity in Quantum Systems”.

\appendix
\section{Elliptic curves and automorphic forms}\label{app:modularforms}
We collect some properties of modular forms for subgroups of
$SL(2,\mathbb Z)$ here. For further reading see for example \cite{Bruinier08,ono2004,Zagier92,koblitz1993,Diamond,schultz2015}.

\subsection{Modular forms}\label{sec:jacobitheta}
We make use of modular forms for the congruence subgroups $\Gamma_0(n)$ and $\Gamma^0(n)$  of $ SL(2,\mathbb Z)$. They are defined as 
\be\begin{aligned}
\Gamma_0(n) = \left\{\begin{pmatrix}a&b\\c&d\end{pmatrix}\in SL(2,\mathbb Z)\big| \, c\equiv0 \; \mod n\right\},\\
\Gamma^0(n) = \left\{\begin{pmatrix}a&b\\c&d\end{pmatrix}\in SL(2,\mathbb Z)\big| \, b\equiv0 \; \mod n\right\},
\end{aligned}\ee
and are related by conjugation with the matrix $\text{diag}(n,1)$. We furthermore define the \emph{principal congruence subgroup} $\Gamma(n)$ as the subgroup of $SL(2,\mathbb Z)\ni A$ with $A\equiv\mathbbm 1\mod n$. A subgroup $\Gamma$ of $\slz$ is called a congruence subgroup if there exists an integer  $n\in\mathbb N$ such that it contains $\Gamma(n)$. The smallest such $n$ is then called the \emph{level} of $\Gamma$.

The Jacobi theta functions $\vartheta_j:\mathbb{H}\to \mathbb{C}$,
$j=2,3,4$, are defined as
\be
\label{Jacobitheta}
\begin{split}
\vartheta_2(\tau)= \sum_{r\in
  \mathbb{Z}+\frac12}q^{r^2/2},\quad 
\vartheta_3(\tau)= \sum_{n\in
  \mathbb{Z}}q^{n^2/2},\quad
\vartheta_4(\tau)= \sum_{n\in 
  \mathbb{Z}} (-1)^nq^{n^2/2},
\end{split}
\ee
with $q=e^{2\pi i\tau}$. These functions transform under $T,S\in SL(2,\mathbb Z)$ as
\begin{alignat}{3}\nonumber
S:\quad& \vartheta_2(-1/\tau)=\sqrt{-i\tau}\vartheta_4(\tau),\quad&&\vartheta_3(-1/\tau)=\sqrt{-i\tau}\vartheta_3(\tau),\quad&&\vartheta_4(-1/\tau)=\sqrt{-i\tau}\vartheta_2(\tau)\\
T:\quad&\vartheta_2(\tau+1)=e^{\frac{\pi i}{4}}\vartheta_2(\tau),\quad &&\vartheta_3(\tau+1)=\vartheta_4(\tau),&&\vartheta_4(\tau+1)=\vartheta_3(\tau). \label{jttransformations}
\end{alignat}
They furthermore satisfy the Jacobi abstruse identity
\begin{equation}\label{jacobiabstruseidentity}
\vartheta_2^4+ \vartheta_4^4= \vartheta_3^4.
\end{equation}
Derivatives of modular functions are described by Ramanujan's differential operator. It increases the holomorphic weight by 2 and it can be explicitly constructed using the theory of Hecke operators \cite{ono2004}. 
For the derivatives of the Jacobi theta functions, one finds
\begin{equation}\begin{aligned}\label{jacobiderivatives}
D\jt_2^4&=\tfrac 16 \jt_2^4\left( E_2+\jt_3^4+\jt_4^4\right),\\
D\jt_3^4&=\tfrac 16 \jt_3^4\left( E_2+\jt_2^4-\jt_4^4\right),\\D\jt_4^4&=\tfrac 16 \jt_4^4\left( E_2-\jt_2^4-\jt_3^4\right),
\end{aligned}\end{equation}
where $D\coloneqq \frac{1}{2\pi\im}\frac{d}{d\tau}=q\frac{d}{dq}$ and $E_2$ is the quasi-modular Eisenstein series \eqref{Ek} of weight 2, transforming as \eqref{E2trafo}. 

The modular lambda function $\lambda=\frac{\jt_2^4}{\jt_3^4}$ is a Hauptmodul for $\Gamma(2)$.  
The Dedekind eta function $\eta: \mathbb H\to \mathbb C$ is defined as the infinite product
\begin{equation}\label{etafunction}
\eta(\tau)=q^{\frac{1}{24}}\prod_{n=1}^{\infty}(1-q^n), \quad q=e^{2\pi i\tau}.
\end{equation} 
It transforms under the generators of $SL(2,\mathbb Z)$ as
\be\begin{aligned}\label{etatransformation}
S: \quad& \eta(-1/\tau)=\sqrt{-i\tau }\, \eta(\tau),\\
T: \quad& \eta(\tau+1)=e^{\frac{\pi i}{12}}\, \eta(\tau),
\end{aligned}\ee
and relates to the Jacobi theta series as $
\eta^{3}=\frac{1}{2}\jt_2\jt_3\jt_4$.
The derivative of $\eta$ is given by $
\eta'=\frac{\pi\im}{12}\eta\, E_2$.

Another class of theta series is provided by the one of the $A_2$ root lattice, $b_{3,j}:\mathbb H\to \mathbb C$, 
\be\label{b3j}
b_{3,j}(\tau) = \sum_{k_1, k_2\in\mathbb Z+\frac j3}q^{k_1^2+k_2^2+k_1k_2}, \quad j\in\{-1,0,1\}.
\ee
It is clear that $b_{3,-1}=b_{3,1}$. The transformation properties under $\slz$ are 
\be 
\label{b3trafos}
\begin{aligned}
S: \quad b_{3,j}\left(-\frac1\tau\right)&=-\frac{i\tau}{\sqrt3}\sum_{l\mod 3}\omega_3^{2jl}\, b_{3,l}(\tau), \\
T: \quad b_{3,j}(\tau+1)&=\omega_3^{j^2} b_{3,j}(\tau).
\end{aligned} \ee 
The $b_{3,j}$ series can be expressed through the Dedekind eta function \eqref{etatransformation} as
\be \begin{aligned}\label{b301eta}
b_{3,0}(\tau)=\frac{\eta(\tfrac\tau3)^3+3\eta(3\tau)^3}{\eta(\tau)},\qquad 
b_{3,1}(\tau)=3\frac{\eta(3\tau)^3}{\eta(\tau)}.
\end{aligned} \ee
It furthermore relates to the quasi-modular Eisenstein series $E_2$ by
\begin{equation}\label{e2b30}
E_2(\tfrac\tau3)-3E_2(\tau)=-2b_{3,0}(\tfrac\tau3)^2.
\end{equation}
A relation to the Jacobi theta functions is given by 
\begin{equation}
b_{3,0}(\tau)=\jt_3(2\tau)\jt_3(6\tau)+\jt_2(2\tau)\jt_2(6\tau).
\end{equation}
Quotients of $\eta$-functions are frequently used to generate bases for the spaces of modular functions for congruence subgroups of $SL(2,\mathbb Z)$. We use the following

\begin{theorem}[\cite{ono2004,gordon1993}] Let $f(\tau)=\prod_{\delta|N}\eta(\delta\tau)^{r_\delta}$ be an $\eta$-quotient with $k=\frac 12\sum_{\delta|N}r_\delta\in \mathbb Z$ and $\sum_{\delta|N}\delta r_\delta\equiv \sum_{\delta|N}\frac{N}{\delta} r_\delta\equiv  0\mod 24$. Then, $f$ is a weakly holomorphic modular form for $\Gamma_0(N)$ with weight $k$.
\label{thm:etaquotients}
\end{theorem}

\subsubsection*{Eisenstein series}\label{sec:eisenstein}
We let $\tau\in \mathbb{H}$ and define $q=e^{2\pi i \tau}$. Then the Eisenstein series $E_k:\mathbb{H}\to \mathbb{C}$ for even $k\geq 2$ are defined as the $q$-series 
\be
\label{Ek}
E_{k}(\tau)=1-\frac{2k}{B_k}\sum_{n=1}^\infty \sigma_{k-1}(n)\,q^n,
\ee
with $\sigma_k(n)=\sum_{d|n} d^k$ the divisor sum. For $k\geq 4$ even, $E_{k}$ is a modular form  of weight $k$ for
$\operatorname{SL}(2,\mathbb{Z})$. On the other hand $E_2$ is a quasi-modular form, which means that the $\operatorname{SL}(2,\mathbb{Z})$ transformation of $E_2$ includes a shift in addition to the weight,
\be 
\label{E2trafo}
E_2\!\left(\frac{a\tau+b}{c\tau+d}\right) =(c\tau+d)^2E_2(\tau)-\frac{6\im}{\pi}c(c\tau+d).
\ee
From the $S$-transformation, we find that 
\begin{equation}\label{e4e6zeros}
E_4(e^{\pi\im/3})=0, \qquad E_6(\im)=0,
\end{equation}
and the zeros are unique in $SL(2,\mathbb Z)\backslash\mathbb H$ according to the valence formula for modular forms on $SL(2,\mathbb Z)$. Any modular form for $SL(2,\mathbb Z)$ can be related to the Jacobi theta functions \eqref{Jacobitheta} by 
\begin{equation}\label{e4e6jacobi}
E_4=\frac12(\jt_2^8+\jt_3^8+\jt_4^8),\qquad E_6=\frac12(\jt_2^4+\jt_3^4)(\jt_3^4+\jt_4^4)(\jt_4^4-\jt_2^4).
\end{equation}
All quasi-modular forms for $SL(2,\mathbb Z)$ can be expressed as polynomials in $E_2$, $E_4$ and $E_6$. The derivatives of the Eisenstein series are quasi-modular,
\begin{equation}\begin{aligned}\label{eisensteinderivative}
E_2'=\frac{2\pi\im}{12}(E_2^2-E_4), \quad
E_4'=\frac{2\pi\im}{3}(E_2 E_4-E_6), \quad
E_6'=\frac{2\pi\im}{2}(E_2 E_6-E_4^2).
\end{aligned}\end{equation}
These equations give the differential ring structure of quasi-modular forms on $\psl$.
With our normalisation \eqref{Ek} the $j$-invariant can be written as 
\begin{equation}\label{je4e6}
j=1728\frac{E_4^3}{E_4^3-E_6^2}=256\frac{(\jt_3^8-\jt_3^4\jt_4^4+\jt_4^8)^3}{\jt_2^8\jt_3^8\jt_4^8}.
\end{equation}

\subsection{Modular curves}\label{sec:modularcurves}
A subgroup $\Gamma$ of $\slz$ is a congruence subgroup if $\Gamma\supset \Gamma(N)$ for some $N\in\mathbb N$, which is called the \emph{level} of $\Gamma$. The  (projective) \emph{index} of a congruence subgroup $\Gamma$ is defined as 
\begin{equation}
	\ind \Gamma=[\psl:\Gamma],
\end{equation}
and it is finite for all $N$. By $\slz$ we strictly mean $\psl$ in the following. In fact, one can prove \cite{Diamond}
\begin{equation}\begin{aligned}\label{indexcongruence}
		\ind \Gamma(N)&=N^3\prod_{p|N}\left(1-\tfrac{1}{p^2}\right),\qquad 
		\ind \Gamma^0(N)&=N\prod_{p|N}\left(1+\tfrac{1}{p}\right),
\end{aligned}\end{equation}
where the sum is over all prime divisors of $N$. It can also be computed in the following way. The volume of the curve $\Gamma\backslash \mathbb H$ is defined as
\begin{equation}\label{volume}
	\vol(\Gamma\backslash\mathbb H)=\int_{\Gamma\backslash\mathbb H}\mathrm d\mu,
\end{equation}
where $\mathrm d\mu=y^{-2}\,\mathrm dx \mathrm dy$ is the hyperbolic metric on $\mathbb H$, with $\tau=x+\im y$. Since $\vol(\slz\backslash\mathbb H)=\frac\pi3$ can easily be computed, the index of any $\Gamma\subseteq\slz$ is then given by
\begin{equation}\label{indexvolume}
	\ind \Gamma=\frac3\pi \vol(\Gamma\backslash\mathbb H).
\end{equation}

Let $\Gamma$ be a congruence subgroup of $\slz$. Cusps of $\Gamma$ are $\Gamma$-equivalence classes of $\mathbb Q\cup \{\infty\}$. Adjoining coordinate charts to the cusps and compactifying gives the modular curve $X(\Gamma)\coloneqq \Gamma\backslash (\mathbb H\cup \mathbb Q\cup \{\im \infty\})$. The isotropy (stabiliser) group of $\infty$ in $\slz$ is the abelian group of translations,
\begin{equation}
	\slz_\infty=\left\{\left(\begin{smallmatrix}1&m\\ 0&1\end{smallmatrix}\right): m\in\mathbb Z\right\}.
\end{equation}
For each cusp $s\in\mathbb Q\cup\{\im \infty\}$ some $\delta_s \in\slz$ maps $s\mapsto \infty$. The \emph{width} of $s$ is defined as
\begin{equation}\label{widthofcusp}
	h_{\Gamma}(s)=\left|\slz_\infty/ (\delta_s \Gamma \delta_s^{-1})_\infty\right|.
\end{equation}
It can be proven that this definition is independent of $\delta_s$. For a fixed group $\Gamma$ it can be viewed as a well-defined function $ \mathbb Q\cup \{\im \infty\}\to \mathbb N_0$. It is straightforward to show that the sum over the widths of all inequivalent cusps $\mathcal C$ is equal to the index \cite{sebbar2001}
\begin{equation}\label{sumwidthindex}
	\sum_{s\in\mathcal C_\Gamma} h_{\Gamma}(s)=\ind \Gamma.
\end{equation}
The width of 0 (which is the level) is the lcm of all widths, and the width of $\infty$ is the gcd of all widths.

Other invariants of modular curves are the elliptic fixed points. A point $\tau\in\mathbb H$ is an \emph{elliptic point} for $\Gamma$ if its isotropy (stabiliser) group is nontrivial. The \emph{period} of $\tau$ is defined as the order of the isotropy group. It can be shown that any congruence subgroup of $\slz$ has only finitely many elliptic points, and the period for any point $\tau\in\mathbb H$ is 1, 2 or 3. 

\subsubsection*{Riemann-Hurwitz formula}
Let $f:X\to Y$ be a nonconstant holomorphic map between compact Riemann surfaces $X,Y$. It has a degree $n\in\mathbb N$, such that $|f^{-1}(y)|=n$ for all but finitely many $y\in Y$. More precisely, for each point $x\in X$ let $e_x\in \mathbb N$ be the ramification degree of $f$ at $x$, i.e. the multiplicity with which $f$ takes $0$ to $0$ as a map in local coordinates, making $f$ an $e_x$-to-1 map around $x$.  Then there exists a positive integer $n$ such that 
\begin{equation}
\sum_{x\in f^{-1}(y)}e_x=n
\end{equation}
for all $y\in Y$.
If $g_X$ and $g_Y$ are the genera of $X$ and $Y$, the Riemann-Hurwitz formula
\begin{equation}\label{RHrs}
2g_X-2=n(2g_Y-2)+\sum_{x\in X}(e_x-1)
\end{equation}
states that the Euler characteristic of $X$ is that of $Y$ multiplied by the degree $n$ of the cover, corrected by contributions from the ramification points. It is obvious that $g_X\geq g_Y$, otherwise $f$ is not holomorphic. 

This allows to compute the genus of a modular curve $X(\Gamma)$ for
any congruence subgroup $\Gamma\subseteq\slz$. For this, let $X=X(\Gamma)$ and $Y=X(1)$. Let $y_2=\slz\cdot \im$, $y_3=\slz\cdot e^{\frac{\pi\im}{3}}$ and $\varepsilon_2$ and $\varepsilon_3$ be the number of elliptic fixed points of period $2$ and $3$ for $X(\Gamma)$, and finally $y_\infty=\slz\cdot \infty$ and $\varepsilon_\infty$ be the number of cusps $X(\Gamma)$. Then $e_x-1$ is only nonzero  when $x\in f^{-1}(y_h)$ for $h=2,3,\infty$.
Since $g_{X(1)}=0$, it  follows from \eqref{RHrs} that 
\begin{equation}\label{riemannhurwitz}
g=1+\frac{n}{12}-\frac{\varepsilon_2}{4}-\frac{\varepsilon_3}{3}-\frac{\varepsilon_\infty}{2},
\end{equation}
where $g=g_{X(\Gamma)}$ and $n=\ind \Gamma$. See, for example,
\cite{Diamond} for a more detailed derivation of this formula.

Conjugacy classes of subgroups of $\slz$ can be classified by the data $(n,\varepsilon_\infty,\varepsilon_2,\varepsilon_3)$ together with the set of widths of the $\varepsilon_\infty$ cusps. For instance, the subgroups of $\slz$ of index $n=6$ have been completely classified \cite{alex2007sur,schultz2015}. There are precisely 22 subgroups that fall into 8 conjugacy classes, and they are listed in  Table \ref{tableindex6}. It has been shown in \cite{wohlfahrt1964} that every subgroup of $\slz$ with index $n \leq 6$ is a congruence subgroup. Examples of \emph{noncongruence} subgroups of $\slz$ of index $7$ have been constructed  already in the 19th century by Fricke. Hauptmoduln for low index subgroups of $\slz$ are studied in more detail in \cite{maier2006,maier2008}.

\begin{table}[h]\begin{center}
		\begin{tabular}{|>{$\displaystyle}l<{$}  >{$\displaystyle}l<{$} >{$\displaystyle}l<{$} >{$\displaystyle}l<{$} >{$\displaystyle}l<{$}  |}
			\hline 
			(\varepsilon_\infty,\varepsilon_2,\varepsilon_3)& \text{cusps}& \text{Hauptmodul}&|\text{conj}|&\\
			\hline
			(1,0,0) & 6& g=1& 1& \\
			(1,0,3)& 6 & j=-27x^3(x^3+16)&2&\\
			(1,4,0)&6& j=27(x^2+4)^3&3& \\
			(2,2,0)&3+3&j=\tfrac{x^3(x+12)^3}{(x+9)^3}&3&\\
			(2,2,0)&4+2& j=\tfrac{x^3(x+8)^3}{(x+4)^2}&3&\\
			(2,2,0)&5+1&j=\tfrac{(x^2+10x+5)^3}{x}&6 &\\
			(3,0,0)& 2+2+2& j=\tfrac{(x^2+192)^3}{(x^2-64)^2}&1&\\
			(3,0,0)&4+1+1& j=\tfrac{(x^2+48)^3}{x^2+64}&3& \\
			\hline 
		\end{tabular}
		\caption{Classification of conjugacy classes of index 6 subgroups of $\slz$ \cite{schultz2015}.  The first column gives the number of cusps and elliptic fixed points of order $2$ and $3$. According to the Riemann-Hurwitz formula \eqref{riemannhurwitz} all but the $(1,0,0)$ one induce genus 0 modular curves, whose Hauptmodul $x$ is expressed in terms of the $j$-invariant. In fact the group with signature $(1,0,0)$ is the only subgroup of $\slz$ of nonzero genus and index $\leq 7$. The second column gives the indices of the $\varepsilon_\infty$ number of  cusps, and the last column counts the number of conjugate groups with the same data. From the expression of $j=\tfrac{p(x)}{q(x)}$ as a rational function in $x$ we learn that $\deg p=n$ gives the index and the difference $\deg p-\deg q=h(\infty)$ gives the width of $\infty$. The roots of $q(x)=0$ then give the cusps in $\mathbb Q$ with widths provided by their corresponding multiplicity.} \label{tableindex6}\end{center}
\end{table}

\subsection{Kodaira classification}\label{sec:kodaira}
Let us consider an elliptic curve in Weierstra{\ss} form,
\begin{equation}\label{weierstrass}
y^2=4x^3-g_2 x-g_3,
\end{equation}
where $g_2$ and $g_3$ are functions of  some parameter $u$. This is the case for all SW curves \eqref{eq:curves}, where $g_2$ and $g_3$ are polynomials of degree 2 and 3 in $u$. The discriminant of the Weierstra{\ss} curve is $\Delta=g_2^3-27g_3^2$. The study of the elliptic curves on the discriminant divisor $\Sigma=\{u\,|\, \Delta(u)=0\}$ is the famous Kodaira classification. At a generic point in $\Sigma$, $g_2$ and $g_3$ do not vanish simultaneously. If we denote by $\ord\, f$ the order of vanishing (order of zero) of a polynomial or power series $f$ at a given point, then on a generic point in $\Sigma$ we have $\ord(g_2,g_3,\Delta)=(0,0,1)$. The various combinations of orders of vanishing of the invariants $g_2$ and $g_3$ are classified \cite{kodaira63II,kodaira63III}, see for a review
\cite{Weigand:2018rez}.
At special points in $\Sigma$ both $g_2=g_3=0$. From \eqref{weierstrass} we see that the curve becomes cuspidal, $y^2=x^3$. However the shape of the curve depends on the orders of vanishing of the respective quantities. Part of the classification is given in Table \ref{kodaira}.   The singular fibres of the Kodaira classification are all realised as curves of physical theories \cite{Martone:2020nsy,Argyres:2020wmq}.

\begin{table}[h]\begin{center}
\begin{tabular}{>{$\displaystyle}l<{$} | >{$\displaystyle}l<{$} |>{$\displaystyle}l<{$} |>{$\displaystyle}l<{$} }
\text{type}& \ord\, g_2&\ord\, g_3&\ord\, \Delta\\
   \hline
   I_0&\geq0&\geq0&0\\
   I_1& 0&0&1\\
   II  &\geq1&1&2 \\
   III& 1& \geq 2 & 3\\
   IV& \geq 2 & 2 & 4\\
   I_m&0&0&m \\
    \end{tabular}
\caption{Part of the Kodaira-Tate table for singular fibres of the Weierstra{\ss} model} \label{kodaira}\end{center}
\end{table}

\section{Additional proof of Matone's relation}\label{sec:matone2}

We can give a second proof of the Matone's relation \eqref{matonepnf}. 
As the behaviour of the polynomials \eqref{pnf} is highly constrained, they allow to give another proof of the $N_f\leq 3$ Matone relation \eqref{matonepnf}. For this, first note that the rhs $X\coloneqq \left(\tfrac{du}{da}\right)^2 \frac{du}{d\tau}$ has mass dimension $\Lambda_{N_f}^4$. In the massless $0\leq N_f\leq 3$ theories, $X$ can be written as $-\frac{16\pi\im}{4-N_f}\frac{\Delta_{N_f}}{Q_{N_f}}$ where $\Delta_{N_f}$ is the physical discriminant and $Q_{N_f}$ a polynomial, both monic in $u$. This fixes the overall normalisation. Since $\Delta_{N_f}$ has degree $N_f+2$ for any mass, dimensional analysis fixes $\deg Q_{N_f}=N_f$.  We are left with determining $Q_{N_f}$ and proving that indeed $Q_{N_f}=\pma_{N_f}$.

The relation 
\begin{equation}\label{Qmatoneansatz}
\frac{du}{d\tau}=-\frac{16\pi\im}{4-N_f}\frac{\Delta_{N_f}}{Q_{N_f}}\left(\frac{da}{du}\right)^2
\end{equation}
is RG invariant since its ingredients are. For instance, both $\frac{du}{d\tau}$ and $\frac{da}{du}$ flow from $N_f$ to $N_f-1$ by decoupling a hypermultiplet of mass $m_i$. This is not true for both $\Delta_{N_f}$ or $Q_{N_f}$, since their degrees as polynomials in $u$ are reduced by $1$ by decoupling a multiplet. However, it is easy to check that 
\begin{equation}\label{discrterm}
\Delta_{N_f}=u^{N_f+2}-u^{N_f+1}\sum_{i=1}^{N_f}m_i^2 +\CO(u^{N_f})
\end{equation}
as $u\to\infty$.
This implies that $\deg \lim_{m_i\to\infty}\frac{\Delta_{N_f}}{m_i^2}=\deg \Delta_{N_f}-1=\deg \Delta_{N_f-1}$. In fact, one can check explicitly that 
\begin{equation}
\lim_{m_i\to\infty}\frac{\Delta_{N_f}}{m_i^2}=-\Delta_{N_f-1}.
\end{equation}
The reason for this is that on the singular locus $\Delta_{N_f}=0$ there is always a singularity $u_*$ that behaves as $u_*\sim m_i^2$ for $m_i\to \infty$, it is the singularity that decouples (see also \cite{Tachikawa13} for a pictorial description). The relation \eqref{Qmatoneansatz} is then RG invariant if and only if  
\begin{equation}\label{qdecoup}
\lim_{m_i\to\infty}\frac{Q_{N_f}}{m_i^2}=-\frac{4-(N_f-1)}{4-N_f}Q_{N_{f-1}},
\end{equation}
for $N_f>0$. This condition does not give enough constraints on $Q_{N_f}$ in general to fully determine it. However,  \eqref{qdecoup} together with $\deg Q_{N_f}=N_f$, $[Q_{N_f}]=\Lambda_{N_f}^{2N_f}$ and the symmetry in the masses (or equivalently, the freedom of choice of $m_i$) restricts $Q_{N_f}$ to be a polynomial  where each coefficient has finitely many terms.  The decoupling \eqref{qdecoup} as well as limits to known examples fix all of these coefficients. For instance, for $N_f=0$ it is known that $Q_1=1$. For $N_f=1$, we can make a general ansatz
\begin{equation}
Q_1=u+c_1m_1^2+c_2 m_1 \Lambda_1+c_3\Lambda_1^2.
\end{equation}
The first term is constrained in that $Q_1$ is monic, and due to units only three $u$-independent terms remain. The decoupling \eqref{qdecoup} to $N_f=0$ implies that $c_1=-\tfrac43$. From the massless limit we find $c_3=0$. On $N_f=1$ there are three AD theories with mass $m_1=\tfrac34\omega_3$ (where $\omega_3^3=1$), for which $Q_{1}$ is known. This gives three additional equations for $c_2$, which all imply $c_2=0$. As anticipated, we find $Q_1=\pma_1$. 

For $N_f=2,3$ there are  more terms in a general ansatz for $Q_{N_f}$, however many are fixed from \eqref{qdecoup}, the other above mentioned constraints and the limits to zero and AD mass. One can check that $Q_2=\pma_2$ and $Q_3=\pma_3$. This gives a proof of \eqref{matonepnf} that is  independent to the one for \eqref{matoneg2g3}.

\section{Beta functions of massive theories}\label{sec:beta}
As discussed in Section \ref{sec:discussion}, our work suggests to generalise the proposal of \cite{Dolan:2005mb, Dolan:2005dw} to conjecture that the beta functions of the massive theories (at least for $N_f=2,3$) are given by
\begin{equation}\label{betaansatz}
	\beta_{N_f}= -\frac{\Delta}{\pma_{N_f}}\frac{1}{u}\frac{d\tau}{du}=\frac{4-N_f}{16\pi i}\frac{1}{u}\left(\frac{du}{da}\right)^2.
\end{equation}
Note that the second formula immediately implies that the beta functions vanish at the AD points, since we know that $\frac{du}{da}(\tau_{\text{AD}})=0$ and $u(\tau_{\text{AD}})$ is finite. This is of course the expected behaviour since these theories are superconformal. Note, however, that this is not the case for the proposal $\beta=-\frac{1}{u}\frac{d\tau}{du}$ given in \cite{Minahan:1996ws, Kanno:1998qj}, which is related to the critique raised in \cite{Carlino:1999tc,Konishi:2000nq}. We also have that, in the modular theories, $\frac{du}{da}$ is a weight $-2$ modular form of the corresponding monodromy group, and therefore we see that this also applies to $\beta_{N_f}$. We can make some further checks of this conjecture in the two cases $N_f=2$ with equal mass and $N_f=3$ with one non-zero mass where we have explicit expressions for all the relevant quantities. Similar analyses could be made for the theories where the mass has been tuned to the AD values, but we leave that for the interested reader. 

Using the formulas of Section \ref{sec:nf2} we find that the proposed beta function of equal mass $N_f=2$ is given by
\begin{equation}
	\beta_2=\frac{1}{8\pi i}\frac{1}{u}\left(\frac{du}{da}\right)^2=\frac{1}{\pi i}\frac{\jt_2^4+\jt_3^4+\sqrt{f_2}}{\jt_4^8+\jt_2^4\jt_3^4+(\jt_2^4+\jt_3^4)\sqrt{f_2}},
\end{equation}
with $f_2=\jt_4^8+16\frac{m^2}{\Lambda_2^2}\jt_2^4\jt_3^4$. For $N_f=3$ with one mass the formulas of Section \ref{sec:nf3} instead gives us
\begin{equation}
	\beta_3=\frac{1}{16\pi i}\frac{1}{u}\left(\frac{du}{da}\right)^2= \frac{1}{2\pi i}\frac{\jt_3^4+\jt_4^4+2\sqrt{f_3}}{2\jt_3^4\jt_4^4+(\jt_3^4+\jt_4^4)\sqrt{f_3}}
\end{equation}
with $f_3=\jt_3^4\jt_4^4+64\frac{m^2}{\Lambda_3^2}\jt_2^8$. Taking the massless limits in the above formulas of course gives back the expressions of \cite{Dolan:2005dw} (here we are disregarding the shift in $u$ for $N_f=3$).

Let us study the behaviour near the special points. For $\tau\to i\infty$ we immediately find 
\begin{equation}
	\begin{aligned}
		\beta_2 &= -\frac{i}{\pi}+\CO(q), \\
		\beta_3 &= -\frac{i}{2\pi}+\CO(q),
	\end{aligned}
\end{equation}
which is the expected behaviour \cite{Dolan:2005dw}. We also need to study the behaviour near the strong coupling singularities. For the two theories under study these singularities are given in Equations \eqref{discrnf2equal} and \eqref{nf3m00sing}, and the transformations to get the relevant dual expressions for $\beta_{N_f}$ are given in the corresponding sections. For $N_f=2$ the behaviour near each of these cusps is then
\begin{equation}\label{beta2D}
	\begin{aligned}
		\beta_{2,-}&=\frac{2i}{\pi}\frac{2m+\Lambda_2}{8m+\Lambda_2}+\CO(q), \\
		\beta_{2,*}&=-\frac{2i}{\pi}\frac{(2m-\Lambda_2)(2m+\Lambda_2)}{8m^2+\Lambda_2^2}+\CO(q^{1/2}), \\
		\beta_{2,+}&=\frac{2i}{\pi}\frac{2m-\Lambda_2}{8m-\Lambda_2}+\CO(q).
	\end{aligned}
\end{equation} 
For $N_f=3$ we instead find
\begin{equation}\label{beta3D}
	\begin{aligned}
		\beta_{3,-}&=\frac{i}{\pi}\frac{16m+\Lambda_3}{16m}+\CO(q^{1/2}), \\
		\beta_{3,*}&=-\frac{i}{2\pi}\frac{(16m-\Lambda_3)(16m+\Lambda_3)}{256m^2+\Lambda^2_3}+\CO(q), \\
		\beta_{3,+}&=\frac{i}{\pi}\frac{16m-\Lambda_3}{16m}+\CO(q^{1/2}).
	\end{aligned}
\end{equation}

We can now make the following observations. We can explicitly see that the beta functions vanish at the AD points by recognising that in $N_f=2$ equal mass we have AD points for $m=\pm\frac{\Lambda_2}{2}$, and tuning the mass to one of these values will merge either $u_+$ or $u_-$ with $u_*$, and we see that the corresponding contributions are exactly the ones that vanish for this specific value. Similarly, in $N_f=3$ with one mass the AD value is $m=\pm\frac{\Lambda_3}{16}$, and this will again merge either $u_+$ or $u_-$ with $u_*$, and the corresponding contributions exactly vanish. The beta functions also seem to blow up for specific values of the mass, these are exactly the values for which the corresponding singularity gets moved to $u=0$, for example if $m=0$ in $N_f=3$ $u_+$ and $u_-$ will meet at $u=0$. This behaviour at $u=0$ is of course already noted in \cite{Dolan:2005dw} and it is claimed that it corresponds to repulsive fixed points.

\providecommand{\href}[2]{#2}\begingroup\raggedright\endgroup

\end{document}